\documentclass[a4paper,11pt]{article}
\usepackage{jheppub} % for details on the use of the package, please
\usepackage[T1]{fontenc} % if needed  
\usepackage{amsmath}  
\usepackage{amssymb}  
\usepackage{latexsym}
\usepackage{enumitem}
\usepackage{lipsum}
\usepackage{graphicx}
\usepackage{subfig} 
\usepackage{empheq}
\usepackage{upgreek} 
\usepackage{cancel}
\usepackage{csquotes}
\usepackage{ytableau}
\usepackage{hyperref}
\usepackage{pifont}
\usepackage{csquotes}
\usepackage{ytableau}
\usepackage{blkarray}
\usepackage{multirow}
\usepackage{lscape}
\usepackage{float}

\usepackage{braket}
\usepackage{amssymb}
\usepackage{amsmath}
\usepackage{amsthm}
\usepackage{mathrsfs}
\usepackage{skak}
\usepackage{framed}
\usepackage{hyperref}
\usepackage{slashed}
\usepackage{array}

\hypersetup{colorlinks=true}
\hypersetup{linkcolor=violet}
\hypersetup{citecolor=violet}
\hypersetup{urlcolor=violet}
\numberwithin{equation}{section}

\usepackage
{datetime}
\usepackage{fancyhdr}
\usepackage{tikz, pgf}
\usetikzlibrary{shapes.misc}
\usetikzlibrary{shapes}
\usetikzlibrary{fit}
\usetikzlibrary{arrows}

\usetikzlibrary{decorations.pathmorphing}	% For Feynman Diagrams
\usetikzlibrary{decorations.markings}
 \usetikzlibrary{patterns}

\tikzset{
    sugra/.style={decorate, decoration={snake}, draw=black},
    scalarphi/.style={dashed,draw=black, postaction={decorate},
        },
    scalarchi/.style={draw=brown}, 
    hwbou/.style={draw=blue, postaction={decorate}, ultra thick
        },
    vector/.style={draw=blue,decorate, decoration={snake}, draw},
	provector/.style={decorate, decoration={snake,amplitude=2.5pt}, draw},
	antivector/.style={decorate, decoration={snake,amplitude=-2.5pt}, draw},
   	 fermion/.style={draw=cyan, postaction={decorate},
        decoration={markings,mark=at position .55 with {\arrow[draw=black]{>}}}},
    fermionbar/.style={draw=cyan, postaction={decorate},
        decoration={markings,mark=at position .55 with {\arrow[draw=black]{<}}}},
    fermionnoarrow/.style={draw=black},
    gluon/.style={decorate, draw=red,
        decoration={coil, amplitude=4pt, segment length=5pt}},
    scalar/.style={dashed,draw=black, postaction={decorate},
        decoration={markings,mark=at position .55 with {\arrow[draw=black]{>}}}},
    scalarbar/.style={dashed,draw=black, postaction={decorate},
        decoration={markings,mark=at position .55 with {\arrow[draw=black]{<}}}},
    electron/.style={draw=black, postaction={decorate},
        decoration={markings,mark=at position .55 with {\arrow[draw=black]{>}}}},
    scalarnoarrow/.style={dashed, draw=black},
    electron/.style={draw=black, postaction={decorate},
        decoration={markings, mark=at position .55 with {\arrow[draw=black]{>}}}},
	bigvector/.style={decorate, decoration={snake, amplitude=4pt}, draw},
    photon/.style={draw=violet, decorate, decoration={snake}, draw},
    higgs/.style={dashed, draw=black, postaction={decorate},
        },	
        goldstone/.style={draw=brown, postaction={decorate},
        },    
          ghost/.style={dashed, draw=magenta, postaction={decorate},
        decoration={markings, mark=at position .55 with {\arrow[draw=black]{>}}}
        },  
          antighost/.style={dashed, draw=magenta, postaction={decorate},
        decoration={markings, mark=at position .55 with {\arrow[draw=black]{<}}}
        }, 
            scalartwo/.style={dashed,draw=brown, postaction={decorate},
        decoration={markings,mark=at position .55 with {\arrow[draw=black]{>}}}},
    scalarbartwo/.style={dashed,draw=brown, postaction={decorate},
        decoration={markings,mark=at position .55 with {\arrow[draw=black]{<}}}}, 
    fermiontwo/.style={draw=purple, postaction={decorate},
        decoration={markings,mark=at position .55 with {\arrow[draw=black]{>}}}},
    fermionbartwo/.style={draw=purple, postaction={decorate},
        decoration={markings,mark=at position .55 with {\arrow[draw=black]{<}}}},    
        mphoton/.style={decorate, decoration={snake}, draw=violet},
        realscalar/.style={draw=black}, 
        fakerealscalar/.style={draw=white}, 
        realscalarone/.style={ draw=black},
    	realscalartwo/.style={draw=brown},    	    pseudoscalar/.style={draw=brown},
        mgluon/.style={decorate, draw=blue,
        	decoration={coil,amplitude=4pt, segment length=5pt}},
         weylfermion/.style={draw=orange, postaction={decorate},
        decoration={markings,mark=at position .55 with {\arrow[draw=black]{>}}}},
         weylfermionbar/.style={draw=orange, postaction={decorate},
        decoration={markings,mark=at position .55 with {\arrow[draw=black]{<}}}}, 
    majorana/.style={draw=cyan, postaction={decorate},
        decoration={markings,mark=at position .55 with {\arrow[draw=black]{><}}}},
    majoranabar/.style={draw=cyan, postaction={decorate},
        decoration={markings,mark=at position .55 with {\arrow[draw=black]{><}}}},    
   	wboson/.style={draw=blue,decorate, decoration={snake,amplitude=4pt}, draw},  
    zboson/.style={draw=violet, decorate, decoration={snake}, draw},   
    lepton/.style={draw=black, postaction={decorate},
        decoration={markings, mark=at position .55 with {\arrow[draw=black]{>}}}},
    leptonbar/.style={draw=black, postaction={decorate},
        decoration={markings, mark=at position .55 with {\arrow[draw=black]{<}}}}, 
    clepton/.style={draw=cyan, postaction={decorate},
        decoration={markings, mark=at position .55 with {\arrow[draw= black]{>}}}},
    cleptonbar/.style={draw=cyan, postaction={decorate},
        decoration={markings, mark=at position .55 with {\arrow[draw=black]{<}}}},   
   nlepton/.style={draw=orange, postaction={decorate},
        decoration={markings, mark=at position .55 with {\arrow[draw=black]{>}}}},
    nleptonbar/.style={draw=orange, postaction={decorate},
        decoration={markings, mark=at position .55 with {\arrow[draw=black]{<}}}},              
        graviton/.style={draw=blue, decorate, decoration={snake, amplitude=4pt}, draw},  
        spinj/.style={draw=red, decorate, decoration={snake, amplitude=4pt}, draw},  
        bgraviton/.style={draw=blue, decorate, decoration={snake, amplitude=4pt}, draw},  
        gravitino/.style={draw=red, postaction={decorate}, 
        decoration={snake,  markings, mark=at position .55 with {\arrow[draw=black]{><}}}},
    	gravitinobar/.style={draw=red, postaction={decorate},
        decoration={snake, markings, mark=at position .55 with {\arrow[draw=black]{><}}} },  
    phir/.style={draw=blue, postaction={decorate},},
   phil/.style={dashed,draw=blue,},
     phiav/.style={draw=cyan, postaction={decorate},},
   phidif/.style={dashed,draw=cyan,},  
    chir/.style={draw=red, postaction={decorate},},
   chil/.style={dashed,draw=red,},  
}

\tikzstyle{block} = [draw, rectangle, 
    minimum height=3em, minimum width=6em]
\usepackage{comment}

 \specialcomment{rudradetails}{%
  \begingroup \color{blue}}{%
  \endgroup}

\excludecomment{rudradetails}

\def\iimg{ {\bf i}}

\def\mathc{{ \mathbb{C} }}
\def\mathr{{ \mathbb{R} }}
\def\mathh{{ \mathbb{H} }}

\hbadness=\maxdimen
\vbadness=\maxdimen

\allowdisplaybreaks[1]

\newcommand{\setseq}[8]{
	\begin{tikzpicture}[node distance=0.5cm]
		\def\lastj{0}
		\node (l\lastj) [] {#2};
		\foreach \j[remember=\j as \lastj] in #1
		{
			\node (l\j) [below of=l\lastj] {\j};
		};

		\def\lastj{0}
		\node (m0) [right of=l0, xshift=2cm] {#4};
		\foreach \j[remember=\j as \lastj] in #3
		{
			\node (m\j) [below of=m\lastj] {\j};
		};
		
		\def\lastj{0}
		\node (r0) [right of=m0, xshift=2cm] {#6};
		\foreach \j[remember=\j as \lastj] in #5
		{
			\node (r\j) [below of=r\lastj] {\j};
		};
		
		\foreach \j/\k in #7
		{
			\draw[->] (l\j) -- (m\k);
		};
		
		\foreach \k/\l in #8
		{
			\draw[->] (m\k) -- (r\l);
		}
		
	\end{tikzpicture}
}

\title{\boldmath Three point interaction of Dirac fermions with higher spin particles and discrete symmetries}

\vspace{10mm}

\author[ \symbishop, \symrook ]{ Kushal Chakraborty}
\author[ \symrook]{,  Aakash Kumar }
\author[ \symrook ]{, Arnab Rudra }
\author[ \symknight ]{, Amey Yeole }
\author[]{\\ }
\vspace*{4.0ex} 
\vspace{10mm}
\affiliation[\symbishop]{National Institute of Theoretical and Computational Sciences,\\
School of Physics and Mandelstam Institute for Theoretical Physics,\\
University of the Witwatersrand, Wits, 2050, South Africa.\\ }
\vspace*{4.0ex} 
\affiliation[\symrook]{Indian Institute of Science Education and Research Bhopal,\\
 Bhopal Bypass Rd, Bhauri, Madhya Pradesh 462066, India.\\ }
\vspace*{4.0ex} 
\affiliation[\symknight]{Physikinstitut der Freie Universit\"{a}t,\\
 Arnimallee 14, 14195 Berlin, Germany.\\ }
\vspace*{4.0ex} 
\vspace*{2.0ex}
\emailAdd{kushal.chakraborty@wits.ac.za}
\emailAdd{aakash19@iiserb.ac.in}
\emailAdd{rudra@iiserb.ac.in} 
\emailAdd{amey.pravin.yeole@fu-berlin.de}  
\abstract{We constructed all possible kinematically allowed three-point interactions of two massless Dirac spinors with massive higher-spin bosons. In any $D$ spacetime, the interactions have been constructed using the projections of the higher spin irreducible representations of $Spin(D-1)$ over the product of two irreducible spinor representations of $Spin(D-2)$. Based on this analysis, we have further classified the space of theories involving two massless Dirac spinors and a single (or multiple) massive higher spin(s) based on the discrete symmetries: $C,\, R,$ and $ T$. We found that in any $D=2m+1/2m$, the interacting theories of a single massive higher spin have a \enquote{$m$} mod $2$ (or $D$ mod $4$) classification.}

\begin{document} 
\maketitle
%\raggedbottom   

\newpage
\section{Introduction}
\label{sec:ckryintro}

Recently, there has been some interest in understanding massive particles with spin greater than 2, commonly known as massive higher spin (MHS) particles. MHS particles are one of the ubiquitous features of any string theory model; in such cases, the masses of these particles are controlled by the string tension, see \cite{Polchinski:1998rq, Polchinski:1998rr, Green:1987mn}. On the other hand, these particles do not appear in the standard model of particle physics. In \cite{Camanho}, the authors pointed out the possible relevance of these particles in a more general context beyond string theory. It was argued that time advancement in higher derivative theories of gravity can be avoided by including the MHS particles. Since then, people have tried to explore these directions \cite{Haring:2023zwu, Caron-Huot:2021rmr, Caron-Huot:2020cmc, Caron-Huot:2020adz}.

Most of the work on MHS to date has focused on the bosonic particles. The fermionic particles are relevant both phenomenologically and conceptually. For example, the rich world of chemistry and most structure formation is due to the Fermi exclusion principle. In the standard model of particle physics, the fermions are responsible for the violation of various discrete symmetries like $C$ (charge conjugation), $R_i$ (spatial reflections)\footnote{$R_i$ is defined as the reflection along any $i^{th}$ spatial directions. $P$ are reflections along all spatial directions. We use $R_i$ and not $P$ because in odd spacetime, $P$ is in the connected part of the rotation group because of $det(P)=+1$. In even spacetime dimensions, $R_i$ and $P$ are connected by an element of the rotation group and both have $det(R_i)=det(P)=-1$.}, and $CR_i$. Before the discovery of $R_i$ and $CR_i$ violation, it was used to be believed that $C,~R_i,$ and $T$ (time reversal) are the symmetries of any theory. $CR_i$ violation is one of the necessary criteria for the matter-antimatter asymmetry in the observed universe. On the phenomenological side, the higher spin particles can provide various beyond-standard model EFTs. For example, one could ask: 1) what is the interaction of higher spin particles with standard model particles, \cite{Criado:2021itq}? 2) How does the beta functions of the standard model change in the presence of higher spin particles? Can such particles be any candidate for dark matter particles, \cite{Alexander:2020gmv}? From the study of bosonic MHS, we know that the presence of spin allows us to write many independent three-point interaction terms. As a result, the possible interactions and the landscape of higher spin interactions are much bigger. One possible tool to look at a smaller landscape is to impose supersymmetry. Supersymmetry always comes with fermionic particles, and the interactions of fermions become relevant.

In order to understand MHS particles, one can start by classifying the three-point functions. In $D=4$, the scattering amplitudes for all masses and spins were constructed in \cite{Arkani-Hamed:2017jhn}. A similar classification was done in the center of mass frame in \cite{JACOB2000774, TRUEMAN1964322} with more recent discussions in \cite{de_Rham_2018, hebbar2022spinningsmatrixbootstrap4d}. Moreover, the CFT techniques have also been used to compute the amplitudes using the CFT correlators \cite{Costa_2011,kravchuk2016countingconformalcorrelators, cuomo2017generalbootstrapequations4d}. The interactions of the MHS with photons and gravitons in arbitrary dimensions were classified in \cite{Chakraborty:2020rxf, Chowdhury:2019kaq}. Given the three-point functions, one would like to understand the scattering amplitudes in arbitrary dimensions. This problem is straightforward in principle. However, the explicit computation is cumbersome because the expression of the higher spin propagators is messy\cite{LPsingh1,LPsingh2}. This problem was circumvented in \cite{Balasubramanian2021}, using the derivative method in the kinetic space. The authors found the explicit expressions for the four particle scattering (of Gravitons and photons) due to the exchange of massive higher spin particles.

Classifying the supersymmetric interaction in higher dimensions remains an open problem. This is partly because the property of fermions depends on spacetime dimensions, and one also needs to understand the interaction of fermions. Thankfully, in the 3+1 dimension, one can use the spinor helicity variables and easily understand the consequences of supersymmetry. This was done in \cite{Balasubramanian2023}. However, one would like to understand supersymmetry in other dimensions \cite{Buchbinder:2021qrg}. In order to do that, one first needs to classify the three-point functions involving fermions. We undertake that job in this work.

In the case of fermions (both massive and massless), the basic building blocks are Majorana/Weyl fermions. On the other hand, it is simpler to classify the interaction of Dirac fermions. The primary reason behind this is twofold: 1) It is possible to define the action of all the discrete symmetries on Dirac fermions, which is not always true for chiral asymmetric theories of Weyl fermions. In such theories, the action of time reversal and reflections is not well defined. 2) A theory of Dirac admits the action of wick rotation. This is because the Dirac fermions are complex representations of the Clifford algebra (see appendix \ref{App:CliffordAlgebra}). In this work, we have classified the three-point function of two massless Dirac fermions and one massive higher-spin bosonic particle. We also write down the action of discrete symmetries $C$, $R$, and $T$ and their action on the three-point functions.

\subsection{Main results}
\label{sec:ckrymainresults}

In this section, we summarise the key results of this paper; the details of how to arrive at these results can be found in the later sections. We divided the discussion into parts based on whether the spacetime dimension is even or odd.
\newline
\newline
\textbf{Even dimensional spacetime:}
\begin{enumerate}
	\item The basic building blocks in this analysis are massless Dirac bilinears of the following form
\begin{equation}
	\begin{split}
		&(\iimg)^{s+\frac{(r+1)(r+2)}{2}}(\partial^{(\nu_1}\cdots\partial^{\nu_s)}\Psi^\dagger\beta)\Gamma^{[\mu_1}\cdots \Gamma^{\mu_r]}\Gamma_\star\Psi,
		\\
	&	(\iimg)^{s+\frac{r(r+1)}{2}}(\partial^{(\nu_1}\cdots\partial^{\nu_s)}\Psi^\dagger\beta)\Gamma^{[\mu_1}\cdots \Gamma^{\mu_r]}\Psi
	\end{split}
\end{equation}
\label{longhandexpressions1}
$r$ is called the rank of bilinear. The factors of $\iimg$ are to make sure that the term is hermitian. Because of the presence(absence) of $\Gamma_\star$ in the first (second) bilinear, we call it Chiral (non-chiral).

	\item A massive higher spin (MHS)\footnote{Unless explicitly mentioned a MHS means a quantum MHS field.} field is represented by the Young-Tableau, 
\begin{align}
 \begin{ytableau}
		\mu_1&\nu_1&\cdots&\nu_{\tilde{s}}\\ \mu_2&\rho_1&\cdots \\ \vdots 
	\end{ytableau} 
\end{align}	
Using the kinematics of the massless and massive Little groups, we show that two massless Dirac spinors couple to a massive higher spin fields only if the corresponding Young tableau is of the following form
\begin{align}
	\Phi^{(\tilde {r},\tilde {s})}\equiv \begin{ytableau}
		\mu_1&\nu_1&\cdots&\nu_{\tilde{s}}\\ \mu_2\\ \vdots \\ \mu_{\tilde{r}}
	\end{ytableau} 
\end{align}
i.e. the number of boxes in any row apart from the first row is at most one. We use the short-hand notation
$\Phi^{(\tilde{r},\tilde{s})}\equiv \Phi^{[\mu_1\cdots\mu_{\tilde{r}}]\{\nu_1\cdots\nu_{\tilde{s}}\}}$. The indices $\tilde{r}$ and $\tilde{s}$ denotes the antisymmetric and symmetric indices in $\Phi^{(\tilde{r},\tilde{s})}$. A flavored higher spin is denoted by $\Phi^{(\tilde{r}_a,\tilde{s}_a)}_a$, where \enquote{$a$} is the flavor index and $\tilde{r}_a,\tilde{s}_a$ holds the same meaning as $\tilde{r}$ and $\tilde{s}$.

	\item  For any massive higher spin (MHS) field other than scalars, there are \textit{four independent couplings} out of which two are non-chiral, and two are chiral. For scalar fields, there are only two independent couplings given by the chiral and non-chiral Yukawa terms. These couplings can be found in table \ref{tab:evencouplings}. In the paper, unless explicitly mentioned, the couplings of any MHS field with the even (or odd) ranked Dirac bilinears are termed as the even (or odd) ranked couplings.
	\item Using the $C,\, R_i,\, T$ theorem, we have shown that the intrinsic parities of any MHS field $\Phi^{(\tilde{r},\tilde{s})}$ must always satisfy (in both even and odd spacetime)
\begin{align}
	\eta_{_C}^{(\tilde{r},\tilde{s})}		\eta_{_R}^{(\tilde{r},\tilde{s})}		\eta_{_T}^{(\tilde{r},\tilde{s})}=+1\label{crt}
\end{align}
 where $C$ is the charge conjugation operation, $R_i$ are reflections along the $i^{th}$ spatial direction, and $T$ is the time reversal operation\footnote{The $C,R_i,T$ transformations of MHS $\Phi^{(\tilde{r},\tilde{s})}$ are summarized as follows
	\begin{align}
		C\Phi^{(\tilde{r},\tilde{s})}C^{-1}=\eta_{_C}^{(\tilde{r},\tilde{s})}\Phi^{(\tilde{r},\tilde{s})}\quad,\quad 	R\Phi^{(\tilde{r},\tilde{s})}R_i^{-1}=\eta_{_R}^{(\tilde{r},\tilde{s})}(\mathcal{R}_i)^{\tilde{r}+\tilde{s}}\Phi^{(\tilde{r},\tilde{s})}\quad,\quad 	T\Phi^{(\tilde{r},\tilde{s})}T^{-1}=\eta_{_T}^{(\tilde{r},\tilde{s})}(\mathcal{T})^{\tilde{r}+\tilde{s}}\Phi^{(\tilde{r},\tilde{s})}
	\end{align}
	where $\mathcal{R}_i$ and $\mathcal{T}$ are the spacetime representations of reflections and time reversal. All the $\eta's$ are $U(1)$ phases.
}.
Equation \eqref{crt} is a higher spacetime analogue of constraints on intrinsic parity, given in equation $5.8.4$ of Weinberg Vol 1 \cite{Weinberg:1995mt}.

	\item  Consider multiple MHS $\Phi^{(\tilde{r}_a,\tilde{s}_a)}_a$ where $a=1,2,\cdots,N$ coupled with a massless Dirac spinor. We have shown that if the MHSs couple only with a chiral (or non-chiral) bilinear of even rank $r=2n$, then the theory will always preserve $C,\, R_i,\, T$, such that the intrinsic parities of every MHS are given by
\begin{align}
	\eta_{_C}^{(\tilde{r}_a,\tilde{s}_a)}=\eta_{_T}^{(\tilde{r}_a,\tilde{s}_a)}=	\eta_{_R}^{(\tilde{r}_a,\tilde{s})}=+1,\qquad\forall ~ a \label{intrinsic}
\end{align}

\item In a $C,R_i,T$ preserving theory of multiple MHS coupled only with a non-chiral bilinear of odd rank $r=2n+1$ and $s$ derivatives, the intrinsic parities of the MHSs must be given by
\begin{align}
	\eta_{_C}^{(\tilde{r}_a,\tilde{s}_a)}=\eta_{_T}^{(\tilde{r}_a,\tilde{s}_a)}=(-1)^{s+n+1}\quad,\quad 	\eta_{_R}^{(\tilde{r}_a,\tilde{s})}=+1,\qquad\forall ~ a \label{intrinsic1}
\end{align}	
Similarly,  In a $C,R_i,T$ preserving theory of these MHS coupled only with a chiral bilinear of odd rank $(r=2n+1)$, the intrinsic parities of the MHSs must be given by
\begin{align}		\eta_{_C}^{(\tilde{r}_a,\tilde{s}_a)}=-\eta_{_T}^{(\tilde{r}_a,\tilde{s}_a)}=(-1)^{s+n+m}\quad,\quad 	\eta_{_R}^{(\tilde{r}_a,\tilde{s}_a)}=-1,	\qquad\qquad \forall ~a \label{intrinsic2}
\end{align}
where $m=\lfloor{D/2}\rfloor$. It would be interesting to derive the same results using the action of $C,\, R_i,\, T$ over the Little group states, i.e., following arguments similar to chapter $2$ of Weinberg Vol 1 \cite{Weinberg:1995mt}.	
		\item Let us consider an interacting theory of a single MHS $\Phi^{(\tilde{r},\tilde{s})}$ interacting with a massless Dirac spinor through all possible three-point couplings. For such theories, we can state the following
	\begin{itemize}
		\item In any $D=2m$, the chiral and non-chiral couplings of a MHS always transform oppositely under $R_i,~CT$. Therefore, an interacting theory with both the\textit{ chiral and non-chiral three-point couplings of a MHS would always break $R_i$ and $CT$}. For example, spinor QED breaks $R_i$ and $CT$ if we also include the coupling with $\Gamma_\star$. This is generally true in even dimensions, even if we keep the coupling constants with the least number of derivatives (often referred to as the minimal coupling). Our result generalizes the statement to three-point couplings of any MHS.
		\item 	In $D=4k$, such a theory must transform in the same way under $CR_i$ and $T$ upto a factor of $\eta_{_T}^{(\tilde{r},\tilde{s})}(-1)^{\tilde{s}+\tilde{n}+1}$, where $\tilde{n}=\lfloor {\tilde{r}/2}\rfloor$. For the unique choice of intrinsic parity $\eta_{_T}^{(\tilde{r},\tilde{s})}=(-1)^{\tilde{s}+\tilde{n}+1}$, the couplings remain invariant under $CR_i$ and $T$. This gives a $D$ mod $4$ classification of such theories based on their transformation under $CR_i$ and $T$.
		\item Similarly, there exists a $D$ mod $4$ classification such that in $D=4k+2$ the theory transforms in the same way under $R_iT$ and $C$, upto a factor of $\eta_{_C}^{(\tilde{r},\tilde{s})}(-1)^{\tilde{s}+\tilde{n}+1}$, where $\tilde{n}=\lfloor {\tilde{r}/2}\rfloor$. For the unique choice of the intrinsic parity $\eta_{_C}^{(\tilde{r},\tilde{s})}=(-1)^{\tilde{s}+\tilde{n}+1}$, the theory is always invariant under $C$ and $R_iT$.
	\end{itemize}
\end{enumerate}
\textbf{Odd dimensional spacetime:}
\begin{enumerate}
	\item  In odd dimensions, there are only two independent bilinears: (i)\textit{ even ranked bilinears}, and (ii) \textit{odd ranked bilinears}. The bilinears are always non-chiral since there is no chirality matrix in odd dimensions.
\item Just like even spacetime, we have classified the three-point interactions of two massless Dirac spinors and a MHS field $\Phi^{(\tilde{r},\tilde{s})}$. For MHS, other than scalars, there are two independent even and odd-ranked couplings. For scalars, the only independent coupling is the Yukawa coupling. The results are summarized in table \ref{tab:oddcouplings}. Unlike even dimensions, in any odd $D=2m+1$, there also exist the self-dual and the anti-self-dual couplings corresponding to the highest rank irreducible representations of the massive Little group $Spin(2m)$. See the section \ref{sec:ckrydiracthreepointodddiminteraction} for elaborate discussion. 

\item Consider a theory of multiple MHS fields $\Phi^{(\tilde{r}_a,\tilde{s}_a)}_a$ coupled with a massless Dirac spinor through bilinears of even rank ($r=2n$) and $s$ derivatives. Such a theory would preserve $C,\, R_i,\,T$ only if the intrinsic parities of every MHS is given by  
\begin{align}
	\eta_{_C}^{(\tilde{r}_a,\tilde{s}_a)}=-\eta_{_T}^{(\tilde{r}_a,\tilde{s}_a)}=	(-1)^{s+n+m+1} \qquad,\qquad \eta_{_R}^{(\tilde{r}_a,\tilde{s}_a)}=-1\qquad \forall ~ a\label{oddint1}
\end{align}
Similarly, in a $C,\, R_i,\,  T$ preserving theory of multiple MHS coupled only with the odd ranked bilinear $(r=2n+1)$, the intrinsic parities must be given by
\begin{align}
	\eta_{_C}^{(\tilde{r}_a,\tilde{s}_a)}=\eta_{_T}^{(\tilde{r}_a,\tilde{s}_a)}=	(-1)^{s+n+1} \qquad,\qquad \eta_{_R}^{(\tilde{r}_a,\tilde{s}_a)}=1\qquad \forall ~ a\label{oddint2}
\end{align}

	\item Let us consider a most general theory of either the $\Phi^{(2\tilde{n},\tilde{s})}$ or the $\Phi^{(2\tilde{n}+1,\tilde{s})}$ MHS coupled with a massless Dirac spinor through both even and odd ranked three-point couplings. In such theories of a single MHS, we show the following
\begin{itemize}
	\item The even and odd ranked couplings of a MHS always transform oppositely under $R_i$ and $CT$. Like even spacetime, \textit{a theory of MHS with both even and odd ranked couplings would always break $R_i$ and $CT$.}
	
	However, (unlike in even dimensions) if we restrict to the minimal coupling, then a theory of a massless Dirac spinor with an MHS preserves $C,\,  R_i,\,  T$. The intrinsic  parities satisfy either \eqref{oddint1} or \eqref{oddint2}, for $a=1$.
	
	\item In $D=4k+1$, a theory of an even ranked MHS $\Phi^{(2\tilde{n},\tilde{s})}$ would preserve $C,~R_iT$ such that the intrinsic parity is given by $\eta_{_C}^{(2\tilde{n},\tilde{s})}=(-1)^{\tilde{s}+\tilde{n}+1}$. Whereas, a theory of an odd ranked MHS $\Phi^{(2\tilde{n}+1,\tilde{s})}$ would preserve $CR_i,~T$ such that the intrinsic parity is given by $\eta_{_T}^{(2\tilde{n},\tilde{s})}=(-1)^{\tilde{s}+\tilde{n}+1}$.
	\item In $D=4k+3$, a theory of an even ranked MHS $\Phi^{(2\tilde{n},\tilde{s})}$ would preserve $CR_i,~T$ such that the intrinsic parity is given by $\eta_{_T}^{(2\tilde{n},\tilde{s})}=(-1)^{\tilde{s}+\tilde{n}+1}$. Whereas, a theory of an odd ranked MHS $\Phi^{(2\tilde{n}+1,\tilde{s})}$ would preserve $C,~R_iT$ such that the intrinsic parity is given by $\eta_{_C}^{(2\tilde{n},\tilde{s})}=(-1)^{\tilde{s}+\tilde{n}+1}$.
\end{itemize}
\end{enumerate}
We can combine the results in even and odd spacetimes in a single statement (also summarized in tables \ref{tab:mod2classification1} and \ref{tab:mod2classification2}.):\\

\textit{In any ${D=2m+1,\,2m}$ spacetime, any theory of a massless Dirac spinor with a single MHS, which includes all the possible three-point couplings, can be classified in a $D$ mod $4$ manner such that }
\begin{itemize}
	\item For an even ranked MHS $\Phi^{(2\tilde{n},\tilde{s})}$, the theory preserves $C,\, R_iT$ in $D=4k+1,\, 4k+2$ with the unique choice of ${\eta_{_C}^{(2\tilde{n},\tilde{s})}=(-1)^{\tilde{s}+\tilde{n}+1}}$. Whereas, the theory preserves $CR_i,~T$ in $D=4k,\, 4k+3$ with the unique choice of ${\eta_{_T}^{(2\tilde{n},\tilde{s})}=(-1)^{\tilde{s}+\tilde{n}+1}}$.
	\item For an odd ranked MHS $\Phi^{(2\tilde{n}+1,\tilde{s})}$, the theory preserves $C,\, R_iT$ in $D=4k+3,\, 4k+2$ with the unique choice of ${\eta_{_C}^{(2\tilde{n}+1,\tilde{s})}=(-1)^{\tilde{s}+\tilde{n}+1}} $. Whereas, the theory preserves $CR_i,~T$ in $D=4k,\, 4k+1$ with the unique choice of ${\eta_{_T}^{(2\tilde{n}+1,\tilde{s})}=(-1)^{\tilde{s}+\tilde{n}+1}} $.
\end{itemize}
\textbf{Organization of the paper : }We have divided the rest of the paper as follows: We encourage the reader first to read section \ref{sec:notations} to make themselves comfortable with the notations and conventions used throughout the paper. The basics of the Clifford algebra and the convention we follow are summarized in appendix \ref{App:CliffordAlgebra} and \ref{App:CB}.  In section \ref{sec:ckrydirecspinorreview}, we review Dirac spinors as the representations of complex Clifford algebra in any $D$ spacetime dimension. In section \ref{sec:ckrydirecbilinear}, we classify all the independent massless Dirac bilinears in even and odd spacetime dimensions by constraining them with the internal axial and vector symmetries. The group theoretical explanation for the construction of Dirac bilinears is discussed in appendix \ref{app:productofirreps}.  In section \ref{sec:MHS}, we review the massive higher spin representations of the Lorentz group and their Young tableau notations. In section \ref{sec:ckrydiracthreepointinteraction}, we couple the bilinears constructed in section \ref{sec:ckrydirecbilinear} with massive higher spin fields. We construct the couplings by taking projections of the bosonic irreducible representations of the $Spin(D-1)$ over the product of spinor representations transforming under $Spin(D-2)$. The branching (projection) rules are explained in appendix \ref{app:Branchingrules} and \ref{app:Projectionofirreps}.  In section \ref{sec:ckrydiraccrt}, we first introduce the action of the discrete transformations $C,\,R_i,\, T$ over the spinors and MHS. In section \ref{sec:MultipleMHS}, we calculate the action of these transformations over the spinor bilinears and the three-point couplings constructed in section \ref{sec:ckrydiracthreepointinteraction}. In section \ref{sec:unconventionalCRT}, we discuss the couplings for multiple Dirac spinors and the action of $CR_iT$. In section \ref{sec:FutureDirections}, we conclude along with some future directions.

\section{A lightning review of Dirac spinors}
\label{sec:ckrydirecspinorreview}

In this section, we summarize the basic concepts we follow in the rest of the paper. We denote the Dirac spinors by $\Psi^\alpha(x)$. These are the complex representations of the Spin group $Spin(D)$ based on the complex vector space $\mathbb{C}^D$.\footnote{For our purpose, we work with the complex representations of the real Clifford algebras which are isomorphic to the irreps of the complex Clifford algebra.}. In $D$ spacetime dimensions,
$\Psi^\alpha(x)\in \mathbb{C}^N$ where $N=2^{\lfloor{\frac{D}{2}}\rfloor}$. From the first course in Quantum field theory, we know that the starting point to study fermions is the Clifford algebra, which consists the gamma matrices that satisfy
\begin{align}
	\{	\Gamma^\mu,\Gamma^\nu\}=2\eta^{\mu\nu}\label{cliffordalgebra}
\end{align}
We work in the mostly positive signature of the spacetime metric $\eta^{\mu\nu}=(-,+,+,\cdots,+)$. We assume that the observables, e.g. scattering amplitudes and cross-sections are independent of the choice of the metric signature. From these Clifford generators, we can construct the Lorentz generators. In the spinor representation, the Lorentz group generators are $(J^{\mu\nu})^\alpha{}_\beta$ which can be constructed using the gamma matrices
\begin{align}
	(	J^{\mu\nu})^\alpha{}_\beta=-\frac{\iimg}{4}[\Gamma^\mu,\Gamma^\nu]^\alpha{}_\beta
\end{align}
In case of fermions, we know that $2\pi$ rotation is not the same as the identity element. Mathematically, we look at the double cover of the Lorentz group, which is known as the spin group. A Dirac spinor transforms under the complex representation of the spin group. The representation matrices are given by 
\begin{equation}
	d(\Lambda)=\exp(\iimg\, \omega_{\mu\nu}J^{\mu\nu})\qquad,\qquad \Lambda(\omega)\in Spin(D) 
\end{equation}
A longer summary of Clifford algebra, Pin, and Spin groups is given in appendix \ref{App:CliffordAlgebra}. The spinor group $Spin(D)$ is independent of whether the metric signature is Minkowski or Euclidean. Therefore, the Dirac spinors in the Minkowski and the Euclidean signatures have the same degrees of freedom. As a result, it is simpler to study a theory of Dirac spinor as we can go back and forth from the Lorentzian signature to Euclidean signature using Wick rotation. The Clifford elements satisfy identity
\begin{align}
	d(\Lambda)\, \Gamma^{\mu}\,  d(\Lambda)^{-1}=\Lambda^{\mu}{}_\nu \Gamma^{\nu}
\end{align}
In even $D=2m$ spacetime, the Clifford basis is spanned by $\{\mathbf{1},\Gamma^0,\Gamma^1,\cdots,\Gamma^{2m-1}\}$ where $\Gamma^0$ is anti-hermitian whereas $\Gamma^i$ are hermitian $\forall \, i=1,\cdots,2m-1$. The hermitian property can also be written in a compact form as
\begin{align}
	\beta \Gamma^\mu \beta^{-1}=-(\Gamma^{\mu})^\dagger\label{paritymatrix}
\end{align}
where $\beta=\iimg \Gamma^0$ is the hermitian operator known as the parity matrix.
We can also define the hermitian volume element also known as the chirality matrix
\begin{align}
	\Gamma_\star=(-\iimg )^{m+1}\Gamma^0\cdots\Gamma^{D-1}
\end{align}
In odd $D=2m+1$ dimensions, we can choose $\Gamma^{2m+1}$ as either $+\Gamma_\star$ or $-\Gamma_\star$ which gives rise to two in-equivalent representations\footnote{There does not exist any matrix in the Clifford algebra which can map $\Gamma_\star$ to $-\Gamma_\star$ keeping all other gamma matrices invariant} of the Clifford algebra. Interestingly, both the signs gives rise to two \enquote{equivalent} spinor representations generated by  $	J^{\mu(2m+1)}_{\pm}=-\frac{\iimg}{4}\left[\Gamma^\mu,\pm \Gamma_*\right]$ which can be mapped to each other by $\Gamma_\star$. Hence, the choice of $\Gamma_\star$ doesn't alter the physics of the spinors.

Any rank $r$ element of the Clifford algebra is defined as $\Gamma^{(r)}\equiv\Gamma^{\mu_1\mu_2\cdots\mu_r}$, where $\mu_1,\mu_2,\cdots,\mu_r$ are all anti-symmetric. There exists a duality between the rank $r$ and $D-r$ elements of the Clifford algebra in even dimension
\begin{align}	\Gamma^{\mu_{1}\cdots\mu_{r}}\Gamma_\star=-(-\iimg)^{m+1}\frac{1}{(D-r)!}\epsilon^{\mu_{r}\cdots\mu_{1}\nu_{1}\cdots\nu_{D-r}}\Gamma_{\nu_{1}\cdots\nu_{D-r}}.\label{evengammarelation}
\end{align}
  In even dimension, there exist two rank $r$ elements : $\Gamma^{(r)}$ and $\Gamma^{(r)}\Gamma_\star$. Whereas in odd dimensions there is a unique one rank $r$ element: $\Gamma^{(r)}$. $\Gamma^{(r)}$ is dual to the rank $D-r$ elements 
  \begin{align}
  	\Gamma^{\mu_{1}\cdots\mu_{r}}= \iimg^{m+1}\frac{1}{(D-r)!}\epsilon^{\mu_{1}\cdots\mu_{D}}\Gamma_{\mu_{D}\cdots\mu_{r+1}}\label{oddgammarelation}
  \end{align}

\section{Dirac bilinears in even and odd spacetimes}
\label{sec:ckrydirecbilinear}
The interactions in a quantum field theory are Lorentz scalars constructed out of the quantum fields. Any scalar that is formed out of spinor fields will necessarily have even number of spinor fields. We begin with terms which are constructed out of two Dirac spinors which are representations of Lorentz group. We call them Dirac spinor bilinears. We are interested in bilinears which transform covariantly as a spacetime tensor of the Lorentz group. In general, such a spinor bilinear\footnote{At this point, we only consider the bilinears for Dirac spinors of the same flavor. We will talk about the multiple flavors Dirac bilinears in the section \ref{sec:unconventionalCRT}.}  is given by $ \Psi^\dagger M^{(r)}\Psi $, where $M^{(r)}$ is an object with $r$ spacetime vector indices and two spinor indices, such that it satisfies the following equation 
\begin{align}
	d(\Lambda)^\dagger M^{(r)} d(\Lambda)\longrightarrow (\Lambda)^{(r)}M^{(r)}
\end{align}
where $(\Lambda)^{(r)}$ denotes the $r$ factors of $\Lambda^\mu{}_\nu$. In case of Dirac spinors, the matrix $M^{(r)}$ is given by $\beta\Gamma^{(r)}$. Using equation \eqref{paritymatrix}, it can be checked that the parity matrix $\beta$ maps the representation $d(\Lambda)$ to it's hermitian conjugate $d(\Lambda)^\dagger$ such that
\begin{align}
	\beta \, d(\Lambda)\, \beta^{-1}=d(\Lambda^{-1})^\dagger
\end{align}
In even $D=2m$ spacetime, the most general bilinears for a single Dirac spinor are given by
\begin{align}
\textit{Non-Chiral}\qquad :&\qquad 	\bar{\Psi}\Gamma^{(r)}\Psi\\ \textit{Chiral}\qquad :&\qquad  \bar{\Psi}\Gamma^{(r)}\Gamma_\star\Psi
\end{align}
where $\bar{\Psi}\equiv \Psi^\dagger\beta$ is the Dirac adjoint spinor which transforms as $d(\Lambda)^{-1}$ under the $Spin(2m)$. The rank $0$ Dirac bilinears gives the Dirac mass term. As we will see later, it is always possible to add a Dirac mass term without breaking any discrete $(C,R_i,T)$ symmetry in a theory of Dirac spinors in even spacetime dimensions. 

For three-point couplings, we only consider the Dirac spinors to be massless. As discussed in \ref{app:productofirreps}, the one particle states of a Dirac spinor transform under the direct sum of the two inequivalent fundamental spinor representations $\rho$ and $\tilde{\rho}$ of the Little group $Spin(2m-2)$.  All the independent massless Dirac bilinears are governed by the irreducible representations in the product of two massless Little group states. The complete group theoretical construction of the irreducible representations in the product of spinor states of $Spin(2m-2)$ is discussed in appendix \ref{app:productofevenirrpes}. In any $D=2m/2m+1$ spacetime, the product of one particle spinor states transforms under the irreducible representations of $Spin(D-2)\otimes Spin(D-2)$ and generates irreducible antisymmetric tensors of rank $0\leq r\leq m-1$. Thus, any irreducible Dirac bilinear can have a maximum rank of $m-1$.

The interactions of these bilinears with the massive higher-spinning Bosons might involve derivatives acting over the spinors. In order to manifest invariance under the Spin group, we first write down the most general bilinears, which involve $s$ number of derivatives and a rank $r$ Clifford element
\begin{align}
	\mathbb{M}^{(r,s)}= (\iimg)^{s+\frac{r(r+1)}{2}}(\partial^{(s)}\Psi^\dagger\beta)\Gamma^{(r)}\Psi
	\qquad,\qquad
	(\mathbb{M}^{(r,s)})^\dagger= \mathbb{M}^{(r,s)} \label{DiracBilinears1}
\end{align}
and
\begin{equation}
	\mathbb{N}^{(r,s)}= (\iimg)^{s+\frac{(r+1)(r+2)}{2}}(\partial^{(s)}\Psi^\dagger\beta)\Gamma^{(r)}\Gamma_\star\Psi
	\qquad,\qquad
	(\mathbb{N}^{(r,s)})^\dagger= \mathbb{N}^{(r,s)}\label{DiracBilinears2}
\end{equation}
These are compact notations to write down the expressions given in section \eqref{longhandexpressions1}.  We have inserted the factors of $\iimg$ appropriately to make them hermitian; these factors will be different in other conventions of the metric \eqref{cliffordalgebra}. Since $\Psi$ are complex-valued, the bilinears are invariant under the phase transformation $\mathcal{U} (\theta)= \exp(\iimg \theta )$. There exists another field redefinition known as the axial rotations
\begin{align}
	\Psi \longrightarrow \mathcal{U}_\star (\theta)\Psi
	\qquad,\qquad
	\mathcal{U}_\star (\theta)= \exp(\iimg \Gamma_\star \theta)
\end{align}
Since $\Gamma_\star$ anti-commutes with $\beta$ and $\Gamma^{\mu}$, the axial rotations mix the even rank bilinears among themselves but keep the odd rank bilinears invariant. The action of this transformation on the bilinears is as follows
\begin{equation}
	\begin{split}
		\forall s\quad,\quad r\in 2\mathbb{Z}+1
		\qquad & :\qquad \mathbb{M}^{(r,s)}\longrightarrow \mathbb{M}^{(r,s)}
		\\
		\forall s\quad,\quad r\in 2\mathbb{Z}+1
		\qquad & :\qquad \mathbb{N}^{(r,s)}\longrightarrow \mathbb{N}^{(r,s)}
		\\
		\forall s\quad,\quad r\in 2\mathbb{Z}
		\qquad & :\qquad \mathbb{M}^{(r,s)}\longrightarrow \mathbb{M}^{(r,s)}\cos 2\theta+\mathbb{N}^{(r,s)}\sin 2\theta
		\\
		\forall s\quad,\quad r\in 2\mathbb{Z}
		\qquad & :\qquad \mathbb{N}^{(r,s)}\longrightarrow - \mathbb{M}^{(r,s)}\sin 2\theta+\mathbb{N}^{(r,s)}\cos 2\theta
	\end{split}
\label{axialoversinglebilinears}
\end{equation}
Later we will explore the consequences due to these field rotations. The space of all the independent bilinears in a chiral asymmetric theory is given by as
\begin{align}
	\mathbb{M}^{(n,s)} \quad,\quad \mathbb{N}^{(n,s)}
	\qquad,\qquad n,\, s\in \mathbb{\mathbb{Z}}	
\label{evenbilinears}
\end{align}

In odd $D=2m+1$ spacetime, the one particle states of the massless Dirac spinors transform under the direct sum $\rho\oplus \rho$, where $\rho$ is the unique spinor representations of the Little group $Spin(2m-1)$. Since there is no chirality matrix in odd spacetime dimensions, the only possible Dirac bilinear, which transforms as a rank $r$ tensor of the Lorentz group are
\begin{align}
		\mathbb{M}^{(r,s)} \qquad \forall\, r,\, s\in \mathbb{Z}
\end{align}
As we have discussed in appendix \ref{app:productofoddirrpes}, all the irreducible bilinears of Dirac spinors in $D=2m+1$ would have rank $0\leq \tilde{r}\leq m-1$. The construction of these bilinears using group theory is also discussed briefly in appendix \ref{app:Projectionofodddimirreps}.
Unlike even spacetime dimensions, the Dirac spinors in odd spacetime only enjoy the vector rotation symmetry $\mathcal{U}(\theta)$ which keeps the bilinear $\mathbb{M}^{(r,s)}$ invariant. Therefore, we cannot flip the sign of the Dirac bilinears in odd spacetime dimensions.
As we will see, this has an important consequence over the signs of interaction terms. Unlike even dimensions, the mass term in odd dimensions might break certain discrete symmetries like $C, R,$ or $T$. Therefore, adding a Dirac mass term in a theory of Dirac spinors in odd spacetime is not always possible. We will discuss the mass term briefly in section \ref{sec:ckrydiraccrt}. For the purpose of three point interactions, the Dirac spinors are always massless.

\section{Massive higher spin representations}\label{sec:MHS}
We denote the MHS field representations of the Lorentz group by $\Phi^{(\tilde{r},\tilde{s})}$,\footnote{We have chosen different notation for the indices of MHS $\Phi^{(\tilde{r},\tilde{s})}$ and the bilinears $\mathbb{M}^{(r,s)}$ because they are representations of different Little groups. All the indices with \enquote{tilde} carry information regarding MHS.}\footnote{Since the Dirac bilinears do not have any mixed symmetric indices, particles other than \eqref{higherspinningboson} cannot couple with the massless Dirac spinors in any dimension. This point is discussed briefly in appendix \ref{app:Projectionofirreps}.}
\begin{align}
	\Phi^{(\tilde {r},\tilde {s})}\equiv \Phi_{[\mu_1\cdots\mu_{\tilde{r}}]\{\nu_1\cdots\nu_{\tilde{s}}\}}
\end{align} 
where $\tilde{s}$ and $\tilde{r}$ are the number of symmetric and antisymmetric spacetime indices such that whenever $\tilde{r}=0\implies \tilde{s}=0$. The MHS follow the equation of motion
\begin{align}
	(\partial^2-\mathbf{m}^2)\Phi^{(\tilde{r},\tilde{s})}=0\label{MHSeqnofmotion}
\end{align}
where $\mathbf{m}$ is the mass of MHS. The one particle states of the MHS bosons transform under the irreducible representations of the massive Little group $Spin(D-1)$ (or $SO(D-1)$). Every index of the MHS field is transverse to its momentum $k_3$. Whereas every pair of indices are traceless. These constraints reduces the on-shell degrees of freedom of the MHS which are same as that of a Young tableau representation of $SO(D-1)$ with same number of indices \cite{das2014lie,georgi2018lie}. In other words, the space of MHS fields which are traceless and transverse to $k_3$ transforms under the Young tableau representations of $SO(D-1)$ which keeps $k_3$ invariant. We can now associate the Young tableaux of $SO(D-1)$ with the MHS fields following equation of motion \eqref{MHSeqnofmotion}.
\begin{align}
	\Phi^{(\tilde {r},\tilde {s})}\equiv \begin{ytableau}
		\mu_1&\nu_1&\cdots&\nu_{\tilde{s}}\\ \mu_2\\ \vdots \\ \mu_{\tilde{r}}
	\end{ytableau}\label{higherspinningboson}
\end{align}
In the usual notation, a Young tableau is denoted by $Y(c_1,c_2,\cdots,c_{\tilde{s}},c_{\tilde{s}+1})$ where $c_i's$ are the number of boxes in a column such that $c_1\geq c_2\geq c_3\geq \cdots\geq c_{\tilde{s}}\geq c_{\tilde{s}+1}\geq 0$. In our notation, $c_1=\tilde{r}$ and $c_2=c_3=\cdots=1$. In the literature, there are multiple ways to define the symmetry properties of the Young tableau. One starts with a general rank $\tilde{r}+\tilde{s}$ tensor. We work in a basis where we first symmetrize $\tilde{s}$ number of indices and then anti-symmetrize $\tilde{r}$ of them. This is same as the basis described by equation $(2.1)$ of \cite{Chakraborty:2020rxf}. Alternately, one can anti-symmetrize $\tilde{r}$ of them followed by symmetrizing $\tilde{s}$ indices. To get acquainted with the notation, we summarize the following:
\begin{itemize}
	\item The scalar fields are denoted by $\Phi^{(0,0)}$.

	\item The completely symmetric traceless fields with $\tilde{s}+1$ indices are denoted by $\Phi^{(1,\tilde{s})}$ ($\tilde{s}\geq 0$). All the $\tilde{s}+1$ indices are symmetrized by construction.
	\item  The completely antisymmetric rank $\tilde{r}\geq 0$ fields are denoted by $\Phi^{(\tilde{r},0)}$.
	\item  The mixed symmetric traceless fields\footnote{Mixed symmetric particles have mixed symmetric and antisymmetric indices. For example: the domino tensor $\Phi^{[\mu_1\mu_2]\nu}\equiv \begin{ytableau}
			\mu_1&\nu\\\mu_2
		\end{ytableau}$ and the Riemann-Weyl tensor $R^{[\mu\nu][\rho\sigma]}\equiv$ \begin{ytableau}
			\mu&\rho\\\nu&\sigma
		\end{ytableau} .} can only be only denoted by $\Phi^{(\tilde{r},\tilde{s})}$ where $\tilde{r}>1,\, \tilde{s}>0$.
\end{itemize}
Since $\Phi^{(\tilde{r},\tilde{s})}$ are massive irreducible representations, the antisymmetric indices $\tilde{r}$ have an upper bound given by the rank of massive Little group $Spin(D-1)$ i.e. $\lfloor{\frac{D-1}{2}}\rfloor \geq\tilde{r}\geq 0$, $\tilde{s}\geq 0$. When $D=2m+1$, the highest rank $m$ representations are further reducible to the direct sum a self dual and an-antiself dual representation. In general the reducibility condition for tensors of $SO(D-1)$ are given by,
\begin{align}
	\widetilde{\Phi}^{[l_1\cdots l_{\tilde{r}}]\{k_1\cdots k_{\tilde{s}}\}}=\pm\frac{(-\iimg)^m}{m!}\epsilon^{l_1\cdots l_{D-1}}	\Phi_{l_{\tilde{r}+1}\cdots l_{D-1}}^{\{k_1\cdots k_{\tilde{s}}\}}\label{reducibility1}
\end{align}
where $l_i,k_i$ are the Little group spatial indices i.e. $1\leq l_i,k_i \leq 2m$ $\forall ~i$. The tensor $\epsilon^{l_1\cdots l_{D-1}}$ is the invariant antisymmetric tensor of $SO(D-1)$. The factor of $(-\iimg)^{m} $ appears because the epsilon tensor squares as $	\epsilon^{l_1\cdots l_{D-1}}\epsilon_{l_{D-1}\cdots l_1}=(-1)^m$ where $m=\lfloor{(D-1)/2}\rfloor$. To manifest $Spin(D-1,1)$ covariance, the duality relation can also be written as
\begin{align}
	\widetilde{\Phi}^{[\mu_1\cdots \mu_{\tilde{r}}]\{\nu_1\cdots \nu_{\tilde{s}}\}}=\pm \frac{(-\iimg)^{m+1}\partial_\mu}{\tilde{r}!\, \mathbf{m}}\epsilon^{\mu\mu_1\cdots \mu_{D-1}}	\Phi_{\mu_{\tilde{r}+1}\cdots \mu_{D-1}}^{\{\nu_1\cdots \nu_{\tilde{s}}\}}\label{reducibility2}
\end{align}
where $\mathbf{m}$ is the mass of the higher spinning boson.  The reducibility condition \eqref{reducibility2} is the solution to the equation of motion \eqref{MHSeqnofmotion}.

In $D=2m+1$ spacetime, we denote the rank $m$ MHS fields by $\Phi^{(m,\tilde{s})}$ and $\widetilde{\Phi}^{(m,\tilde{s})}$ which using the reducibility condition \eqref{reducibility2} follows the duality relation
\begin{align}
	\widetilde{\Phi}^{[\mu_1\cdots \mu_{m}]\{\nu_1\cdots \nu_{\tilde{s}}\}}=\pm \frac{(-\iimg)^{m+1}\partial_\mu}{m!\, \mathbf{m}}\epsilon^{\mu\mu_1\cdots \mu_{2m}}	\Phi_{\mu_{m+1}\cdots \mu_{2m}}^{\{\nu_1\cdots \nu_{\tilde{s}}\}}\label{dualitycovariance}
\end{align}
The self dual and anti self dual irreps are given by
\begin{align}
\text{	Self-dual}\qquad:\qquad \frac{1}{2}(\Phi^{(m,\tilde{s})}+\widetilde{\Phi}^{(m,\tilde{s})})\label{selfdual}\\
\text{Anti-Self-dual}\qquad:\qquad \frac{1}{2}(\Phi^{(m,\tilde{s})}-\widetilde{\Phi}^{(m,\tilde{s})})\label{antiselfdual}
\end{align}
The field $\widetilde{\Phi}$, satisfying equation \eqref{dualitycovariance}, is also a solution to the equation of motion $(\partial^2-\mathbf{m}^2)\Phi^{(m,\tilde{s})}=0$. It follows from \eqref{dualitycovariance} that for $m=even$, the equation will have no solution for real $\Phi^{(m,\tilde{s})}$ because of an $\iimg$ factor on the RHS. Whereas for $m=odd$, it will have solutions for real $\Phi^{(m,\tilde{s})}$.

\section{Three point interactions}
\label{sec:ckrydiracthreepointinteraction}

The research in classifying three-point function of higher spinning particles started in \cite{Chakraborty:2020rxf}.
The authors classified three point function of a higher spin particle with two photons and gravitons in that work. Later, these three point functions were used to compute the tree-level scattering of photons and gravitons due to higher spin exchange in \cite{Balasubramanian2021}.

 In this section, we classify all the three-point couplings of two massless Dirac spinors interacting with massive higher-spinning (MHS) bosons. The couplings are the non-vanishing projections of the $Spin(D-1)$ bosonic irreps over the irreducible representations in the product of two spinor states of $Spin(D-2)$. The branching rules for the projections are described in appendix \ref{app:Branchingrules}. The complete group theoretical construction of the couplings using branching rules is described in \ref{app:Projectionofirreps}. Here we give an analogous on shell approach which is in line with the branching rules.
  
  At the level of representations of the Lorentz group, the projection is achieved by contracting the MHS with the derivatives acting over the spinor bilinears.\footnote{ This procedure has also been discussed in \cite{Chakraborty:2020rxf} for the classifications of interactions between two massless bosons and a MHS.} Since all the bilinears are hermitian, the irreducible MHS which couple with them can always be taken as real with real coupling constants. However, this is not true in odd spacetimes where we can have self dual and anti self dual MHS which can be complex valued depending on the rank of the Lorentz group.
 
 We first write down important properties for the projections of the MHS over the massless Dirac bilinears which simplifies the construction of couplings. These properties can also be understood using the branching rules discussed in appendix \ref{app:Branchingrules}.
 \begin{enumerate}
 	  	\item A coupling vanishes if more than one derivatives acting over the Dirac spinors are contracted with the antisymmetric indices of a MHS. For example in $D=5$ spacetime, the couplings $(\partial^{\mu\nu}\bar{\Psi})\Psi\Phi_{[\mu\nu]}$ and $(\partial^{\mu}\bar{\Psi})(\partial^\nu\Psi)\Phi_{[\mu\nu]}$ vanishes in the momentum space because $k_1^\mu k_1^\nu \Phi_{[\mu\nu]}=-k_1^\mu k_2^\nu \Phi_{[\mu\nu]}=0$.
 	\item The couplings where derivatives contract withing themselves are reducible to couplings without such contracted derivatives. This can again be easily understood in the momentum space where the self contracted derivatives gives rise to Mandelstam factors $k_1\cdot k_2, ~k_2\cdot k_3,$ or $k_1\cdot k_3$. For example, in momentum space the coupling $(\partial^\mu\Psi)(\partial_\mu \Psi)\Phi$ is reduced to $\bar{\Psi}\Psi\Phi$ times the Mandelstam variable $k_1\cdot k_2$.
 	 	\item The couplings where the derivatives are contracted with the gamma matrices of the bilinears either vanishes or reduce to couplings with lesser number of gamma matrices using the equation of motion $(\partial^\mu \bar{\Psi})\Gamma_\mu=0=\Gamma_\mu (\partial^\mu \Psi)$. For example, the coupling $(\partial^\mu\bar{\Psi})\Gamma_\mu\Psi \Phi$ vanishes.\footnote{ If the Dirac spinors were massive, such a coupling would be non-vanishing but still reducible to $\bar{\Psi}\Psi\Phi$.} Whereas the $D=4$ coupling $(\partial^\mu \bar{\Psi})\Gamma_{\mu\nu}\Psi \Phi^{\nu}$ is reducible to the lower rank coupling $(\partial^\mu\bar{\Psi})\Psi\Phi_\mu$.
 	\begin{align}
 		(\partial^\mu \bar{\Psi})\Gamma_{\mu\nu}\Psi \Phi^{\nu}=\frac{1}{2}(\partial^\mu \bar{\Psi})(2 \Gamma_\mu\Gamma_\nu-2\eta_{\mu\nu})\Psi \Phi^{\nu}=-(\partial^\mu\bar{\Psi})\Psi\Phi_\mu
 	\end{align}
 	where in the second equality we have used the equation of motion for the massless Dirac spinor.
 		 \item  To ensure Lorentz invariance of any coupling, the number of indices of MHS $\Phi^{(\tilde{r},\tilde{s})}$ must be equal to the total number of derivatives and gamma's\footnote{Following property $2$ and $3$, the derivatives contracted within themselves or with gamma's are not counted.} in the Dirac bilinear $\mathbb{M}^{(r,s)}$ and $\mathbb{N}^{(r,s)}$. So we obtain the following restriction
 	\begin{align}
 		\tilde{r}+\tilde{s}=r+s\label{r+s}
 	\end{align}
 	\item Since $Spin(D-2)\subset Spin(D-1)$, the rank of the massless Dirac bilinears is always less than or equal to the number of antisymmetric indices in the MHS, i.e., $r\leq \tilde{r}$. This is a direct consequence of the branching rules \eqref{branchingrule1} and \eqref{branchingrule2}. Following property $1$, the non-vanishing projections involve at most one derivative contracted with antisymmetric indices $\tilde{r}$ of MHS\footnote{This point has been discussed while constructing the three point couplings using the branching rules in appendix \ref{app:Projectionofirreps}}. Therefore, the only possible non-vanishing projections of $\Phi^{(\tilde {r},\tilde {s})}$  over the bilinears are of two types
 	\begin{align}
 		\tilde{r}=r,\, \tilde{s}=s\qquad \text{or}\qquad \tilde{r}=r+1,\, \tilde{s}=s-1\label{rsconstraint}
 	\end{align} 
 		\item As discussed in appendix \ref{app:productofirreps} and section \ref{sec:MHS}, the rank of the irreducible Dirac bilinears $\mathbb{M}^{(r,s)},\mathbb{N}^{(r,s)} $ and the irreducible MHS $\Phi^{(\tilde{r},\tilde{s})}$ are always bounded by the rank of the massless and massive Little group, respectively. In any $D=2m/2m+1$ spacetime, these bounds are given by
 		\begin{align}
 		D=2m\qquad&:\qquad 	0\leq r\leq m-1\qquad,\qquad 	0\leq \tilde{r}\leq m-1,\\
 				D=2m+1\qquad&:\qquad 	0\leq r\leq m-1\qquad,\qquad 	0\leq \tilde{r}\leq m,
 		\end{align}
 		  Therefore, any coupling with $r$ and/or $\tilde{r}$ beyond these bounds are reducible using the reducibility conditions \eqref{evengammarelation} and \eqref{reducibility2}.
 \end{enumerate}

   We already know that there are four types of Dirac bilinears in even dimensions: $	\mathbb{M}^{(2n,s)},\, 	\mathbb{M}^{(2n+1,s)},\, 	\mathbb{N}^{(2n,s)},\, \mathbb{N}^{(2n+1,s)}$. In odd dimensions, there are only two independent bilinears:  $	\mathbb{M}^{(2n,s)}$ and $	\mathbb{M}^{(2n+1,s)}$. This implies that the number of couplings of a MHS $\Phi^{(\tilde{r},\tilde{s})}$ for any $\tilde{r}$ and $\tilde{s}$ are finite. We demonstrate how these couplings look in the following sections.
\subsection{Even dimensional spacetime}
\label{sec:ckrydiracthreepointevendiminteraction}

In even $D=2m$ spacetime, the MHS transform under the irreducible bosonic representations of the Little group $Spin(2m-1)$ whereas the massless Dirac spinors transform under the Little group $Spin(2m-2)$. Since both the groups have rank $m-1$, we have $0\leq \tilde{r}\leq m-1,$ and $0\leq  r \leq m-1$. Using the Lorentz invariance condition \eqref{rsconstraint}, any MHS of rank $\tilde{r}\neq 0$ would have four independent couplings corresponding to the projections over $\mathbb{M}^{(\tilde{r},\tilde{s})},~\mathbb{M}^{(\tilde{r}-1,\tilde{s}+1)}$, $\mathbb{N}^{(\tilde{r},\tilde{s})},$ and $\mathbb{N}^{(\tilde{r}-1,\tilde{s}+1)}$. These couplings have been summarized in table \ref{tab:evencouplings}. As described in appendix \ref{app:Projectionofevendimirreps}, the same couplings can be derived using the branching rules.

 For $\tilde{r}=0$, which corresponds to a massive scalar field $\Phi$, there are only two independent couplings given by the Yukawa terms: $\bar{\Psi}\Psi\Phi$ and $\iimg \bar{\Psi}\Gamma_\star\Psi\Phi$. 
	\renewcommand{\arraystretch}{2}
\begin{table}[H]
	\centering
	\begin{tabular}{ ||c||c|c|c|| }
		\hline
		\hline
	S.no.&	Notation            & Three-point functions                     & $\tilde{r}$                    \\
		\hline
		\hline
	$1$&	$ \mathbf{e} ~	\mathbb{M}^{(\tilde{r},\tilde{s})}\Phi^{(\tilde{r},\tilde{s})}$&  $\mathbf{e} ~ (\iimg)^{\tilde{s}+\frac{\tilde{r}(\tilde{r}+1)}{2}}\, (\partial^{\nu_1\cdots\nu_{\tilde{s}}}\bar{\Psi})\Gamma^{\mu_1\cdots\mu_{\tilde{r}}}\, \Psi \, \Phi_{[\mu_1\cdots\mu_{\tilde{r}}]\{\nu_1\cdots\nu_{\tilde{s}} \}}$&$0\leq \tilde{r}\leq m-1 $\\
		\hline
	$2$&	$\mathbf{f}~\mathbb{M}^{(\tilde{r}-1,\tilde{s}+1)}\Phi^{(\tilde{r},\tilde{s})}$&$ \mathbf{f} ~ (\iimg)^{\tilde{s}+1+\frac{\tilde{r}(\tilde{r}-1)}{2}} (\partial^{\nu_1\cdots \nu_{\tilde{s}}}\partial^{\mu_{\tilde{r}}}\bar{\Psi})\Gamma^{\mu_1\cdots\mu_{\tilde{r}-1}}\, \Psi \, \Phi_{[\mu_1\cdots\mu_{\tilde{r}}]\{\nu_1\cdots\nu_{\tilde{s}} \}}$&$1\leq \tilde{r}\leq m-1 $   \\
		\hline
	$3$&	$ \mathbf{g} ~	\mathbb{N}^{(\tilde{r},\tilde{s})}\Phi^{(\tilde{r},\tilde{s})}$&  $\mathbf{g}~ (\iimg)^{\tilde{s}+\frac{(\tilde{r}+1)(\tilde{r}+2)}{2}}\, (\partial^{\nu_1\cdots\nu_{\tilde{s}}}\bar{\Psi})\Gamma^{\mu_1\cdots\mu_{\tilde{r}}}\,\Gamma_\star \, \Psi \, \Phi_{[\mu_1\cdots\mu_{\tilde{r}}]\{\nu_1\cdots\nu_{\tilde{s}} \}}$&$0\leq \tilde{r}\leq m-1 $\\
		\hline
	$4$&	$
		\mathbf{h}~\mathbb{N}^{(\tilde{r}-1,\tilde{s}+1)}\Phi^{(\tilde{r},\tilde{s})}$&$	 \mathbf{h} ~ (\iimg)^{\tilde{s}+1+\frac{\tilde{r}(\tilde{r}+1)}{2}} (\partial^{\nu_1\cdots \nu_{\tilde{s}}}\partial^{\mu_{\tilde{r}}}\bar{\Psi})\Gamma^{\mu_1\cdots\mu_{\tilde{r}-1}}\, \Gamma_\star\Psi \, \Phi_{[\mu_1\cdots\mu_{\tilde{r}}]\{\nu_1\cdots\nu_{\tilde{s}} \}}$&$1\leq \tilde{r}\leq m-1$\\
		\hline
		\hline
	\end{tabular}\caption{Three point couplings of real $\Phi^{(\tilde{r},\tilde{s})}$ for any arbitrary $\tilde{r},\tilde{s}$ in $D=2m$ spacetime. Here $\mathbf{e},\, \mathbf{f},\, \mathbf{g}\, \mathbf{h}$ are the real-valued coupling constants.}	\label{tab:evencouplings}
\end{table}
 In a theory of multiple MHS of flavors $a=1,2,\cdots, N$, the coupling constant depends on $\tilde{r},\tilde{s}$ and the flavor index $a$. The table \ref{tab:evencouplings} gives the couplings for both $\tilde{r}=2\tilde{n}$ and $2\tilde{n}+1$ where $\tilde{n\in \mathbb{Z}}$. For example, in $5+1$ dimensions, for $\tilde{r}=2,\tilde{s}=0$, this gives the following couplings of a massive $2-$form given by 
\begin{equation}
	\begin{split}
			\iimg 	\mathbf{e} \bar{\Psi}\Gamma^{\mu_1\mu_2}\Psi\Phi_{[\mu_1\mu_2]}\qquad&,\qquad \mathbf{f} (\partial^{\mu_{2}}\bar{\Psi})\Gamma^{\mu_1}\Psi\Phi_{[\mu_1\mu_2]}\\ 
			\mathbf{g} \bar{\Psi}\Gamma^{\mu_1\mu_2}\Gamma_\star\Psi\Phi_{[\mu_1\mu_2]}\qquad&,\qquad \mathbf{h} (\partial^{\mu_{2}}\bar{\Psi})\Gamma^{\mu_1}\Gamma_\star\Psi\Phi_{[\mu_1\mu_2]}
	\end{split}
\end{equation}
In any $D=2m$, for $\tilde{r}=\tilde{s}=0$, the only couplings are $\mathbf{e}$ and $\mathbf{g}$ corresponding to the Yukawa term and its chiral counterpart. Similarly for $\tilde{r}=1,\tilde{s}=0$ i.e. massive spin $1$, we get the chiral and non-chiral massive QED couplings $\mathbf{e}\bar{\Psi}\Gamma^\mu\Psi A_\mu$ and  $\mathbf{g}\bar{\Psi}\Gamma^\mu\Gamma_\star\Psi A_\mu$. The corresponding non-minimal couplings are given by $\mathbf{f}(\partial^\mu \bar{\Psi})\Psi A_\mu$ and  $\mathbf{h}(\partial^\mu \bar{\Psi})\Gamma_\star\Psi A_\mu$. In section \ref{sec:MultipleMHS}, we will see that the minimal and non-minimal couplings together always break the reflection symmetry.

All the couplings in table \ref{tab:evencouplings} are for $ \tilde{r}\leq m-1$. A natural question that may arise is the following: {\it What happens to the three-point function of the MHS whose one particle states are reducible representations of $Spin(2m-1)$, i.e., with $\tilde{r}>m-1$ ?}. In all such cases, the three-point function can always be reduced to a three-point function of lower rank. Here, we consider an example to explain it. Consider a higher rank coupling of $\Phi^{(\tilde{r},\tilde{s})}$ where $m-1<\tilde{r}\leq 2m$ with the $\mathbb{M}^{(r,s)}$ for $\tilde{r}=r,\, \tilde{s}=s$ given by
	      \begin{align}
	      	(\partial^{\nu_1\cdots \nu_{\tilde{s}}}\bar\Psi)\, \Gamma^{\mu_1\cdots \mu_{\tilde{r}}}\, \Psi\, \Phi_{[\mu_1\cdots \mu_{\tilde{r}}]\{\nu_1\cdots\nu_{\tilde{s}}\}}\label{higherrankcoupling}
	      \end{align}
	      Using \eqref{reducibility2}, the MHS with $\tilde{r}> m-1$ is reducible to the lower rank irreducible tensor with  $\tilde{r}-2m+1$ antisymmetric indices,
	 \begin{align}
	 	\Phi^{[\mu_1\cdots \mu_{\tilde{r}}]\{\nu_1\cdots\nu_{\tilde{s}}\}}\propto\partial_{\mu}\epsilon^{\mu_1\cdots\mu_{\tilde{r}}\mu_{\tilde{r}+1}\cdots\mu_{2m-1}\mu}\Phi_{[\mu_{\tilde{r}+1}\cdots\mu_{2m-1}]}^{\{\nu_1\cdots\nu_{\tilde{s}}\}}\label{evendimcovariantdualtiy}
	 \end{align}
	 where $\epsilon^{\mu_1\cdots\mu_{2m-1}\mu}$ the invariant epsilon tensor of $SO(2m-1,1)$.
	 Equation \ref{evendimcovariantdualtiy} is the covariant form of the following duality condition at the level of the Little Group
	 \begin{align}
	 	\Phi^{[l_1\cdots l_{\tilde{r}}]\{ k_1\cdots k_{\tilde{s}}\}}\propto \epsilon^{l_1\cdots l_{\tilde{r}}l_{\tilde{r}+1}\cdots l_{2m-1}}\Phi_{[l_{\tilde{r}+1}\cdots l_{2m-1}]}^{\{k_1\cdots k_{\tilde{s}}\}}
	 \end{align}
	 where $l_i,k_i$ are the little group $SO(2m-1)$ indices running over the spatial values $1\leq l_i,k_i\leq 2m-1$. The proportionality factor includes $\iimg$ and permutation factors among $\tilde{r}$, $m$. Using \eqref{evendimcovariantdualtiy} and  \eqref{evengammarelation}, the rank $\tilde{r}$ gamma's in \eqref{higherrankcoupling} can be reduced to rank $2m-\tilde{r}$ gamma's. This relates three-point function with the rank $\tilde{r}>m-1$ (given in \eqref{higherrankcoupling}) to a three-point function with rank $2m-r$, given by (upto some factors of $\iimg$ which can be restored through hermiticity)
	 \begin{align}
	 	(\partial^{\nu_1\cdots \nu_{\tilde{s}}}\bar\Psi)\,	\Gamma^{\rho_{{\tilde{r}+1}}\cdots\rho_{2m-1}\mu}\Gamma_\star\, \Psi\, \partial_\mu\Phi_{[\rho_{\tilde{r}+1}\cdots\rho_{2m-1}]\{\nu_1\cdots\nu_{\tilde{s}}\}}
	 \end{align}
	 We can further reduce the coupling to a lower rank by shifting the derivative over $\Phi$ to the spinor field $\Psi$ and using the Dirac equation of motion. For example consider the following higher rank coupling in $D=4$ for $\tilde{r}=2$ and $\tilde{s}=1$, $(\partial^{\nu}	\bar{\Psi})\Gamma^{[\mu_1\mu_2]}\Psi\Phi_{[\mu_1\mu_2]\nu}$.
	 After the Little group reduction and using the equation of motion  $\partial^\mu\Gamma_\mu\Psi=0$, we can write the coupling as $	(	\partial^{\mu\nu}	\bar{\Psi})\, \Psi\, \tilde\Phi_{\{\mu\nu\}}$.
	
\subsubsection{A theory of single MHS}	
 Consider a single MHS $\Phi^{(\tilde{r},\tilde{s})}$. We know it couples to the following Dirac bilinears
\begin{equation}
	\mathbb{M}^{(\tilde{r},\tilde{s})}
	\quad,\quad 	
	\mathbb{M}^{(\tilde{r}-1,\tilde{s}+1)}
	\quad,\quad 	
	\mathbb{N}^{(\tilde{r},\tilde{s})}
	\quad,\quad 	
	\mathbb{N}^{(\tilde{r}-1,\tilde{s}+1)}
\end{equation}
For any $\tilde{r}$, two are even rank bilinears, and the other two are odd rank bilinears. We already mentioned the axial field redefinition symmetry $\mathcal{U}_\star(\theta)$ in \eqref{axialoversinglebilinears}. Now, we discuss its consequences.
Due to these axial rotations, we can set the coefficient of $ \mathbb{N}^{(2n,s)}$ to be zero\footnote{Once we have done this, we can no longer use this transformation, apart from $\theta=\frac{f\pi}{2}, \, f\in \mathbb{Z}$.}. This implies that in a theory of $\Phi^{(2\tilde{n},\tilde{s})}$ (or $\Phi^{(2\tilde{n}+1,\tilde{s})}$), we can set the coupling constant of $\mathbb{N}^{(2\tilde{n},\tilde{s})}\Phi^{(2\tilde{n},\tilde{s})}$ (or $\mathbb{N}^{(2\tilde{n},\tilde{s}+1)}\Phi^{(2\tilde{n}+1,\tilde{s})}$) to be zero. Therefore, in a theory of a single MHS, the only independent bilinears are 
\begin{equation}
	\mathbb{M}^{(2n,s)}
	\quad,\quad 	
	\mathbb{M}^{(2n+1,s)}
	\quad,\quad 	
	\mathbb{N}^{(2n+1,s)}\qquad ,\quad  n,s\in \mathbb{Z}
\end{equation}
Moreover, under $\mathcal{U}_\star(\theta)$ for $\theta=(2f+1)\frac{\pi}{2}~\forall \, f\in \mathbb{Z}$, the even ranked bilinears pick up a sign whereas the odd ranked bilinears remain invariant,
\begin{align}
	\mathbb{M}^{(2n,s)}\longrightarrow - \mathbb{M}^{(2n,s)}\quad ,\quad \mathbb{M}^{(2n+1,s)}\longrightarrow  \mathbb{M}^{(2n+1,s)}\quad ,\quad \mathbb{N}^{(2n+1,s)}\longrightarrow  \mathbb{N}^{(2n+1,s)}
\end{align}
This implies that we can always flip the sign of even-ranked non-chiral bilinears using discrete axial rotations without altering the signs of the odd-ranked chiral and non-chiral bilinears. This has an important consequence on the signs of interaction terms in a theory of Dirac spinors. For example, the mass term corresponds to the rank $0$ bilinear, which does not have a physical significance in even dimensions. Also, in the standard model, the sign of the massive spin $1$ ($\tilde{r}=1,\tilde{s}=0$) coupling $\mathbf{f} (\partial^\mu\bar{\Psi}) \Psi \Phi_{\mu}$ is not physical. Whereas, the sign of the chiral coupling $\mathbf{g}	\bar{\Psi} \Gamma^\mu \Gamma_\star\Psi \Phi_{\mu}$ has physical significance. Similarly, the sign of the spin $0$ ($\tilde{r}=\tilde{s}=0$) Yukawa coupling $ \mathbf{e}   \bar{\Psi}\Psi \Phi$ is not physical.

We saw that, in a theory of only one MHS, we can get rid of $\mathbb{N}^{(2n,s)}$ using axial transformations. However, in the presence of multiple MHS, it is not always possible to get rid of all the couplings with $\mathbb{N}^{(2n,s)}$. For example, consider the following Yukawa couplings of two distinguishable spin $0$ massive scalars in any $D=2m$,
\begin{align}
	\mathcal{L}^{int}=\mathbf{g}_1\bar{\Psi}\Psi\Phi_1+\mathbf{f}_1\iimg\bar{\Psi}\Gamma_\star\Psi\Phi_1+\mathbf{g}_2\bar{\Psi}\Psi\Phi_2+\mathbf{f}_2\iimg\bar{\Psi}\Gamma_\star\Psi\Phi_2
\end{align} 
Under the chiral rotation $\mathcal{U}_\star(\theta)$, the Lagrangian transforms as follows
\begin{align}
	\mathcal{L}^{int}\xrightarrow{\mathcal{U}_\star(\theta)}&
	(\mathbf{g}_1\cos(2\theta)-f_1\sin(2\theta))\bar{\Psi}\Psi\Phi_1+ (\mathbf{g}_1\sin(2\theta)+f_1\cos(2\theta))\iimg\bar{\Psi}\Psi\Phi_1+\nonumber\\& (\mathbf{g}_2\cos(2\theta)-f_2\sin(2\theta))\bar{\Psi}\Psi\Phi_2+ (\mathbf{g}_2\sin(2\theta)+f_2\cos(2\theta))\iimg\bar{\Psi}\Psi\Phi_2
	\label{twomhsint}	
\end{align}
The chiral couplings of both $\Phi_1$ and $\Phi_2$ can be set to $0$ only if $\frac{\mathbf{f}_1}{\mathbf{g}_1}=\frac{\mathbf{f}_2}{\mathbf{g}_2}$; which is not true in general. Therefore, in the theory of multiple MHS, we must consider the couplings of MHS with all four bilinears. As discussed, in a theory of single MHS, we can flip the signs of the even-ranked bilinear couplings that are irrelevant (i.e., not physical). In the case of multiple higher spins, only the ratio of the sign of even ranked bilinears is physical. For example, in \eqref{twomhsint}, the sign of $\mathbf{g}_1/\mathbf{g}_2$ ( or $\mathbf{f}_1/\mathbf{f}_2$) cannot be changed using the discrete field redefinitions. 
	  
\subsection{Odd dimensional spacetime}
\label{sec:ckrydiracthreepointodddiminteraction}
In odd $D=2m+1$ spacetime, the one particle states of the massless spinors transform under $Spin(2m-1)$. As discussed in appendix \ref{app:productofoddirrpes}, the irreducible antisymmetric tensors in the product of two spinor states of $Spin(2m-1)$ has a maximum rank of $m-1$. On the other hand, the one particle states of the MHS transforms under $Spin(2m)$, which has a maximum rank of $m$. By virtue of the Lorentz invariance condition \eqref{rsconstraint}, any MHS $\Phi^{(\tilde{r},\tilde{s})}$ with rank $0<\tilde{r}<m$ would have two independent couplings corresponding to its projection over the spinor bilinears $\mathbb{M}^{(\tilde{r},\tilde{s})}$ and $\mathbb{M}^{(\tilde{r}-1,\tilde{s}+1)}$. For rank $\tilde{r}=0$, i.e., massive scalar, we get the only possible coupling, the Yukawa term $\bar{\Psi}\Psi\Phi$. The couplings for $\tilde{r}<m$ are summarized in table \ref{tab:oddcouplings}. These are same as the couplings derived using branching rules in appendix \ref{app:Projectionofodddimirreps}.
\renewcommand{\arraystretch}{2}
\begin{table}[H]
	\centering
	\begin{tabular}{ ||c||c|c|c|| }
		\hline
		\hline
	S.no.&	Notation            & Three-point functions                   & $\tilde{r} $                    \\
		\hline
		\hline
	$1$&	$\mathbf{e} ~	\mathbb{M}^{(\tilde{r},\tilde{s})}\Phi^{(\tilde{r},\tilde{s})}$& $\mathbf{e} ~ (\iimg)^{\tilde{s}+\frac{\tilde{r}(\tilde{r}+1)}{2}}\, (\partial^{\nu_1\cdots\nu_{\tilde{s}}}\bar{\Psi})\Gamma^{\mu_1\cdots\mu_{\tilde{r}}}\, \Psi \, \Phi_{[\mu_1\cdots\mu_{\tilde{r}}]\{\nu_1\cdots\nu_{\tilde{s}} \}}$&$0\leq \tilde{r}\leq m-1$
		\\ 
		\hline
	$2$&	$\mathbf{f}~\mathbb{M}^{(\tilde{r}-1,\tilde{s}+1)}\Phi^{(\tilde{r},\tilde{s})}$&$ \mathbf{f} ~ \iimg^{\tilde{s}+1+\frac{\tilde{r}(\tilde{r}-1)}{2}} (\partial^{\nu_1\cdots\nu_{\tilde{s}}}\partial^{\mu_{\tilde{r}}}\bar{\Psi})\Gamma^{\mu_1\cdots\mu_{\tilde{r}-1}}\, \Psi \, \Phi_{[\mu_1\cdots\mu_{\tilde{r}}]\{\nu_1\cdots\nu_{\tilde{s}} \}}$&$1\leq \tilde{r}\leq m$\\
		\hline
		\hline
	\end{tabular}\caption{Three point couplings of real $\Phi^{(\tilde{r},\tilde{s})}$ in $D=2m+1$ spacetime.}	\label{tab:oddcouplings}
\end{table}
For rank $m$ MHS, the one particle states decomposes into a sum of self-dual and anti-self-dual given by \eqref{selfdual} and \eqref{antiselfdual}. These massive irreps are new in odd spacetime, which were not present in the even spacetime which gives rise to two independent couplings. One of the couplings is given by row $2$ of table \ref{tab:oddcouplings} for $\tilde{r}=m$. The other coupling is also given by row $2$ by replacing $\Phi^{(\tilde{r},\tilde{s})}$ with $\tilde{\Phi}^{(\tilde{r},\tilde{s})}$. For now we can denote the coupling constant for $\tilde{\Phi}^{(\tilde{r},\tilde{s})}$ by $\tilde{\mathbf{f}}$. 
%The coupling constants of the self dual and the anti-self dual couplings are then given by  
%
%
% Consider the rank $m$ MHS representations $\Phi^{(m,\tilde{s})}$ and $\widetilde{\Phi}^{(m,\tilde{s})}$ which are dual to each other using \eqref{dualitycovariance}. These MHS can be projected over the Dirac bilinears of rank $m-1$ in the following way
%\begin{align}
% \frac{1}{2}{\mathbf{d}}~\mathbb{M}^{(m-1,\tilde{s}+1)}{\Phi}^{(m,\tilde{s})}+\frac{1}{2}\tilde{\mathbf{d}}~\mathbb{M}^{(m-1,\tilde{s}+1)}\widetilde{\Phi}^{(m,\tilde{s})}\label{hodgedualcouplings}
%\end{align}
Then using the duality condition \eqref{dualitycovariance}, the self dual coupling have coupling constant $\frac{1}{2}(\mathbf{f}+\tilde{\mathbf{f}})$. Whereas, the anti self dual coupling have coupling constant $\frac{1}{2}(\mathbf{f}-\tilde{\mathbf{f}})$. For $m=odd$, both the self-dual and anti-self-dual MHS are real. Therefore, their couplings are hermitian, and the coupling constants are also real. Whereas for $m=even$, both the self-dual and anti self-dual MHS are complex. Therefore, their couplings must also involve their hermitian conjugates, and the coupling constants would generally be complex.

For $\tilde{r}=1,\tilde{s}=0$, table \ref{tab:oddcouplings} gives the minimal and non-minimal QED couplings $\mathbf{e}\, \iimg \bar{\Psi}\Gamma^\mu \Psi A_\mu$ and $\mathbf{f}\,  (\partial^\mu\bar{\Psi}) \Psi A_\mu$. The minimal and non-minimal couplings together always break the reflection symmetry. Similar arguments can be made for any MHS. For a fixed $\tilde{s}$, the coupling $\mathbf{e}$ is the minimal coupling when $\tilde{r}=2\tilde{n}$, $\tilde{n}\in \mathbb{Z}$. Whereas the coupling $\mathbf{f}$ is the minimal coupling when $\tilde{r}=2\tilde{n}+1$, $\tilde{n}\in \mathbb{Z}$. The breaking of the reflections when the minimal and non-minimal couplings are present in a Lagrangian will be shown explicitly in section \ref{sec:SMHSodd}.

As discussed at the end of sec \ref{sec:ckrydirecbilinear}, we cannot flip the signs of the bilinears in odd dimensions. Therefore, the signs of the coupling constants $\mathbf{e},\, \mathbf{f},\,  \tilde{\mathbf{f}}$ have physical significance. For example, the mass terms in odd dimensions are always odd under reflections.

In any $D$ spacetime, the tables \ref{tab:evencouplings} and \ref{tab:oddcouplings} describe all the possible couplings quadratic in massless Dirac spinor and a massive higher spinning boson. The counting of the couplings goes as follows
\begin{itemize}
	\item  In even dimensions, for $\tilde{r}\neq 0$, there are four independent couplings; two chiral and two non-chiral.  For $\tilde{r}=\tilde{s}=0$, there are only two independent couplings.\footnote{ Since $\tilde{r}=0$ implies $\tilde{s}=0$, there are no couplings for $\tilde{r}=0,\tilde{s}\neq 0$.}
	\item In odd dimensions,  for $\tilde{r}\neq 0$, there are two independent couplings.  For $\tilde{r}=\tilde{s}=0$, there is only one independent coupling.
\end{itemize}
In any $D$ spacetime, the MHS have a maximum rank $\lfloor {(D-1)/2}\rfloor$ governed by the  Little group $SO(D-1)$. Since $SO(D-1)\subset SO(D+1)$, the couplings in $D+2$ spacetime upto rank $\lfloor {(D-1)/2}\rfloor $ are same as that in $D$ spacetime.

\section{$C,R_i,T$ transformations}
\label{sec:ckrydiraccrt}

Any relativistic QFT is known to always remain invariant under an antilinear-antiunitary discrete transformation known as the $CRT$ transformation \cite{PhysRev.82.914,Weinberg:1995mt,10.1063/1.3060063,osti_4338493,Luders:1954zz,LUDERS19571,76b9be9a-dc2a-31f1-86b5-18215df947a7,CPTwang,RevModPhys.88.035001,Stone_2022,10.1143/PTP.74.610}. We start by pointing out that the name $CRT$ is a bit of misleading. The name suggests that the $ CRT $ is a set of three consecutive discrete transformations: a time reversal transformation $ T $, the reflections $ R $, and the charge conjugation $ C $. However, there exist QFTs where not all of them are individually well-defined. For example, in $D=10$, the type $IIB$ supergravity is chiral asymmetric, which does not have well-defined $R$ and $T$ but has a well-defined $C$ and $RT\equiv C^{-1}(CRT)$. Therefore, depending on the spacetime dimension and the spectrum of the QFT, only a smaller subset of $C, R,$ and $T$ might be well-defined.

\begin{figure}[H]
	\begin{center}
		\begin{tikzpicture}[scale=1.7]
			
			\filldraw[color=white] (0,0) circle [radius=2.5];
			\draw[ultra thick, color=black] (0,0) circle [radius=3.2];
			
\draw [dashed, ultra thick, blue][<->] (0,-3.4)--(0,3.4);
\draw [dashed, ultra thick, red][<->] (-3.4,0)--(3.4,0);
\draw (3.8,0) node{ $\bf t_0$ axis}; 
\draw (0,3.6) node{ $\bf t_1$ axis}; 
			
			\begin{scope}[shift={(0,2.5)}]
				\draw (.08,.15) node{ $Cl(1_+,1_-)\simeq \mathbb{R}$}; 
				\draw (0,-.15) node{ $t_0=+1,\, t_1=-1$}; 
			\end{scope}

			\begin{scope}[shift={(1.5,1.4)}]
				\draw (0,.15) node{ $Cl(2_+,1_-)\simeq \mathbb{R}\oplus \mathbb{R}$}; 
				\draw (0,-.15) node{ $t_0=+1,\, t_1=-1$}; 
			\end{scope}

			\begin{scope}[shift={(2.15,0)}]
				\draw (0,.15) node{ $Cl(3_+,1_-)\simeq \mathbb{R}$}; 
				\draw (0,-.15) node{ $t_0=+1,\, t_1=- 1$}; 
			\end{scope}
			
			\begin{scope}[shift={(1.67,-1.4)}]
				\draw (0,.15) node{ $Cl(4_+,1_-)\simeq \mathbb{C}$}; 
				\draw (0,-.15) node{ $t_0=+1,\, t_1=+1$}; 
			\end{scope}

			\begin{scope}[shift={(0,-2.5)}]
				\draw (.08,.15) node{ $Cl(5_+,1_-)\simeq \mathbb{H}$}; 
				\draw (0,-.15) node{ $t_0=-1 ,\, t_1=+1$}; 
			\end{scope}

			\begin{scope}[shift={(-1.65,-1.4)}]
				\draw (0,.15) node{ $Cl(6_+,1_-)\simeq \mathbb{H}\oplus \mathbb{H}$}; 
				\draw (0,-.15) node{ $t_0=-1,\, t_1=+1$}; 
			\end{scope}
			
			\begin{scope}[shift={(-2.13,0)}]
				\draw (0,.15) node{ $Cl(7_+,1_-)\simeq \mathbb{H}$}; 
				\draw (0,-.15) node{ $t_0=-1,\, t_1=-1$}; 
			\end{scope}
			
			\begin{scope}[shift={(-1.75,1.4)}]
				\draw (0,.15) node{$Cl(0,1_-)\simeq \mathbb{C}$}; 
				\draw (0,-.15) node{ $t_0=-1,\, t_1=-1$}; 
			\end{scope}

			\draw (0,-3.65) node{ $Cl(s,t)\quad:\quad \{\Gamma^\mu, \Gamma^\nu\}=2\eta^{\mu \nu} $};

		\end{tikzpicture}

	\end{center}
	\caption{Real Clifford algebra}
	\label{fig:t0t1}	
\end{figure}
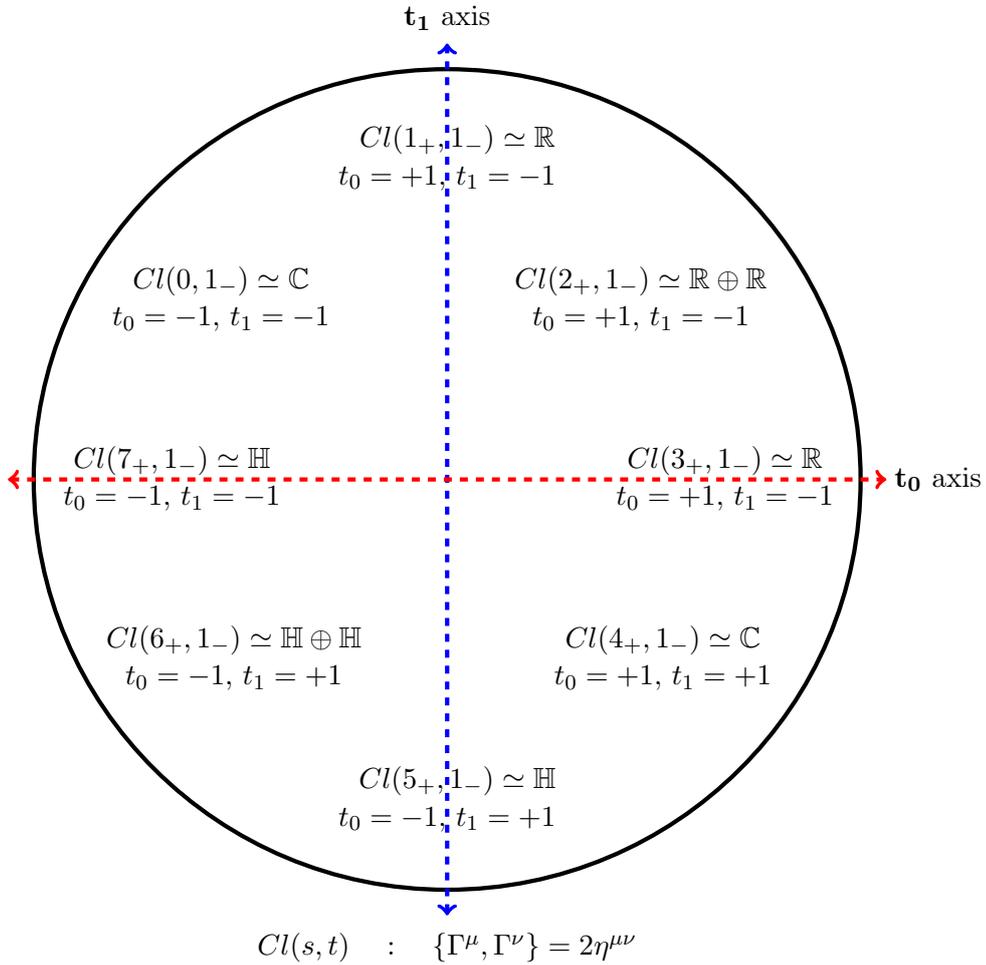

In this paper, we restrict ourselves to Dirac representations, which makes our work simple. For Dirac representations, $C, R,$ and $T$ are individually well defined \cite{Stone_2022, Wan:2023nqe, Zohar/SciPostPhys.8.4.062, 10.1143/PTP.74.610}. In this section, we verify that the three-point functions constructed in the previous section are invariant under $CRT$. We also do a detailed analysis of these couplings under different combinations of $C,\,R,\,T$, which are (i) $C,\, RT$, (ii) $R,\, CT$, and (iii) $T, \, CR$. We show that the couplings have a mod $2$ classification based on their transformation under these combinations. We denote the reflections along the $i^{th}$ spatial direction by $R_i$. The action of the reflection and time reversal on the Clifford algebra elements are as follows 
\begin{align}
	\Gamma^i\Gamma^\mu \Gamma^i
	=-(\mathcal{R}_i)^\mu{}_\nu \Gamma^\nu\qquad,&\qquad
	\Gamma^0\Gamma^\mu \Gamma^0 =\mathcal{T}^\mu{}_\nu \Gamma^\nu
\end{align}
where $\mathcal{R}_i$ and $\mathcal{T}$ are linear-unitary spacetime representation of $R_i$ and $T$ given by the diagonal matrices
\begin{align}
	(\mathcal{R}_i)^{\mu}{}_\nu=\begin{cases}
		+1&~~ \forall~ \mu=\nu\neq i,\\
		-1&~~ \forall~ \mu=\nu= i,\\
		0&~~\forall ~\mu\neq \nu
	\end{cases}\qquad,\qquad 	(\mathcal{T})^{\mu}{}_\nu=\begin{cases}
		+1&~~ \forall~ \mu=\nu\neq 0,\\
		-1&~~ \forall ~\mu=\nu= 0,\\
		0&~~\forall ~\mu\neq \nu
	\end{cases}
\end{align}
For future purposes, it will be useful to mention the following identities; these identities capture the symmetric and conjugation properties of Clifford algebra elements
\begin{align}
	\mathcal{C}\Gamma^\mu \mathcal{C}^{-1}=t_0t_1(\Gamma^\mu)^T\qquad,&\qquad \mathcal{B}\Gamma^\mu \mathcal{B}^{-1}=-t_0t_1(\Gamma^\mu)^*
\end{align}
The matrices $\mathcal{C}$ and $\mathcal{B}$ are called the charge conjugation and the complex conjugation matrices. For symmetric and conjugation properties of $\mathcal{C}$ and $\mathcal{B}$, we have followed the same conventions as \cite{Freedman:2012zz}; the readers can check that reference for the derivation of the above identities. The parameters $t_0,\, t_1$ are either $+1$ or $-1$ depending on the spacetime dimension, summarized in figure \ref{fig:t0t1}. We emphasize that the charge conjugation matrix $\mathcal{C}$ is different from the charge conjugation operator $C$. In Euclidean spacetime, $\mathcal{C}$ is the spinor representation of the operator $C$. 
%It has been historical baggage to call $\mathcal{C}$ as the charge conjugation matrix in Minkowski spacetime.

We first derive all the possible transformations of spinors and bosons under $C, R_i, T$, which preserve the kinetic term of the action. In later sections, we look at the action of these transformations over the interactions and the mass terms. The kinetic term for Dirac spinors in every dimension is given by $\bar{\Psi}\Gamma^\mu(\partial_\mu\Psi)$. In a theory of scalars, we know that the $C,\,R_i,\,T$ transformations, which keep the kinetic term invariant, always keep the MHS term invariant. However, this is not true for Dirac spinors. The Dirac mass term might break certain symmetries depending on the spacetime dimensions. At the end of this section, we will discuss the transformations of the Dirac mass term in even and odd dimensions.

We define the "conventional" charge conjugation transformation\footnote{The unconventional $C,\,R_i,\,T$ is discussed in section \ref{sec:unconventionalCRT}.} as a linear transformation (in the Hilbert space) which maps a one-particle state without changing its Little group quantum numbers to its antiparticle state (see \cite{Weinberg:1995mt}). The action of charge conjugation over the Dirac spinor field is such that it maps the spinor to it is conjugate up to the complex conjugation matrix $\mathcal{B}^{-1}$,
\begin{align}
	C\Psi(x) C^{-1} & =\zeta_{_C}\mathcal{B}^{-1} \Psi^*(x)
\label{Cspinor}
\end{align}
$\zeta_{_C}\in U(1)$ is known as the charge conjugation intrinsic parity. Since $\Psi$ have the internal symmetries $\mathcal{U}(\theta)$ , the intrinsic parity $\zeta_{_C}$ can always be chosen as $\pm 1$.
We emphasize that the charge conjugation operation is linear over the Hilbert space but antilinear over the field space. Next, we look at the action over the higher spinning bosonic fields. The charge conjugation maps the MHS $\Phi^{(\tilde{r},\tilde{s})}$ to it's conjugate field ${\Phi^{(\tilde{r},\tilde{s})}}^*$,
\begin{align}
	C\Phi^{(\tilde{r},\tilde{s})}(x)C^{-1}=\eta_{_C}^{(\tilde{r},\tilde{s})}{\Phi^{(\tilde{r},\tilde{s})}}^*(x)\label{Cboson}
\end{align}
where the intrinsic parity $\eta^{(\tilde{r},\tilde{s})}_{_C}$ of MHS which can be chosen is $+1$ or $-1$ depending on $\tilde{r}$ and $\tilde{s}$. For real MHS ${\Phi^{(\tilde{r},\tilde{s})}}^*=\Phi^{(\tilde{r},\tilde{s})}$, the charge conjugation transformation is trivial up to the intrinsic parity which squares to $+1$. Reflections, on the other hand, act on the space coordinates such that $	R_i\,:\, x^\mu\rightarrow (\mathcal{R}_i)^\mu{}_\nu x^\nu$. The reflection transformations of the generators of the Spin group are given by
\begin{align}
	R_i\iimg J^{\mu\nu}R_{i}^{-1}=\iimg(\mathcal{R}_i)^\mu{}_\rho(\mathcal{R}_i)^\nu{}_\sigma J^{\rho\sigma}\qquad ,\qquad 	R_i\iimg P^{\mu}R_{i}^{-1}=\iimg(\mathcal{R}_i)^\mu{}_\nu P^{\nu}
\end{align}
which shows that $R_i$ is a linear unitary operator over the Hilbert space. The conventional form of reflection transformations over the spinor fields, which preserve the kinetic term of the action, is given by
\begin{align}
	R_i\Psi(x) R_i^{-1} & =\zeta_{_R}\gamma^i \Psi (\mathcal{R}_ix)\label{Rspinor}
\end{align}
where $\zeta_{_R}$ is the reflection intrinsic parity which can be a phase $e^{\iimg \theta}$. At the level of Dirac bilinears and their couplings, these phases do not contribute. Reflections over the MHS act as follows
\begin{align}
	R_i\Phi^{(\tilde{r},\tilde{s})}(x) R_i^{-1} & =\eta_{_R}^{(\tilde{r},\tilde{s})}(\mathcal{R}_i)^{\tilde{r}+\tilde{s}} \Phi^{(\tilde{r},\tilde{s})} (\mathcal{R}_ix)\label{Rboson}
\end{align}
where $(\mathcal{R}_i)^{\tilde{r}+\tilde{s}}$ denotes the $\tilde{r}+\tilde{s}$ factors of $( \mathcal{R}_i)^\mu{}_\nu$ and $\eta_{_R}^{(\tilde{r},\tilde{s})}$ denotes the MHS intrinsic parity under reflection.  For example, the transformation of symmetric rank $2$ MHS $\Phi^{(1,1)}\equiv\Phi^{\{\mu\nu\}}(x)$ would be given by
\begin{align}
	R_i\Phi^{\{\mu\nu\}}(x) R_i^{-1} & =\eta_{_R}^{(1,1)}(\mathcal{R}_i)^{\mu}{}_{\rho}\, (\mathcal{R}_i)^{\nu}{}_\sigma \,
	\Phi^{\{\rho\sigma\}} (\mathcal{R}_ix)\label{reflectionexample}
\end{align}
 For real MHS $\Phi^{\tilde{r},\tilde{s}}$, we can always choose $\eta_{_R}^{(\tilde{r},\tilde{s})}$ as $\pm1$ depending on $\tilde{r},\tilde{s}$. There also exists space inversions along all the spatial directions 
\begin{align}
	\mathcal{P}^{\mu}{}_\nu=\begin{cases}
		+1&~~ \forall~ \mu=\nu= 0,\\
		-1&~~ \forall~ \mu=\nu\neq 0,\\
		0&~~\forall ~\mu\neq \nu
	\end{cases}
\end{align}
where $\mathcal{P}$ are the space-time representations of the unitary parity operators $P$. In odd dimensions, parity operation is connected to the identity by the rotations in spatial planes. In even spacetime dimensions, parity is a combination of reflections along all the spatial directions given by $	P=R_1R_2\cdots R_{2m-1}$. The transformation of a Dirac spinor under parity is given by
\begin{align}
	P\psi(x)P^{-1}=\zeta_{_P}\gamma^0\gamma_\star\psi(\mathcal{P}x)
\end{align}
where $\zeta_{_P}$ is the intrinsic parity, which can be a phase. The Time reversal transformations flips the time coordinate as $	T~:~ x^\mu \rightarrow \mathcal{T}^\mu{}_\nu x^\nu$. The time reversal transformations of the generators of the Spin group are given by
\begin{align}
	T\iimg J^{\mu\nu}T^{-1}=\iimg\mathcal{T}^\mu{}_\rho\mathcal{T}^\nu{}_\sigma J^{\rho\sigma}\qquad ,\qquad 	T\iimg P^{\mu}T^{-1}=\iimg\mathcal{T}^\mu{}_\nu P^{\nu}
\end{align}
Since there are no negative energy states (energies less than the vacuum state are not possible), the second equality implies that $T$ must be an antilinear-antiunitary operator. The time reversal transformation preserves the kinetic term only if the spinors and the MHS must transform as follows
\begin{align}
	T\Psi(x) T^{-1}     & =\zeta_{_T}\mathcal{C}\Psi (\mathcal{T}x)\label{Tspinor}\\
	T\Phi^{(\tilde{r},\tilde{s})}(x) T^{-1}     & =\eta_{_T}^{(\tilde{r},\tilde{s})}(\mathcal{T})^{\tilde{r}+\tilde{s}}\Phi^{(\tilde{r},\tilde{s})} (\mathcal{T}x)\label{Tboson}
\end{align}
where $(\mathcal{T})^{\tilde{r}+\tilde{s}}$ denotes the $\tilde{r}+\tilde{s}$ factors of $ \mathcal{T}^\mu{}_\nu$, whereas $\zeta_{_T}$ and $\eta_{_T}^{(\tilde{r},\tilde{s})}$ are the time reversal intrinsic parity for spinors and MHS. An example of this notation is discussed in equation \eqref{reflectionexample}. Since $T$ is antilinear, we can always choose the intrinsic parity for Dirac spinors as $\pm1$.\footnote{The intrinsic parities $\zeta_{_C}$ and $\zeta_{_T}$ cannot always be chosen as $\pm1$ simultaneously.} Moreover, the intrinsic parity for MHS also depends on $\tilde{r}$ and $\tilde{s}$.

In even spacetimes, together with the intrinsic parities $\zeta_{_C},\, \zeta_{_R},\, \zeta_{_T}$, the $C, R_i, T$ transformations of spinors  are unique up to factors of $\iimg$ and/or $\Gamma_\star$. These factors can be incorporated separately using the vector $\mathcal{U}(\theta)$ or the axial $\mathcal{U}_\star(\theta)$ internal symmetries. We can redefine the $C,R_i,T$ transformations upto chiral phases such that
\begin{align}
	C\qquad:&\qquad \Psi(x)\rightarrow \zeta_{_C}\, e^{\iimg \theta_{_C}\Gamma_\star}\,\mathcal{B}^{-1}\, \Psi^*(x)\label{Cspinorm}\\
	R_i\qquad:&\qquad \Psi(x)\rightarrow \zeta_{_R}\, e^{\iimg \theta_{_R}\Gamma_\star}\,\Gamma^i\, \Psi(\mathcal{R}_ix),\label{Rspinorm}\\
	T\qquad:&\qquad \Psi(x)\rightarrow \zeta_{_T}\, e^{\iimg \theta_{_T}\Gamma_\star}\,\mathcal{C}\, \Psi(\mathcal{T}x)\label{Tspinorm}
\end{align}
where $\theta_{_C},\theta_{_R},\theta_{_T}$ are chiral angles corresponding to $C,\,R_i,\,T$ which might or might not be equal to each other.
It is easy to check that these modified $C,\,R_i,\,T$ also preserve the Dirac kinetic term. Let us see how the Dirac mass terms transform under the $C,\,R_i,\,T$ transformations. In any $D$ dimension, the Dirac mass term is given by $\bar{\Psi}\Psi$ which under charge conjugation \eqref{Cspinor} would transform as follows
\begin{align}
	C\qquad:\qquad	\bar{\Psi}\Psi\rightarrow \Psi^T\mathcal{B}\beta\mathcal{B}^{-1}\Psi^*=-t_0t_1\Psi^T (\beta)^*\Psi^*=-t_0t_1\Psi^\dagger\beta\Psi=-t_0t_1\bar{\Psi}\Psi
\end{align}
where in the second equality we have used $\mathcal{B}\Gamma^\mu\mathcal{B}^{-1}=-t_0t_1$. Using figure \ref{fig:t0t1}, we know that in odd dimensions $t_0t_1=(-1)^m$. Therefore, the mass term preserves $C$ in $D=4k+3$ but breaks it in $4k+1$. In even dimensions, we can always work with the modified $C$ in \eqref{Cspinorm} and choose $\theta_{_C}$ such that $\cos(2\theta_{_C})t_0t_1=-1$ and $\sin(2\theta_{_C})=0$. Therefore, the mass term always preserves $C$ in any even spacetime. The action of reflections \eqref{Rspinor} over the mass term is as follows
\begin{align}
	R_i\qquad:\qquad \bar{\Psi}\Psi\rightarrow \Psi^\dagger \Gamma^i\beta \Gamma^i\Psi=-\bar{\Psi}\Psi
\end{align}
Again, we cannot absorb the minus sign in odd dimensions because the only field space symmetry is $\mathcal{U}(\theta)$, which keeps the mass term invariant. Therefore, the mass term breaks $R_i$ in any odd dimension. In even dimension, we can use the redefined reflections in \eqref{Rspinorm} and choose $\theta_{_R}$ such that $\cos(2\theta_{_R})=-1$ and $\sin(2\theta_{_R})=0$ and the mass term always remain invariant. The action of time reversal in \eqref{Tspinor} over the mass term is as follows
\begin{align}
	T\qquad;\qquad 	\bar{\Psi}\Psi\rightarrow \Psi^\dagger \mathcal{C}\beta^*\mathcal{C}^{-1}\Psi=t_0t_1\bar{\Psi}\Psi
\end{align}
where we have used $\mathcal{C}\beta^*\mathcal{C}^{-1}=t_0t_1\beta$ using \eqref{CBproperty}. In odd dimensions, $t_0t_1=(-1)^{m+1}$ which means that the mass term breaks $T$ in $D=4k+3$ and preserves it in $D=4k+1$. In even dimension, similar to charge conjugation and reflection, we can absorb the sign $t_0t_1$ by choosing $\theta_{_T}$ such that $\cos(2\theta_{_T})t_0t_1=+1$ and $\sin(2\theta_{_T})=0$. Therefore, the mass term always preserves $T$ in any even dimensions. The $C,\,R_i,\,T$ transformation of the mass term are summarized in table \ref{tab:CRTmassterm}.
	\renewcommand{\arraystretch}{1.5}
\begin{table}[H]
	\centering
	\begin{tabular}{ ||c||c|c|c|| }
		\hline
		\hline
		$D$ mod $4$            & $ C$                   & $ R_i$                            & $ T$                                    \\
		\hline
		$0$ & $\checkmark$     & $\checkmark$ & $\checkmark$  \\
		\hline
		$1$ & $\times$     & $\times$ & $\checkmark$  \\
		\hline
		$2$ & $\checkmark$     & $\checkmark$ & $\checkmark$  \\
		\hline
		$3$ & $\checkmark$     & $\times$ & $\times$   \\
		\hline
		\hline
	\end{tabular}\caption{Dirac mass term under $C,R_i,T$ in $D$ mod $4$ spacetime.}	\label{tab:CRTmassterm}
\end{table}
Table \ref{tab:CRTmassterm} shows that the mass term can always be added to a theory of Dirac spinors in even spacetime dimensions without breaking $C,\,R_i,\,T$ transformation.  However, it should be emphasized that the mass term breaks the axial transformation and generates its counterpart chiral mass term $\iimg \bar{\Psi}\Gamma_\star\Psi$.

\section{$C,R_i,T$ and three point interactions}
\label{sec:MultipleMHS}
The $C,\,R,\,T$ transformations of interacting theories have played a vital role in understanding the nature of a theory. For example, in the standard model of particle physics, the weak forces violate the $CR$ transformation, which, on the other hand, is preserved by the strong and electromagnetic interactions. All the discrete $C,\,R,$ and $T$ transformations are preserved in QED. Moreover, every relativistic quantum field theory preserves $CRT$ transformation. The important aspect of what discrete transformations are preserved by a theory (the interacting part of the Lagrangian) is that it fixes the intrinsic parity of the particles. In other words, for an interaction to preserve a certain discrete symmetry, the corresponding intrinsic parity of the particles must remain conserved. For example, in QED, the intrinsic parity of the gauge boson under $C,\,R,\,T$ are fixed as $\eta_{_C}=-\eta_{_R}=\eta_{_T}=-1$. Whereas, in Yukawa theory or in the Standard Model, the intrinsic parity of the spin $0$ boson or the Higgs boson is $\eta_{_C}=\eta_{_R}=\eta_{_T}=1$. However, these theories only involve bosons up to spin $1$. Not much is known about the $C,\,R,\,T$ (or $C,\,R_i,\,T$) transformations of interacting theories MHS particles. In this section, we discuss the $C,\, R_i,\, T$ transformations of the Dirac couplings constructed in section \ref{sec:ckrydiracthreepointinteraction}. We aim to fix the intrinsic parities of the MHS interacting with a massless Dirac spinor through the three-point couplings. As a warm up exercise, one can work with $1+1$ spacetime dimension model where a spin $1$ boson interacts with a Dirac spinor interacts through the following three point couplings
\begin{align}
\mathcal{L}^{int}=\mathbf{e}~	\iimg \bar{\Psi}\Gamma^\mu \Psi A_\mu+\mathbf{f}~	\iimg \bar{\Psi}\Gamma^\mu \Gamma_\star\Psi A_\mu +\mathbf{g}~ \iimg \bar{\Psi} (\partial^{\mu}\Psi) A_\mu +\mathbf{h}~ \bar{\Psi}\Gamma_\star (\partial^{\mu}\Psi) A_\mu\label{Lintspin1}
\end{align}
As discussed earlier, only the even ranked couplings $\mathbf{g}$ and $\mathbf{h}$ are sensitive to the chiral transformations $\mathcal{U}_\star(\theta)$. Therefore the chiral angles in \eqref{Cspinorm}-\eqref{Tspinorm} can always be tuned such that the even ranked couplings are always invariant under $C,~R,~T$. The $C,~R,~T$ transformations of the odd ranked couplings are always independent of the choice of chiral angles. As we know, the $QED$ coupling $\mathbf{e}$ always preserves $C,R,T$ with the choice of intrinsic parities $\eta_{_C}=-\eta_{_R}=\eta_{_T}=-1$. With this choice of intrinsic parities, the chiral coupling $\mathbf{f}$ always breaks $R$ and $T$ but preserves $C$.
\begin{align}
		\mathbf{f}\xrightarrow{C} \mathbf{f}\qquad,&\qquad 	\mathbf{f}\xrightarrow{R} -\mathbf{f}\qquad,\qquad 	\mathbf{f}\xrightarrow{T} -\mathbf{f}
\end{align}
Therefore, an interacting theory of massive spin $1$ boson given by \eqref{Lintspin1} preserves $C$ and $RT$. A higher spin boson in $1+1$ spacetime has the same degrees of freedom as the spin $1$ boson. So the same results hold true for any spin $J$ massive boson with the unique choice of intrinsic parity $\eta_{_C}=\eta_{_R}\eta_{_T}=(-1)^J$. However, the statement changes for the same theory in $3+1$ spacetime. Since the charge conjugation matrix $\mathcal{C}$ commutes with $\Gamma_\star$ in $3+1$, whereas it anticommutes with $\Gamma_\star$ in $1+1$ spacetime. As compared to $1+1$ spacetime, there's an extra minus sign under $C$ and $T$ transformation of the odd ranked bilinears implying that a similar theory in $3+1$ dimensions must remain invariant under $CR$ and $T$. The intrinsic parity of any spin $J$ massive boson is given by $\eta_{_T}=\eta_{_C}\eta_{_R}=(-1)^{J}$. 

In odd spacetime, there are no chiral transformations. So both the even and the odd ranked couplings play a vital role. However, the similar arguments can be made in similar theories in $D$ and $D+2$ spacetime. Using \eqref{CBproperty} with the choice of signs in figure \ref{fig:t0t1}, the charge conjugation matrix $\mathcal{C}$ satisfies different algebras with the clifford generators $\Gamma^\mu$,
\begin{align}
	\mathcal{C}\Gamma^\mu\mathcal{C}^{-1}=-(\Gamma^\mu)^T\qquad,\qquad D=3,7\\
		\mathcal{C}\Gamma^\mu\mathcal{C}^{-1}=+(\Gamma^\mu)^T\qquad,\qquad D=5,9
\end{align}
This implies that the $C$ and $T$ transformations of any even ranked couplings (the bilinears also have a $\beta=\iimg \Gamma^0$ matrix) would be opposite in $D=3,7$ as compared to that in $D=5,9$ spacetime. So if a theory is $C,~RT$ is invariant in $D=Odd$ spacetime, it should be $CR,~T$ invariant in $D+2$ spacetime.

We will show these results explicitly for any MHS in what follows next. In even spacetime, we use the $C,~R_i,$ and $T$ transformations independent of the chiral phases given by \eqref{Cspinor}, \eqref{Rspinor}, and \eqref{Tspinor} respectively. As discussed above, the results must remain independent of the chiral phases. We first summarize the transformations of the bilinears $\mathbb{M}^{(r,s)}$ and $\mathbb{N}^{(r,s)}$ in table \ref{tab:CRTIDirac} in any $D=2m/2m+1$ spacetime.
\renewcommand{\arraystretch}{2}
\begin{table}[H]
	\centering
	\begin{tabular}{ ||c||c|c|c|| }
		\hline
		\hline
		Bilinears            & $ C$                   & $ R_i$                            & $ T$                                    \\
		\hline
		$\mathbb{M}^{(r,s)}$ & $(-1)^{s+1}t_1t_r$     & $(-1)^{r+1}(\mathcal{R}_i)^{r+s}$ & $(-1)^{r+s}t_1t_r(\mathcal{T})^{r+s}$   \\
		\hline
		$\mathbb{N}^{(r,s)}$ & $(-1)^{r+s+m+1}t_1t_r$ & $(-1)^{r}(\mathcal{R}_i)^{r+s}$   & $(-1)^{s+m+1}t_1t_r(\mathcal{T})^{r+s}$ \\
		\hline
		\hline
	\end{tabular}\caption{$C,R_i,T$ transformation properties of bilinears in any $D=2m/2m+1$ spacetime. See appendix \ref{App:CB} for the definition of $t_1,t_r$.}	\label{tab:CRTIDirac}
\end{table}
The transformations depend on the rank $r$, the spacetime dimension $m=\lfloor{D/2}\rfloor$, and the number of derivative $s$ in the Dirac bilinears. For arbitrary $r$ and $s$, all possible couplings of the bilinears and the MHS $\Phi^{(\tilde{r},\tilde{s})}$ are summarized in tables \ref{tab:evencouplings} and \ref{tab:oddcouplings}. Using equation \eqref{Cboson}, \eqref{Rboson}, and \eqref{Tboson}, we know that the $C, R_i, T$ transformation of the MHS depends only on their corresponding intrinsic parities, and $\tilde{r}+\tilde{s}$ factors of $\mathcal{R}_i$ or $\mathcal{T}$. But since Lorentz invariance implies $\tilde{r}+\tilde{s}=r+s$, the factors of $\mathcal{R}_i, \, \mathcal{T}$ from the MHS must exactly cancel the $\mathcal{R}_i,\, \mathcal{T}$ factors from the bilinears in table \ref{tab:CRTIDirac}. Moreover, the full $CRT$ transformations of the bilinears $\forall ~ r\in \mathbb{Z}$ are always independent on the spacetime dimension
\begin{align}
	CR_iT\qquad&:\qquad \mathbb{M}^{(r,s)}\longrightarrow (\mathcal{R}_i)^{r+s}(\mathcal{T})^{r+s}\mathbb{M}^{(r,s)} \\
	CR_iT\qquad&:\qquad \mathbb{N}^{(r,s)}\longrightarrow (\mathcal{R}_i)^{r+s}(\mathcal{T})^{r+s}\mathbb{N}^{(r,s)} 
\end{align}
In even dimensions, these bilinears couple with real MHS. In odd $D=2m+1$ dimensions, all the real MHS $\Phi^{(\tilde{r},\tilde{s})}$ for $\tilde{r}<m$ also interact with Dirac bilinear. For both these cases, $C,\,R_i,\,T$ are well defined separately. Under $CR_iT$, the real massive higher spins $\Phi^{(\tilde{r},\tilde{s})}$ transforms upto a factor of $\eta_{_C}^{(\tilde{r},\tilde{s})}\, \eta_{_R}^{(\tilde{r},\tilde{s})}\, \eta_{_T}^{(\tilde{r},\tilde{s})}\,(\mathcal{R}_i\mathcal{T})^{\tilde{r}+\tilde{s}}$. Therefore, for the couplings to be invariant under $CR_iT$, the intrinsic parities must satisfy
\begin{align}
	\eta_{_C}^{(\tilde{r},\tilde{s})}\,\eta_{_R}^{(\tilde{r},\tilde{s})}\, \eta_{_T}^{(\tilde{r},\tilde{s})}=1,\qquad \qquad \forall ~\tilde{r},\tilde{s}
\end{align}
This is a generalization of the result on the constraints intrinsic parities, given in equation $5.8.4$ of Weinberg Vol 1 \cite{Weinberg:1995mt}. Since $\eta_{_C}^{(\tilde{r},\tilde{s})},\, \eta_{_R}^{(\tilde{r},\tilde{s})},\, \eta_{_T}^{(\tilde{r},\tilde{s})}$ are the intrinsic parities of the real MHS which always equals to $\pm1$, they must satisfy
\begin{align}
	\eta_{_C}^{(\tilde{r},\tilde{s})}=\eta_{_R}^{(\tilde{r},\tilde{s})}\,\eta_{_T}^{(\tilde{r},\tilde{s})}\qquad,\qquad 	\eta_{_R}^{(\tilde{r},\tilde{s})}=\eta_{_C}^{(\tilde{r},\tilde{s})}\,\eta_{_T}^{(\tilde{r},\tilde{s})}\qquad,\qquad 	\eta_{_T}^{(\tilde{r},\tilde{s})}=\eta_{_C}^{(\tilde{r},\tilde{s})}\, \eta_{_R}^{(\tilde{r},\tilde{s})} \label{etaconstraints}
\end{align}
In odd dimensions there also exists couplings for $\tilde{r}=m$ corresponding to the MHS $\Phi^{(m,\tilde{s})}$ and $\widetilde{\Phi}^{(m,\tilde{s})}$ which are dual to each other by the virtue of \eqref{dualitycovariance}. Since the epsilon tensor in any dimensions flips under both $R_i$ and $T$, the self-dual and the anti-self-dual couplings for the MHS $\Phi^{(m,\tilde{s})}\pm \widetilde{\Phi}^{(m,\tilde{s})}$ transform into each other under $R_i$ and $T$. For $m=odd$, the MHS is real, which transforms trivially under charge conjugation. In such a case, $C,\, R_iT$ is well defined. For $m=even$, the MHS is complex and transforms non-trivially under charge conjugation as per \eqref{Cboson}. In such a case, the self-dual and anti-self-dual couplings are invariant only under $CR_iT$. The following sections only consider the $C,\,R_i,\,T$ transformations of real MHS coupled with a massless Dirac spinor.

\subsection{Even dimensional spacetime}
 Let us first discuss multiple MHSs in even spacetime, where they can have chiral couplings. As discussed in section \ref{sec:ckrydirecbilinear}, in a theory of single MHS, the even-ranked chiral couplings with $\mathbb{N}^{(2n,s)}$ are dependent on the even ranked non-chiral couplings with $\mathbb{M}^{(2n,s)}$ through chiral transformations  $\mathcal{U}_\star(\theta):\Psi\rightarrow e^{\iimg \theta\Gamma_\star}\Psi$. However, in a theory of multiple MHS, it is not possible to make such a statement unless the coupling constants are related. We start with multiple MHSs $\Phi^{(\tilde{r}_a,\tilde{s}_a)}_a$ where $a=1,2,\cdots,N$, for any arbitrary $\tilde{r}_a,\tilde{s}_a$, coupled with massless Dirac spinors of the same flavor. Multiple flavor Dirac couplings will be discussed in section \ref{sec:unconventionalCRT}. As discussed, the most general three-point interactions of the MHSs are given by
\begin{align}
	\mathcal{L}^{int}=	\mathcal{L}^{int}_{_1}+	\mathcal{L}^{int}_{_2}+\cdots +\mathcal{L}^{int}_{_N}	+\mathcal{L}^{int}_{_{mix}}=\sum_{a=1}^{N}\mathcal{L}^{int}_{_a}+\mathcal{L}^{int}_{_{mix}}\label{Lint}
\end{align}
where $	\mathcal{L}^{int}_{_a}$ involves all possible three point couplings of  $\Phi^{(\tilde{r}_a,\tilde{s}_a)}_a$ with a massless Dirac fermion and $\mathcal{L}^{int}_{_{mix}}$ involves the interactions between all $\Phi^{(\tilde{r}_a,\tilde{s}_a)}_a$. For the purpose of this paper, we keep $\mathcal{L}^{int}_{_{mix}}=0$. In $\mathcal{L}^{int}_{_a}$, the interactions are of two categories: $\mathbb{M}^{(r,s)}\Phi^{(\tilde{r}_a,\tilde{s}_a)}_a$ and  $\mathbb{N}^{(r,s)}\Phi^{(\tilde{r}_a,\tilde{s}_a)}_a$ which are subjected to the condition $r+s=\tilde{r}_a+\tilde{s}_a$ to manifest Lorentz invariance of the coupling. As discussed earlier, see \eqref{rsconstraint}, there are only two possible solutions: $r=\tilde{r}_a,~ s=\tilde{s}_a$ and $r=\tilde{r}_a-1,~s=\tilde{s}_a+1$. This implies that for any arbitrary $\Phi^{(\tilde{r}_a,\tilde{s}_a)}_a$, the most general three-point coupling with a massless Dirac spinor is given by
\begin{equation}
	\begin{split}
		\mathcal{L}^{int}_{_a}&=\mathbf{e}_a\, \mathbb{M}^{(\tilde{r}_a,\tilde{s}_a)}\Phi^{(\tilde{r}_a,\tilde{s}_a)}_a+ \mathbf{f}_a\, \mathbb{M}^{(\tilde{r}_a-1,\tilde{s}_a+1)}\Phi^{(\tilde{r}_a,\tilde{s}_a)}_a\\
		&+\mathbf{g}_a\, \mathbb{N}^{(\tilde{r}_a,\tilde{s}_a)}\Phi^{(\tilde{r}_a,\tilde{s}_a)}_a+\mathbf{h}_a\, \mathbb{N}^{(\tilde{r}_a-1,\tilde{s}_a+1)}\Phi^{(\tilde{r}_a,\tilde{s}_a)}_a
	\end{split}
\end{equation}\label{Linta}
where $\mathbf{e}_a,\mathbf{f}_a,\mathbf{g}_a,\mathbf{h}_a$ are the couplings constants for $a^{th}$ MHS. Note that if all the bosons are massive scalars, i.e., $\tilde{r}_a=\tilde{s}_a=0$, then the only non-vanishing couplings are $\mathbf{e}_a $ and $\mathbf{g}_a$. Using table \ref{tab:CRTIDirac}, we can write down the $C,\,R_i,\,T$ transformations of the couplings in $\mathcal{L}^{int}_{_a}$. These are summarized in table \ref{tab:LiCRT}.
\renewcommand{\arraystretch}{2}
\begin{table}[H]
	\begin{tabular}{ ||c||c|c|c|c|| }
		\hline
		\hline
		& $\mathbf{e}_a$                   & $\mathbf{f}_a$                            & $\mathbf{g}_a$    &$\mathbf{h}_a$                                \\
		\hline
		\hline
		$C$ & $(-1)^{\tilde{s}_a+1}t_1t_{\tilde{r}_a}$ $\eta_{_C}^{a}$   &$(-1)^{\tilde{s}_a}t_1t_{\tilde{r}_a -1}$ $\eta_{_C}^{a}$ &$(-1)^{\tilde{r}_a+\tilde{s}_a+m+1}t_1t_{\tilde{r}_a}$ $\eta_{_C}^{a}$ & $(-1)^{\tilde{r}_a+\tilde{s}_a+m+1}t_1t_{\tilde{r}_a-1}$$\eta_{_C}^{a}$\\
		\hline
		$R_i$ & $(-1)^{\tilde{r}_a+1}$  $\eta_{_R}^{a}$  &$(-1)^{\tilde{r}_a}$  $\eta_{_R}^{a}$ &$(-1)^{\tilde{r}_a}$  $\eta_{_R}^{a}$ &$(-1)^{\tilde{r}_a+1}$  $\eta_{_R}^{a}$\\
		\hline
		$T$ &  $(-1)^{\tilde{r}_a+\tilde{s}_a}t_1t_{\tilde{r}_a}$   $\eta_{_T}^{a}$  & $(-1)^{\tilde{r}_a+\tilde{s}_a}t_1t_{\tilde{r}_a-1}$  $\eta_{_T}^{a}$ &$(-1)^{\tilde{s}_a+m+1}t_1t_{\tilde{r}_a}$  $\eta_{_T}^{a}$ & $(-1)^{\tilde{s}_a+m}t_1t_{\tilde{r}_a-1}$  $\eta_{_T}^{a}$\\
		\hline
		\hline
	\end{tabular}\caption{Transformation of \eqref{Linta} under $C,R_i,T$ in $D=2m$ spacetime. The intrinsic parities of $a^{th}$ MHS are  $\eta_{_C}^{a}\equiv \eta_{_C}^{(\tilde{r}_a,\tilde{s}_a)}$,  $\eta_{_R}^{a}\equiv \eta_{_R}^{(\tilde{r}_a,\tilde{s}_a)}$,  $\eta_{_T}^{a}\equiv \eta_{_T}^{(\tilde{r}_a,\tilde{s}_a)}$.}	\label{tab:LiCRT}
\end{table}
Since $CR_iT$ is always a symmetry, every MHS satisfies $\eta_{_C}^a\eta_{_R}^a\eta_{_T}^a=+1,~ \forall ~a$. For any arbitrary $\tilde{r} $, table \ref{tab:LiCRT} gives that under $R_i,~CT$, the couplings $\mathbf{f}_a,\mathbf{g}_a$ always transforms differently as compared to $\mathbf{e}_a,\mathbf{h}_a,$ $~\forall ~a$. Therefore, $\sum_{a}\mathcal{L}^{int}_{_a}$ would always break $R_i,~ CT$. Let us check what happens under $CR_i, ~T$ and $C,~R_iT$. For $CR_i, \,T$ to be a symmetry of $\sum_{a}\mathcal{L}^{int}_{_a}$, every coupling must remain invariant under it, which implies that the factors in table \ref{tab:LiCRT} must satisfy
\begin{align}
	(-1)^{\tilde{r}_a+\tilde{s}_a}t_1t_{\tilde{r}_a}=(-1)^{\tilde{r}_a+\tilde{s}_a}t_1t_{\tilde{r}_a-1}=(-1)^{\tilde{s}_a+m+1}t_1t_{\tilde{r}_a}=(-1)^{\tilde{s}_a+m}t_1t_{\tilde{r}_a-1}
\end{align}
The first equality holds true only if $t_{\tilde{r}_a}=t_{\tilde{r}_a-1}$, whereas the last equality holds true only if $t_{\tilde{r}_a}=-t_{\tilde{r}_a-1}$. Since $t_{\tilde{r}}\neq 0$, this implies that the chiral couplings would always break $CT,~R_i$. Therefore, $\sum_{a}\mathcal{L}^{int}_{_a}$, for generic value of the coupling constants, always breaks $CT,~R_i$.\\

Similarly, for $C,R_iT$ to be a symmetry of $\sum_{a}\mathcal{L}^{int}_{_a}$, it must satisfy
\begin{align}
	(-1)^{\tilde{s}_a+1}t_1t_{\tilde{r}_a}=(-1)^{\tilde{s}_a}t_1t_{\tilde{r}_a -1}=(-1)^{\tilde{r}_a+\tilde{s}_a+m+1}t_1t_{\tilde{r}_a}=(-1)^{\tilde{r}_a+\tilde{s}_a+m+1}t_1t_{\tilde{r}_a-1}
\end{align}
Again, the first equality holds true only if $t_{\tilde{r}_a}=-t_{\tilde{r}_a-1}$, whereas the last equality holds true only if $t_{\tilde{r}_a}=t_{\tilde{r}_a-1}$. Since $t_{\tilde{r}}\neq 0$, this implies that if all the chiral couplings are present, the theory would always break $C,\, R_iT$. Therefore, $\mathcal{L}^{int}_{_a}$ always breaks $C,R_iT$ for $a=1,2,\cdots,N$. Hence, in even spacetime, the interactions of the type \eqref{Lint} would always break $C,\, R_i,\, T$ or any combination of them but always preserves $CR_iT$ such that $\eta_{_C}^a\eta_{_R}^a\eta_{_T}^a=+1~\forall ~a$.

\subsubsection{Even or odd ranked couplings}
Now, let us consider a specific type of theory where multiple MHSs are coupled only with an even or an odd ranked massless spinor bilinears. For example, consider the interaction with only the even ranked non-chiral bilinear $\mathbb{M}^{(2n,s)}$. Any MHS $\Phi^{(\tilde{r}_a,\tilde{s}_a)}_a$ would have a coupling which is of the form $\mathbb{M}^{(2n,s)}\Phi^{(\tilde{r}_a,\tilde{s}_a)}_a$ such that $2n+s=\tilde{r}_a+\tilde{s}_a$. As we know, there can only be two possible solutions: $2n=\tilde{r}_a,~s=\tilde{s}_a$ or $2n+1=\tilde{r}_a,~s-1=\tilde{s}_a$ $\forall ~a$. Under $C,R_i,T$ defined in \eqref{Cspinor}, \eqref{Rspinor}, \eqref{Tspinor}, any coupling would transform upto factors of $\eta_{_C}^{(\tilde{r}_a,\tilde{s}_a)}(-1)^{s+n+1}t_0t_1$, $-\eta_{_R}^{(\tilde{r}_a,\tilde{s}_a)}$, and $\eta_{_T}^{(\tilde{r}_a,\tilde{s}_a)}(-1)^{s+n}t_0t_1$, respectively. However, since Dirac spinors here are massless, we can use the modified $C,\,R_i,\,T$ transformations in \eqref{Cspinorm}-\eqref{Tspinorm} and suitably choose the chiral angles $\theta_{_C},\theta_{_R},\theta_{_T}$ such that $\mathbb{M}^{(2n,s)}$ is always invariant under $C,\, R_i,\, T$. This implies that in a $C,\, R_i,\, T$ invariant theory of only even ranked bilinears, the intrinsic parity of the MHS must be given by
\begin{align}
	\eta_{_C}^{(\tilde{r}_a,\tilde{s}_a)}=\eta_{_R}^{(\tilde{r}_a,\tilde{s}_a)}=\eta_{_T}^{(\tilde{r}_a,\tilde{s}_a)}=+1\qquad \forall ~a
\end{align}
The same choices can be made for a theory of MHSs coupled only with even ranked chiral bilinear $\mathbb{N}^{(2n,s)}$. These results are consistent with the already known intrinsic parities of the spin $0$ massive bosons in Yukawa coupling $\bar{\Psi}\Psi\Phi$. In our notation the MHS spin $0$ boson $\Phi$ is denoted by $\Phi^{(0,0)}$. The $C,R_i,T$ intrinsic parities of $\Phi$ are all $+1$, 
\begin{align}
	\eta_{_C}^{(0,0)}=	\eta_{_R}^{(0,0)}=	\eta_{_T}^{(0,0)}=+1
\end{align}
However, if the MHSs are coupled with an odd-ranked bilinear, then it is not possible to absorb the signs using chiral phase redefinitions. Consider a theory of multiple MHS $\Phi^{(\tilde{r}_a,\tilde{s}_a)}_a$ coupled only with odd ranked chiral bilinear $\mathbb{M}^{(2n+1,s)}$ such that $2n+s+1=\tilde{r}_a+\tilde{s}_a$. Using table \ref{tab:CRTIDirac}, in order  to preserve $C,R_i,T$, the intrinsic parities of the MHSs coupled with rank $r=2n+1$ bilinear must be given by
\begin{align}
	\eta_{_C}^{(\tilde{r}_a,\tilde{s}_a)}=\eta_{_T}^{(\tilde{r}_a,\tilde{s}_a)}=(-1)^{s+n+1}\qquad,\qquad \eta_{_R}^{(\tilde{r}_a,\tilde{s}_a)}=+1\qquad \forall ~a
\end{align}
This is consistent with the already known intrinsic parities of a spin $1$ boson in massive QED. In $3+1$ dimensions, the $C,R_i,T$ invariant massive QED coupling is given by $\bar{\Psi}\Gamma^\mu\Psi A_\mu$. Where the spin $1$ field $A_\mu$ in our notations corresponds to $\tilde{r}=1$ and $\tilde{s}=0$ MHS. Since $n=0$ and $s=0$ for the rank $1$ bilinear $\bar{\Psi}\Gamma^\mu\Psi$, the intrinsic parity of the spin $1$ boson is given by
\begin{align}
	\eta_{_C}^{(1,0)}=\eta_{_T}^{(1,0)}=-1\qquad,\qquad \eta_{_R}^{(1,0)}=+1
\end{align} 
Similarly, consider a theory of MHS  $\Phi^{(\tilde{r}_a,\tilde{s}_a)}_a$ coupled only with the odd ranked chiral bilinear $\mathbb{N}^{(2n+1,s)}$. Using table \ref{tab:CRTIDirac}, we can fix the intrinsic parities of the MHSs such that the theory preserves $C,R_i,T$,
\begin{align}
	\eta_{_C}^{(\tilde{r}_a,\tilde{s}_a)}=-\eta_{_T}^{(\tilde{r}_a,\tilde{s}_a)}=(-1)^{s+n+m}\qquad,\qquad \eta_{_R}^{(\tilde{r}_a,\tilde{s}_a)}=-1\qquad \forall ~\tilde{r}_a,\tilde{s}_a
\end{align}
where $n=\lfloor{r/2}\rfloor$, $r$ is the rank of the bilinear, and $m=\lfloor {D/2}\rfloor $.
\subsubsection{Single MHS theory}
Let us now focus on the case when a Dirac spinor is coupled to a single MHS\footnote{These are good EFTs at a scale much lower than the mass of the MHS. Otherwise, it leads to causality violation \cite{Camanho}.}. As discussed in the beginning of this section and in section \ref{sec:ckrydirecbilinear}, using the axial rotation, in case of only one MHS in \eqref{Lint} i.e. $a=1$, either $\mathbb{N}^{(r,s)}$ or $\mathbb{N}^{(r-1,s+1)}$ can be set to zero depending on whether $r$ is even or odd. In such a case, the intrinsic parities can be fixed such that there is a $m$ mod $2$ classification, which we will discuss now. In general there are two types of MHS: the even ranked $\Phi^{(2\tilde{n},\tilde{s})}$ and the odd ranked $\Phi^{(2\tilde{n}+1,\tilde{s})}$. Consider first the even ranked MHS $\Phi^{(2\tilde{n},\tilde{s})}$. As discussed, we can do an axial transformation and set the coefficient of $\mathbb{N}^{(2\tilde{n},\tilde{s})}$ to zero. In such a case, the interacting Lagrangian of $\Phi^{(2\tilde{n},\tilde{s})}$ can be written as
\begin{align}
	\mathcal{L}^{int}=\mathbf{e}~ \mathbb{M}^{(2\tilde{n},\tilde{s})}\Phi^{(2\tilde{n},\tilde{s})}+ \mathbf{f}~ \mathbb{M}^{(2\tilde{n}-1,\tilde{s}+1)}\Phi^{(2\tilde{n},\tilde{s})}+\mathbf{h}~ \mathbb{N}^{(2\tilde{n}-1,\tilde{s}+1)}\Phi^{(2\tilde{n},\tilde{s})}\label{L2n}
\end{align}
Using table \ref{tab:LiCRT}, the $C,\, R_i,\, T$ transformations of the Lagrangian in \eqref{L2n} can be summarized in table \ref{tab:L2nCRT}.
\renewcommand{\arraystretch}{2}
\begin{table}[H]
	\centering
	\begin{tabular}{ ||c||c|c|c|| }
		\hline
		\hline
		& $\mathbf{e}$                  & $\mathbf{f}$                             &$\mathbf{h}$                                \\
		\hline
		\hline
		$C$ & $(-1)^{\tilde{s}+\tilde{n}+1}t_0t_1$ $\eta_{_C}^{(2\tilde{n},\tilde{s})}$   &$(-1)^{\tilde{s}+\tilde{n}+1}$ $\eta_{_C}^{(2\tilde{n},\tilde{s})}$  & $(-1)^{\tilde{s}+m+\tilde{n}}$$\eta_{_C}^{(2\tilde{n},\tilde{s})}$\\
		\hline
		$R_i$ & $-\eta_{_R}^{(2\tilde{n},\tilde{s})}$  &$+\eta_{_R}^{(2\tilde{n},\tilde{s})}$ &$-\eta_{_R}^{(2\tilde{n},\tilde{s})}$\\
		\hline
		$T$ &  $(-1)^{\tilde{s}+\tilde{n}}t_0t_1$   $\eta_{_T}^{(2\tilde{n},\tilde{s})}$  & $(-1)^{\tilde{s}+\tilde{n}+1}$  $\eta_{_T}^{(2\tilde{n},\tilde{s})}$ & $(-1)^{\tilde{s}+m+\tilde{n}+1}$ $\eta_{_T}^{(2\tilde{n},\tilde{s})}$\\
		\hline
		\hline
	\end{tabular}\caption{$C,R_i,T$ Transformation of the couplings of $\Phi^{(2\tilde{n},\tilde{s})}$ given by \eqref{L2n} in $D=2m$. }	\label{tab:L2nCRT}
\end{table}
Table \ref{tab:L2nCRT} shows that $R_i,CT$ is always broken. However, we can have $CR_i,T$ and $C,~R_iT$ transformations. First, consider the $CR_i,\, T$ transformations in table \ref{tab:L2nCRT}. In a single MHS theory, the sign of the bilinear $\mathbb{M}^{2\tilde{n},\tilde{s}}$ can always be flipped, keeping the signs of $\mathbb{M}^{(2\tilde{n}-1,\tilde{s})}$ and $\mathbb{N}^{(2\tilde{n}-1,\tilde{s})}$ unchanged, by redefining the $C,\, R_i,\, T$ operations upto a chiral phase $ \exp{\big(\iimg(2f+1)\frac{\pi}{2}\Gamma_\star\big)}$. Therefore, for $CR_i,T$ to be a symmetry of \eqref{L2n}, the couplings $\mathbf{f}$ and $\mathbf{h}$ must transform in the same way,
\begin{align}
	(-1)^{\tilde{s}+\tilde{n}+1} = (-1)^{\tilde{s}+m+\tilde{n}+1}\label{L2nconstraint1}
\end{align}
For a MHS $\Phi^{(2\tilde{n},\tilde{s})}$ with any arbitrary $\tilde{n}$ and $\tilde{s}$, equation \eqref{L2nconstraint1} holds true only if $m=Even$. Therefore, for the choice of intrinsic parity $\eta_{_T}^{(2\tilde{n},\tilde{s})}=(-1)^{\tilde{s}+\tilde{n}+1}$, the $CR_i,T$ transformations are symmetry of \eqref{L2n} only in $D=4k$ spacetime.

Similarly, we can consider $C,~R_iT$ transformations in table \ref{tab:L2nCRT}. Again using axial rotations $\Psi\rightarrow \exp{\big(\iimg(2f+1)\frac{\pi}{2}\Gamma_\star\big)}\Psi$, the sign of the coupling $\mathbf{e}$ can be fixed as $+1$ without altering the signs of the couplings $\mathbf{f}$ and $\mathbf{h}$. This implies that for $C,R_iT$ to be a symmetry of $\eqref{L2n}$ it must satisfy
\begin{align}
	(-1)^{\tilde{s}+\tilde{n}+1}= (-1)^{\tilde{s}+m+\tilde{n}}\label{L2nconstraint2}
\end{align}
For a MHS $\Phi^{(2\tilde{n},\tilde{s})}$ with any arbitrary $\tilde{n}$ and $\tilde{s}$, equation \eqref{L2nconstraint2} holds true only if $m=Odd$. Therefore, for the choice of intrinsic parity $\eta_{_C}^{(2\tilde{n},\tilde{s})}=(-1)^{\tilde{s}+\tilde{n}+1}$, the $C,R_iT$ transformations are the symmetry of \eqref{L2n} only in $D=4k+2$ spacetime. 

Now let us consider the other possibility of MHS which is the odd ranked $\Phi^{(2\tilde{n}+1,\tilde{s})}$. Using table \ref{tab:evencouplings}, it can couple with a massless Dirac spinor in three possible ways: $\mathbb{M}^{(2\tilde{n}+1,\tilde{s})}\Phi^{(2\tilde{n}+1,\tilde{s})}$, $\mathbb{M}^{(2\tilde{n},\tilde{s}+1)}\Phi^{(2\tilde{n}+1,\tilde{s})}$, and $\mathbb{N}^{(2\tilde{n}+1,\tilde{s})}\Phi^{(2\tilde{n}+1,\tilde{s})}$. These correspond to the couplings $\mathbf{e}$, $\mathbf{f}$, and $\mathbf{g}$ in the Lagrangian \eqref{Lint}. The coupling $\mathbf{h}$ can be set to zero by axial transformation. The interaction Lagrangian for $\Phi^{(2\tilde{n}+1,\tilde{s})}$ is then given by
\begin{align}
	\mathcal{L}^{int}=\mathbf{e}\, \mathbb{M}^{(2\tilde{n}+1,\tilde{s})}\Phi^{(2\tilde{n}+1,\tilde{s})}+ \mathbf{f}\, \mathbb{M}^{(2\tilde{n},\tilde{s}+1)}\Phi^{(2\tilde{n}+1,\tilde{s})}+\mathbf{g}\, \mathbb{N}^{(2\tilde{n}+1,\tilde{s})}\Phi^{(2\tilde{n}+1,\tilde{s})}\label{L2n+1}
\end{align}
Using table \ref{tab:LiCRT}, the $C,R_i,T$ transformation of \eqref{L2n+1} are summarized in table \ref{tab:L2n+1CRT}. 
\renewcommand{\arraystretch}{2}
\begin{table}[H]
	\centering
	\begin{tabular}{ ||c||c|c|c|| }
		\hline
		\hline
		& $\mathbf{e}$                  & $\mathbf{f}$                             &$\mathbf{g}$                                \\
		\hline
		\hline
		$C$ & $(-1)^{\tilde{s}+\tilde{n}+1}$ $\eta_{_C}^{(2\tilde{n}+1,\tilde{s})}$   &$(-1)^{\tilde{s}+\tilde{n}}t_0t_1$ $\eta_{_C}^{(2n+1,\tilde{s})}$  & $(-1)^{\tilde{s}+m+\tilde{n}}$$\eta_{_C}^{(2\tilde{n}+1,\tilde{s})}$\\
		\hline
		$R_i$ & $+\eta_{_R}^{(2\tilde{n}+1,\tilde{s})}$  &$-\eta_{_R}^{(2\tilde{n}+1,\tilde{s})}$ &$-\eta_{_R}^{(2\tilde{n}+1,\tilde{s})}$\\
		\hline
		$T$ &  $(-1)^{\tilde{s}+\tilde{n}+1}$   $\eta_{_T}^{(2\tilde{n}+1,\tilde{s})}$  & $(-1)^{\tilde{s}+\tilde{n}+1}t_0t_1$  $\eta_{_T}^{(2\tilde{n}+1,\tilde{s})}$ & $(-1)^{\tilde{s}+m+\tilde{n}+1}$ $\eta_{_T}^{(2\tilde{n}+1,\tilde{s})}$\\
		\hline
		\hline
	\end{tabular}\caption{$C,R_i,T$ Transformation of the couplings of $\Phi^{(2\tilde{n}+1,\tilde{s})}$ given by \eqref{L2n} in $D=2m$ spacetime. }	\label{tab:L2n+1CRT}
\end{table}
Again, we can see from table \ref{tab:L2n+1CRT} that $R_i,\, CT$ is always violated by \eqref{L2n+1}. Like in the case of even ranked $\Phi^{(2\tilde{n},\tilde{s})}$, we can consider the action of $CR_i,T$ and $C,R_iT$.
Consider the $CR_i,T$ transformation of $\mathcal{L}^{int}$ in \eqref{L2n+1}.
Similar to even-ranked MHS, we can do an axial rotation to fix the sign of $\mathbf{f}$, corresponding to the coupling of odd-ranked MHS with the even-ranked bilinear to be $+1$. For $CR_i,\, T$ to be the symmetry, the couplings $\mathbf{e}$ and $\mathbf{g}$ must transform in the same way, i.e., the corresponding factors must be the same
\begin{align}
	(-1)^{\tilde{s}+\tilde{n}+1}= (-1)^{\tilde{s}+m+\tilde{n}+1}\label{L2n+1constraint1}
\end{align}
For arbitrary $\tilde{s}$ and $\tilde{n}$, the \eqref{L2n+1constraint1} holds true only if $m=even$. Therefore, for the choice of the intrinsic parity $\eta_{_T}^{(2\tilde{n}+1,\tilde{s})}=(-1)^{\tilde{s}+\tilde{n}+1}$, the most general theory of odd ranked MHS $\Phi^{(2\tilde{n}+1,\tilde{s})}$ coupled with a massless Dirac spinor through \eqref{L2n+1} would remain invariant under $CR_i,T$ only in $D=4k$ spacetime.

Consider the $C,\, R_iT$ transformations of \eqref{L2n+1}. We can fix the sign $\mathbf{f}$ to be $+1$ using the axial transformations. For $C,~R_iT$ to be a symmetry, the couplings $\mathbf{e}$ and $\mathbf{g}$ must transform in the same way i.e.
\begin{align}
	(-1)^{\tilde{s}+\tilde{n}+1}=(-1)^{\tilde{s}+m+\tilde{n}}\label{L2n+1constraint2}
\end{align}
For any arbitrary $\tilde{s}$ and $\tilde{n}$, equation \eqref{L2n+1constraint2} holds true only if $m=Odd$. Therefore, for the choice of the intrinsic parity $\eta_{_C}^{(2\tilde{n}+1,\tilde{s})}=(-1)^{\tilde{s}+\tilde{n}+1}$, the most general theory of odd ranked MHS $\Phi^{(2\tilde{n}+1,\tilde{s})}$ coupled with a massless Dirac spinor through \eqref{L2n+1} would remain invariant under $C,R_iT$ only in $D=4k+2$ spacetime.

Using the results for even ranked $\Phi^{(2\tilde{n},\tilde{s})}$ and odd ranked $\Phi^{(2\tilde{n}+1,\tilde{s})}$, we conclude the following statements for a theory of massless Dirac spinors with a single MHS interacting through three-point couplings
\begin{enumerate}
	[leftmargin=*,labelindent=0em]
	\item In any even dimension, the chiral and non-chiral couplings transform oppositely under reflection $R_i$. Therefore, a theory with both the couplings always violates $R_i$. This also implies that in a theory where $R_i$ is a symmetry, the chiral and non-chiral interactions of a MHS cannot be present simultaneously.
	\item  In $D=4k$, the chiral and non-chiral couplings transforms oppositely under $C$ but in the same way under $T$ upto a factor of $(-1)^{\tilde{s}+\tilde{n}+1}\eta_{_T}^{(\tilde{r},\tilde{s})}$ where $\tilde{n}=\lfloor{\tilde{r}/2}\rfloor$. Therefore, a theory with both couplings will always transform in the same way under $CR_i$ and $T$. This gives a mod $4$ classification of $D=4k$ theories with both the chiral and non-chiral couplings. If we choose the intrinsic parity of MHS as $\eta_{_T}^{(\tilde{r},\tilde{s})}=(-1)^{\tilde{s}+\tilde{n}+1}$ where $\tilde{n}=\lfloor{\tilde{r}/2}\rfloor$ then such theories will always preserve $CR_i$ and $T$.
	\item In $D=4k+2$, the chiral and non-chiral couplings transforms oppositely under $T$ but in the same way under $C$ upto a factor of $(-1)^{\tilde{s}+\tilde{n}+1}\eta_{_C}^{(\tilde{r},\tilde{s})}$ where $\tilde{n}=\lfloor{\tilde{r}/2}\rfloor$. Therefore, a theory with both couplings will always transform in the same way under $R_iT$ and $C$. This gives a mod $4$ classification of $D=4k+2$ theories with both the chiral and non-chiral couplings. If we choose the intrinsic parity of MHS as $\eta_{_C}^{(\tilde{r},\tilde{s})}=(-1)^{\tilde{s}+\tilde{n}+1}$ where $\tilde{n}=\lfloor{\tilde{r}/2}\rfloor$ then such theories will always preserve $R_iT$ and $C$.
	
	The mod $4$ classification under $C,\, R_iT$ and $T,\, CR_i$ is summarized in table \ref{tab:mod4classificationeven}. We emphasize that the classifications in $D=4k$ and $D=4k+2$ are different. In $D=4k$, the classification can only be done under $CR_i,\,T$ , whereas in $D=4k+2$, it can only be done under $ C,\, R_iT$.
	\renewcommand{\arraystretch}{1.5}
	\begin{table}[H]
		\centering
		\begin{tabular}{ ||c||c|c|c|| }
			\hline
			\hline
				\multicolumn{4}{||c||}{$\Phi^{(\tilde{r},\tilde{s})}$ $\forall ~\tilde{r},\tilde{s}$}\\
			\hline	
			\hline
			$D$ mod $4$            & $ C, \, R_iT$                   & $ R_i, \, CT$                            & $ T, \, CR_i$                                    \\
			\hline
			$0$ & $\times$     & $\times$ & $\checkmark$  \\
			\hline
			$2$ & $\checkmark$     & $\times$ & $\times$   \\
			\hline
			\hline
		\end{tabular}\caption{$D$ mod $4$ classification of theories with both chiral and non-chiral three point couplings of a MHS with a massless Dirac spinor in $D=2m$ spacetime.}	\label{tab:mod4classificationeven}
	\end{table}
\end{enumerate}
\subsection{Odd dimensional spacetime}
In odd spacetime, there are no chiral couplings. The only independent couplings with a MHS $\Phi^{(\tilde{r}_a,\tilde{s}_a)}$ are of two types $\mathbb{M}^{(\tilde{r}_a,\tilde{s}_a)}\Phi^{(\tilde{r}_a,\tilde{s}_a)}$ and $\mathbb{M}^{(\tilde{r}_a-1,\tilde{s}_a+1)}\Phi^{(\tilde{r}_a,\tilde{s}_a)}$. Consider now $N$ number of MHS i.e. $a=1,2,\cdots, N$ interacting with a massless Dirac spinor through
\begin{align}
	\mathcal{L}^{int}=	\sum_{a=1}^{N}	\mathcal{L}^{int}_{_a}=\sum_{a=1}^{N}\bigg(\mathbf{e}_a \mathbb{M}^{(\tilde{r}_a,\tilde{s}_a)}\Phi^{(\tilde{r}_a,\tilde{s}_a)}_a+ \mathbf{f}_a \mathbb{M}^{(\tilde{r}_a-1,\tilde{s}_a+1)}\Phi^{(\tilde{r}_a,\tilde{s}_a)}_a\bigg)\label{Lintodd}
\end{align}
 Using table \ref{tab:CRTIDirac}, the $C,R_i,T$ transformations of $\mathcal{L}^{int}_{_a}$ is summarized in table \ref{tab:LiCRTodd}.
\renewcommand{\arraystretch}{2}
\begin{table}[H]
	\centering
	\begin{tabular}{ ||c||c|c||| }
		\hline
		\hline
		& $\mathbf{e}_a$                   & $\mathbf{f}_a$               \\
		\hline
		\hline
		$C$ & $(-1)^{\tilde{s}_a+1}t_1t_{\tilde{r}_a}$ $\eta_{_C}^{a}$   &$(-1)^{\tilde{s}_a}t_1t_{\tilde{r}_a -1}$ $\eta_{_C}^{a}$\\
		\hline
		$R_i$ & $(-1)^{\tilde{r}_a+1}$  $\eta_{_R}^{a}$  &$(-1)^{\tilde{r}_a}$  $\eta_{_R}^{a}$ \\
		\hline
		$T$ &  $(-1)^{\tilde{r}_a+\tilde{s}_a}t_1t_{\tilde{r}_a}$   $\eta_{_T}^{a}$  & $(-1)^{\tilde{r}_a+\tilde{s}_a}t_1t_{\tilde{r}_a-1}$  $\eta_{_T}^{a}$ \\
		\hline
		\hline
	\end{tabular}\caption{Transformation of \eqref{Lintodd} under $C,R_i,T$ in $D=2m+1$ spacetime. The intrinsic parities of $a^{th}$ MHS are  $\eta_{_C}^{a}\equiv \eta_{_C}^{(\tilde{r}_a,\tilde{s}_a)}$,  $\eta_{_R}^{a}\equiv \eta_{_R}^{(\tilde{r}_a,\tilde{s}_a)}$,  $\eta_{_T}^{a}\equiv \eta_{_T}^{(\tilde{r}_a,\tilde{s}_a)}$.}	\label{tab:LiCRTodd}
\end{table}
Again, it is clear from table \ref{tab:LiCRTodd} that $\mathcal{L}^{int}$ would always violate $R_i,CT$ for any $N$. For $CR_i,~T$ to be a symmetry, the couplings $\mathbf{e}_a$ and $\mathbf{f}_a$ must transform similarly. This implies that $t_{\tilde{r}_a}=t_{\tilde{r}_a -1}$ and $\eta_{_T}^a=(-1)^{\tilde{r}_a+\tilde{s}_a}t_1t_{\tilde{r}_a}~\forall ~a$. Since $t_{2\tilde{n}}=(-1)^{\tilde{n}}t_0$ and $t_{2\tilde{n}+1}=(-1)^{\tilde{n}}t_1$, we get $t_0t_1=+1$  for $\tilde{r}_a=2\tilde{n}_a+1$, and $t_0t_1=-1$ for  $\tilde{r}_a=2\tilde{n}_a$. Using figure \ref{fig:t0t1}, we know that $t_0t_1=+1$ for $m=2k$ and $t_0t_1=-1$ for $m=2k+1$. This implies that in any odd dimension, $CR_i,\, T$ is a symmetry only if all the MHS in \eqref{Lintodd} are either odd or even ranked. If $\mathcal{L}^{int}$ in \eqref{Lintodd} has both odd and even ranked MHS, then it must break $C,\,R_i,\,T$ and is only invariant under $CR_iT$ such that $\eta_{_C}^a\eta_{_R}^a\eta_{_T}^a=+1$ $\forall ~a$.

Similarly, for $C,R_iT$ to be symmetry, we must have $t_{\tilde{r}_a}=-t_{\tilde{r}_a -1}$ and $\eta_{_C}^a=(-1)^{\tilde{s}_a+1}t_1t_{\tilde{r}_a}~\forall ~a$. Again using $t_{2\tilde{n}}=(-1)^{\tilde{n}}t_0$ and $t_{2\tilde{n}+1}=(-1)^{\tilde{n}}t_1$, this gives $t_0t_1=-1$ for $\tilde{r}_a=2\tilde{n}_a+1$, and $t_0t_1=+1$ for $\tilde{r}_a=2\tilde{n}_a$. Since $t_0t_1=-1$ when $m=2k+1$, and $t_0t_1=+1$ when $m=2k$, it implies that in any odd dimensions $C,~ R_iT$ is broken if both the even and odd ranked MHS are present in the Lagrangian. Therefore, in the presence of both odd and even ranked MHS, the Lagrangian in \eqref{Lintodd} is invariant only under $CR_iT$ such that $\eta_{_C}^a\eta_{_R}^a\eta_{_T}^a=+1$.
\subsubsection{Even or odd ranked couplings}
Similar to even spacetime, we can consider the multiple MHS theory coupled only with the even-ranked or odd-ranked Dirac bilinears. The coupling of any flavor \enquote{$a$} MHS $\Phi^{(\tilde{r}_a,\tilde{s}_a)}_a$ with the even ranked bilinear is given by  $\mathbb{M}^{(2n,s)}\Phi^{(\tilde{r}_a,\tilde{s}_a)}_a$ such that $2n+s=\tilde{r}_a+\tilde{s}_a$. Using table \ref{tab:CRTIDirac}, such a coupling would transform under $C,R_i,$ and $T$ upto factor of $(-1)^{s+n+m+1}\eta_{_C}^{(\tilde{r}_a,\tilde{s}_a)}$, $-\eta_{_R}^{(\tilde{r}_a,\tilde{s}_a)}$, and $(-1)^{s+n+m}\eta_{_T}^{(\tilde{r}_a,\tilde{s}_a)}$, respectively. Unlike even dimensions, we cannot redefine the $C,\, R_i,\, T$ transformations up to chiral phases and absorb all these factors. Therefore, in a $C,\, R_i,\, T$ preserving theory of multiple MHS coupled only with $r=2n$ ranked bilinear of a massless Dirac spinor, the intrinsic parities must be given by
\begin{align}
	\eta_{_C}^{(\tilde{r}_a,\tilde{s}_a)}=-\eta_{_T}^{(\tilde{r}_a,\tilde{s}_a)}=	(-1)^{s+n+m+1} \qquad,\qquad \eta_{_R}^{(\tilde{r}_a,\tilde{s}_a)}=-1\qquad \forall ~ a
\end{align}
Similarly, in a theory of multiple MHS $\Phi^{(\tilde{r}_a,\tilde{s}_a)}_a$ coupled only with an odd ranked bilinear $\mathbb{M}^{(2n+1,s)}$ of a massless Dirac bilinear, every coupling transforms upto factor of $(-1)^{s+n+1}\eta_{_C}^{(\tilde{r}_a,\tilde{s}_a)}$, $\eta_{_R}^{(\tilde{r}_a,\tilde{s}_a)}$, and $(-1)^{s+n+1}\eta_{_T}^{(\tilde{r}_a,\tilde{s}_a)}$ under $C,\, R_i,$ and $T$ respectively. For such a theory to preserve $C,\, R_i,\, T$, the intrinsic parity of any MHS coupled with $r=2n+1$ ranked massless Dirac spinor must be given by
\begin{align}
	\eta_{_C}^{(\tilde{r}_a,\tilde{s}_a)}=\eta_{_T}^{(\tilde{r}_a,\tilde{s}_a)}=	(-1)^{s+n+1} \qquad,\qquad \eta_{_R}^{(\tilde{r}_a,\tilde{s}_a)}=1\qquad \forall ~ a
\end{align}
\subsubsection{Single MHS theory}\label{sec:SMHSodd}
Consider a most general theory of single even ranked MHS $\Phi^{(2\tilde{n},\tilde{s})}$ given by
\begin{align}
	\mathcal{L}^{int}=\mathbf{e}\, \mathbb{M}^{(2\tilde{n},\tilde{s})}\Phi^{(2\tilde{n},\tilde{s})}+\mathbf{f}\, \mathbb{M}^{(2\tilde{n}-1,\tilde{s}+1)}\Phi^{(2\tilde{n},\tilde{s})}
\end{align}
Using table \ref{tab:LiCRTodd} for $a=1$ and $\tilde{r}=2\tilde{n}$, we can conclude that such a theory would always violate $R_i,~CT$. The theory preserves $CR_i,T$ if both the couplings transform in the same way i.e. $(-1)^{\tilde{s}+\tilde{n}+m}=(-1)^{\tilde{s}+\tilde{n}+1}$ and the intrinsic parity is given by $\eta_{_T}^{(2\tilde{n},\tilde{s})}=(-1)^{\tilde{s}+\tilde{n}+1}$. This is possible only if $m=odd$, i.e., in $D=4k+3$ spacetime. Whereas the theory preserves $C,~R_iT$ if $(-1)^{\tilde{s}+\tilde{n}+m}=(-1)^{\tilde{s}+\tilde{n}}$ and the intrinsic parity is given by $\eta_{_C}^{(2\tilde{n},\tilde{s})}=(-1)^{\tilde{s}+\tilde{n}+1}$. This is possible only if $m=even$, i.e., in $D=4k+1$ spacetime.

Now consider a most general theory of an odd-ranked single MHS $\Phi^{(2\tilde{n}+1,\tilde{s})}$ given by
\begin{align}
	\mathcal{L}^{int}=\mathbf{e}\, \mathbb{M}^{(2\tilde{n}+1,\tilde{s})}\Phi^{(2\tilde{n}+1,\tilde{s})}+\mathbf{f}\, \mathbb{M}^{(2\tilde{n},\tilde{s}+1)}\Phi^{(2\tilde{n}+1,\tilde{s})}
\end{align}
Using table \ref{tab:LiCRTodd}, we know that the theory must always violate $R_i,~CT$. The theory preserves $CR_i,~T$ only if the couplings transform in the same way i.e. $(-1)^{\tilde{s}+\tilde{n}+1}=(-1)^{\tilde{s}+\tilde{n}+m+1}$ and the intrinsic parity satisfies $\eta_{_T}^{(2\tilde{n}+1,\tilde{s})}=(-1)^{\tilde{s}+\tilde{n}+1}$. This is possible only if $m=even$, i.e., the spacetime is $D=4k+1$ dimensional. Whereas, the theory would preserve $C,~R_iT$ only if the intrinsic parity is given by $\eta_{_C}^{(2\tilde{n}+1,\tilde{s})}=(-1)^{\tilde{s}+\tilde{n}+1}$, and $(-1)^{\tilde{s}+\tilde{n}+1}=(-1)^{\tilde{s}+\tilde{n}+m}$. This holds only if $m=odd$, i.e., the spacetime is $D=4k+3$ dimensional.

In $D=4k+1$ and $D=4k+3$, we can summarize the results for the even ranked $\Phi^{(2\tilde{n},\tilde{s})}$ or the odd ranked $\Phi^{(2\tilde{n}+1,\tilde{s})}$ coupled with a massless Dirac spinor through three-point couplings as follows
\begin{enumerate}
	\item In $D=4k+1$, a theory of an even ranked MHS $\Phi^{(2\tilde{n},\tilde{s})}$ would preserve only $C,~R_iT$ such that the intrinsic parity is given by $\eta_{_C}^{(2\tilde{n},\tilde{s})}=(-1)^{\tilde{s}+\tilde{n}+1}$. Whereas, a theory of an odd ranked MHS $\Phi^{(2\tilde{n}+1,\tilde{s})}$ would preserve only $CR_i,~T$ such that the intrinsic parity is given by $\eta_{_T}^{(2\tilde{n}+1,\tilde{s})}=(-1)^{\tilde{s}+\tilde{n}+1}$.
	\item In $D=4k+3$, a theory of an even ranked MHS $\Phi^{(2\tilde{n},\tilde{s})}$ would preserve only $CR_i,~T$ such that the intrinsic parity is given by $\eta_{_T}^{(2\tilde{n},\tilde{s})}=(-1)^{\tilde{s}+\tilde{n}+1}$. Whereas, a theory of an odd ranked MHS $\Phi^{(2\tilde{n}+1,\tilde{s})}$ would preserve only $C,~R_iT$ such that the intrinsic parity is given by $\eta_{_C}^{(2\tilde{n}+1,\tilde{s})}=(-1)^{\tilde{s}+\tilde{n}+1}$.
\end{enumerate}
These results have been summarized in a tabular form in table \ref{tab:mod4classificationodd}.
\begin{table}[H]
	\begin{minipage}{0.45\linewidth}
			\renewcommand{\arraystretch}{1.5}
		\begin{tabular}{ ||c||c|c|c|| }
			\hline
			\hline
			\multicolumn{4}{||c||}{$\Phi^{(2n,\tilde{s})}$}\\
			\hline	
			\hline
			$D$ mod $4$            & $ C, \, R_iT$                   & $ R_i, \, CT$                            & $ T, \, CR_i$                                    \\
			\hline
			$1$ & $\checkmark$    & $\times$ &$\times$   \\
			\hline
			$3$ & $\times$    & $\times$ & $\checkmark$  \\
			\hline
			\hline
		\end{tabular}
	\end{minipage}\hfill
	\begin{minipage}{0.45\linewidth}
			\renewcommand{\arraystretch}{1.5}
		\begin{tabular}{ ||c||c|c|c|| }
		\hline
		\hline
		\multicolumn{4}{||c||}{$\Phi^{(2n+1,\tilde{s})}$}\\
		\hline	
		\hline
		$D$ mod $4$            & $ C, \, R_iT$                   & $ R_i, \, CT$                            & $ T, \, CR_i$                                    \\
		\hline
		$1$ & $\times$    & $\times$ &$\checkmark$   \\
		\hline
		$3$ & $\checkmark$    & $\times$ & $\times$  \\
		\hline
		\hline
	\end{tabular}
	\end{minipage}\caption{$D$ mod $4$ classification of theories with three point couplings of an even ranked MHS $\Phi^{(2\tilde{n},\tilde{s})}$ (left) and an odd ranked MHS $\Phi^{(2\tilde{n}+1,\tilde{s})}$ (right) with massless Dirac spinor in $D=2m+1$ mod $4$ spacetime.}	\label{tab:mod4classificationodd}
\end{table}
\subsection{An $m$ mod $2$ classification}
The results from the even and the odd spacetimes can be merged such that there exists a mod $2$ classification of the theories of an even or an odd ranked MHS coupled with a massless Dirac spinor through all possible three-point couplings.

Let us start with an even ranked MHS $\Phi^{(2\tilde{n},\tilde{s})}$ coupled with a massless Dirac spinor. The interaction in $D=2m$ is governed by \eqref{L2n} and in $D=2m+1$ is governed by \eqref{L2n+1} for $a=1,~\tilde{r}=2\tilde{n}$. It is $D$ mod $4$ classifications as summarized in tables \ref{tab:mod4classificationeven} and \ref{tab:mod4classificationodd} can be clubbed in to give a nice $D$ mod $4$ classification as summarized in table \ref{tab:mod2classification1}. The three point couplings of $\Phi^{(2\tilde{n},\tilde{s})}$ with a massless Dirac spinor for any arbitrary $\tilde{n}$ and $\tilde{s}$ preserves $C,R_iT$ in $D=1,2,$ mod $4$ with a unique choice of intrinsic parity $\eta_{_C}^{(2\tilde{n},\tilde{s})}=(-1)^{\tilde{s}+\tilde{n}+1}$. Whereas it preserves $CR_i, ~T$ in $D=3,4,$ mod $4$ spacetime with a unique choice of intrinsic parity $\eta_{_T}^{(2\tilde{n},\tilde{s})}=(-1)^{\tilde{s}+\tilde{n}+1}$.\\

Similarly, using tables \ref{tab:mod4classificationeven} and \ref{tab:mod4classificationodd}, the odd ranked MHS $\Phi^{(2\tilde{n}+1,\tilde{s})}$ coupled with a massless Dirac spinor through \eqref{L2n} and \eqref{Lintodd} preserves $C,~R_iT$ in $D=2,3,$ mod $4$ spacetime. Whereas it preserves $CR_i,~T$ in $D=4,5$ mod $4$ spacetime. This again gives a $m$ mod $2$ classification of the theory of three-point couplings of an odd-ranked MHS with a massless Dirac spinor as summarized in table \ref{tab:mod2classification2}. For even ranked MHS $\Phi^{(2\tilde{n},\tilde{s})}$, the theory has the same symmetries (either $CR_i,~T$ or $C,~R_iT$) in $m=2k$ and $m=2k+1$. In any $D=2m-1,2m$ spacetime, the theory of even ranked MHS in $m=2k$ will always remain invariant under $T,~CR_i$, whereas in $m=2k+1$ it will always remain under $C~R_iT$.For odd ranked MHS $\Phi^{(2\tilde{n}+1,\tilde{s})}$, the theory has different symmetries in $m=2k$ and $m=2k+1$. In $D=2m+1,2m$, the theory of odd ranked MHS in $m=2k$ will always remain invariant under $T,~CR_i$, whereas in $m=2k+1$ it will always remain invariant under $C,~R_iT$. This shows that the theory of an even or odd ranked MHS coupled with massless Dirac spinors can be classified into a $m$ mod $2$ (or $D$ mod $4$) manner.
\begin{table}[H]
	\begin{minipage}{0.45\linewidth}
		\centering
			\renewcommand{\arraystretch}{1.5}
	\begin{tabular}{ ||c||c|c|c|| }
		\hline
		\hline
			\multicolumn{4}{||c||}{$\Phi^{(2\tilde{n},\tilde{s})}$ $\forall ~\tilde{n},\tilde{s}$}\\
		\hline	
		\hline
		$D$  mod $4$          & $ C, \, R_iT$                   & $ R_i, \, CT$                            & $ T, \, CR_i$                                    \\
		\hline
		\hline
			$0$ & $\times$    & $\times$ &$\checkmark$   \\
		\hline
		$1$ & $\checkmark$    & $\times$ & $\times$  \\
		\hline
		$2$ & $\checkmark$    & $\times$ &$\times$   \\
		\hline
		$3$ & $\times$    & $\times$ & $\checkmark$  \\
		\hline
		\hline
	\end{tabular}\caption{\enquote{$m$} mod $2$ classification of a theory of a real even ranked MHS $\Phi^{(2\tilde{n},\tilde{s})}$ interacting with a massless Dirac spinor through all possible three point couplings in $D=2m+1/2m$ spacetime.}	\label{tab:mod2classification1}
		\end{minipage}\hfill
	\begin{minipage}{0.45\linewidth}
	\centering
		\renewcommand{\arraystretch}{1.5}
		\begin{tabular}{ ||c||c|c|c|| }
		\hline
		\hline
		\multicolumn{4}{||c||}{$\Phi^{(2\tilde{n}+1,\tilde{s})}$ $\forall ~\tilde{n},\tilde{s}$}\\
		\hline	
		\hline
		$D$  mod $4$          & $ C, \, R_iT$                   & $ R_i, \, CT$                            & $ T, \, CR_i$                                    \\
		\hline
		\hline
		$0$ & $\times$    & $\times$ &$\checkmark$   \\
		\hline
		$1$ & $\times$    & $\times$ &$\checkmark$   \\
		\hline
		$2$ & $\checkmark$    & $\times$ & $\times$  \\
		\hline
		$3$ & $\checkmark$    & $\times$ &$\times$   \\
		\hline
		\hline
	\end{tabular}\caption{\enquote{$m$} mod $2$ classification of a theory of a real odd ranked MHS $\Phi^{(2\tilde{n}+1,\tilde{s})}$ interacting with a massless Dirac spinor through all possible three point couplings in $D=2m+1/2m$ spacetime.}	\label{tab:mod2classification2}
	\end{minipage}
\end{table}
Using table \ref{tab:mod2classification1}, it can be stated that $R_i$ and $CT$ are never a symmetry in an interacting theory of an even ranked MHS coupled with massless Dirac spinors through three-point couplings in any $D$ dimension. One of the couplings (either chiral or non-chiral) would always violate $R_i$ and $CT$.  
Moreover, under dimensional compactification, we observe that the $C$ and $T$ of $D=2m+1$ go to $CR_i$ and $TR_i$ of $D=2m$. Whereas $C$ and $T$ of $D=2m$ goes to $C$ and $T$ of $D=2m-1$ respectively. Like even ranked MHS, we can use table \ref{tab:mod2classification2} to come across similar remarks. Again, It is clear that $R_i$ and $CT$ are never a symmetry.
Under dimensional compactification, we observe that the $C$ and $T$ of $D=2m+1$, respectively, goes to $C$ and $T$ of $D=2m$. Whereas $C$ and $T$ of $D=2m$ goes to $CR_i$ and $R_iT$ of $D=2m-1$ respectively. The dimensional reduction for $C,\,R_i,\,T$ in the case of both even ranked and odd ranked MHS are summarized in figure \ref{tab:dimred}. A formal proof for the dimensional reduction of $C,\, R_i, \, T$ in interacting theories is left to be done for the future. 
	\begin{figure}[H]
		\begin{minipage}{0.45\linewidth}
			\centering
			\renewcommand{\arraystretch}{0.5}
	\begin{figure}[H]
		\begin{center}
			\setseq{{$C$,$T$,$CR_i$,$R_iT$,$R_i$,$CT$,$CR_iT$}}{{$D=2m+1$}}{{$C$,$T$,$CR_i$,$R_iT$,$R_i$,$CT$,$CR_iT$}}{{$D=2m$}}{{$C$,$T$,$CR_i$,$R_iT$,$R_i$,$CT$,$CR_iT$}}{{$D=2m-1$}}{{$C$/$CR_i$,$R_i$/$R_i$,$T$/$R_iT$, $CR_i$/$C$, $R_iT$/$T$, $CT$/$CT$, $CR_iT$/$CR_iT$}}{{$C$/$C$, $R_i$/$R_i$, $T$/$T$, $CR_i$/$CR_i$, $R_iT$/$R_iT$, $CT$/$CT$, $CR_iT$/$CR_iT$}}
		\end{center}
	\end{figure}
	\end{minipage}\hfill
\begin{minipage}{0.45\linewidth}
\centering
\renewcommand{\arraystretch}{1.5}
\begin{figure}[H]
	\begin{center}
		\setseq{{$C$,$T$,$CR_i$,$R_iT$,$R_i$,$CT$,$CR_iT$}}{{$D=2m+1$}}{{$C$,$T$,$CR_i$,$R_iT$,$R_i$,$CT$,$CR_iT$}}{{$D=2m$}}{{$C$,$T$,$CR_i$,$R_iT$,$R_i$,$CT$,$CR_iT$}}{{$D=2m-1$}}{{$C$/$C$, $R_i$/$R_i$, $T$/$T$, $CR_i$/$CR_i$, $R_iT$/$R_iT$, $CT$/$CT$, $CR_iT$/$CR_iT$}}{{$C$/$CR_i$,$R_i$/$R_i$,$T$/$R_iT$, $CR_i$/$C$, $R_iT$/$T$, $CT$/$CT$, $CR_iT$/$CR_iT$}}
	\end{center}
\end{figure}
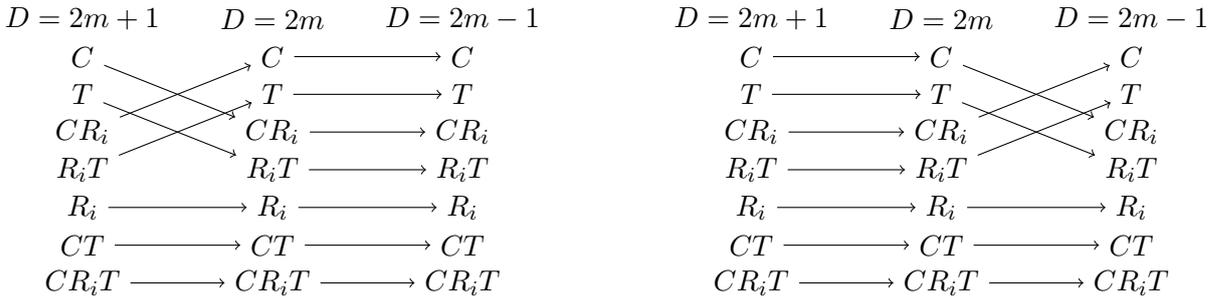
	\end{minipage}\caption{Dimensional reduction of $C,R_i,T$ in a theory of an even ranked real MHS $\Phi^{(2\tilde{n},\tilde{s})}$ (left) and an odd ranked MHS $\Phi^{(2\tilde{n}+1,\tilde{s})}$ (right) interacting with a massless Dirac spinor through all possible three point couplings.}	\label{tab:dimred}
\end{figure}
 	\section{Multiple flavors and unconventional $CRT$}\label{sec:unconventionalCRT}
 	Till now, we have only discussed the three-point interactions of a massless Dirac spinor and multiple MHS. For multiple spinors, constructing the Dirac bilinears and their couplings is also important from different points of view. For example, the standard model of particle physics has fermions of different flavors like electrons and neutrinos. Any theory of multiple spinors and bosons has a richer possibility of the discrete $R_i,\, T$ transformations and the internal symmetries in the field space. For example, consider the simplest theory of two different massive Dirac spinors in even dimensions
 \begin{align}
 	\bar{\Psi}_1\Gamma^\mu (\partial_\mu\Psi_1)+\bar{\Psi}_2\Gamma^\mu (\partial_\mu\Psi_2)+\mathbf{m}(\bar{\Psi}_1\Psi_1+\bar{\Psi}_2\Psi_2)+\tilde{\mathbf{m}}(\bar{\Psi}_1\Psi_2+\bar{\Psi}_2\Psi_1)
 \end{align}
 The theory is also invariant under the field space rotations $\Psi_1\leftrightarrow\Psi_2$, which would not be present in a theory of a single Dirac spinor. The theory is also invariant under the conventional $C,\, R_i,\, T$ transformations defined in equations \eqref{Cspinor}, \eqref{Rspinor}, and \eqref{Tspinor} for each spinor. However,  these discrete transformations were constructed assuming they keep the kinetic term invariant and can only transform the field to itself or its antiparticle. However, there is no good rationale for not considering the $C,\, R_i,\, T$ transformations, which mix the fields among themselves. This section will discuss how one can construct the Dirac bilinears in any dimension for multiple flavors. Then we comment on the unconventional $C,\, R_i,\, T$ transformations (see chapter 2, appendix Weinberg vol 1 \cite{Weinberg:1995mt}) of these bilinears. We leave some of the details to the readers to figure out. 
 
 Consider a spectrum of finite $N$ number of Dirac spinors $\Psi_a$, where $a$ is the flavor index, which runs from $1$ to $N$. In even dimensions, the most general rank $r$ Dirac bilinears with $s$ number of derivatives are 
 	\begin{align}
 	\mathbb{M}_{ab}^{(r,s)}&\equiv (\iimg)^{s+\frac{r(r+1)}{2}} \bar{\Psi}_a\Gamma^{(r)}(\partial^{(s)}\Psi_b)\quad,\qquad (	\mathbb{M}_{ab}^{(r,s)})^\dagger=	\mathbb{M}_{ba}^{(r,s)},\label{multiplenonchiralbilinear}\\
 	\mathbb{N}_{ab}^{(r,s)}&\equiv(\iimg)^{s+\frac{(r+1)(r+2)}{2}} \bar{\Psi}_a\Gamma^{(r)}\Gamma_\star(\partial^{(s)}\Psi_b)\quad,\qquad (	\mathbb{N}_{ab}^{(r,s)})^\dagger=	\mathbb{N}_{ba}^{(r,s)}\label{multiplechiralbilinear},
 	\end{align}
 	In odd spacetime, the only possible bilinears are $\mathbb{M}^{(r,s)}_{ab}$. In even spacetime, under the axial transformation $\mathcal{U}_\star(\theta_a)$ for $\theta_a=\theta_b=(2f+1)\frac{\pi}{2}$ $\forall ~a,b$, the even ranked bilinears picks up a sign, whereas the odd ranked bilinears remain invariant $\forall ~a,b$
 \begin{align}
 	\mathbb{M}^{(2n,s)}_{ab}\longrightarrow 	-\mathbb{M}^{(2n,s)}_{ab}\qquad,&\qquad 	\mathbb{N}^{(2n,s)}_{ab}\longrightarrow 	-\mathbb{N}^{(2n,s)}_{ab},\\
 	\mathbb{M}^{(2n+1,s)}_{ab}\longrightarrow 	\mathbb{M}^{(2n+1,s)}_{ab}\qquad,&\qquad 	\mathbb{N}^{(2n+1,s)}_{ab}\longrightarrow 	\mathbb{N}^{(2n+1,s)}_{ab},
 \end{align}
 In even spacetime, there are total $4\times\frac{N(N+1)}{2}$ number of independent bilinears, where $4N$ number of bilinears are of same flavors and $4\times\frac{N(N-1)}{2}$ number of bilinears are of different flavors. The factor of $4$ appears because there are even and odd ranked bilinears for both chiral and non-chiral. If the theory has an internal flavor symmetry, the number of independent terms might reduce. In general, these bilinears couple with the complex-valued MHS $\Phi^{(\tilde{r},\tilde{s})}$ in the following ways
 \begin{align}
 \mathbf{g}_{ab}\mathbb{M}^{(r,s)}_{ab}\Phi^{(\tilde{r},\tilde{s})}+\mathbf{g}^*_{ab} \mathbb{M}^{(r,s)}_{ba}(\Phi^{(\tilde{r},\tilde{s})})^\dagger\qquad,\qquad \forall\quad r+s=\tilde{r}+\tilde{s},\label{multiplecomplexcoupling1}\\
\mathbf{h}_{ab}  \mathbb{N}^{(r,s)}_{ab}\Phi^{(\tilde{r},\tilde{s})}+\mathbf{h}^*_{ab} \mathbb{N}^{(r,s)}_{ba}(\Phi^{(\tilde{r},\tilde{s})})^\dagger\qquad ,\qquad \forall \quad r+s=\tilde{r}+\tilde{s},\label{multiplecomplexcoupling2}
 \end{align}
where $\mathbf{g}_{ab}$ and $\mathbf{h}_{ab}$ are the coupling constants dependent on the flavour $a,b$. There are only two possible solutions of $r+s=\tilde{r}+\tilde{s}$ as discussed in \eqref{rsconstraint}.
 For particular cases of real $\Phi^{(\tilde{r},\tilde{s})}_{ab}$ satisfying $\Phi^{(\tilde{r},\tilde{s})}=(\Phi^{(\tilde{r},\tilde{s})})^\dagger$, the couplings are given by
 \begin{align}
 	( \mathbb{M}^{(r,s)}_{ab}+ \mathbb{M}^{(r,s)}_{ba})\Phi^{(\tilde{r},\tilde{s})}\qquad,\qquad \forall\quad r+s=\tilde{r}+\tilde{s},\label{multiplerealcoupling1}\\
 (	\mathbb{N}^{(r,s)}_{ab}+ \mathbb{N}^{(r,s)}_{ba})\Phi^{(\tilde{r},\tilde{s})}\qquad ,\qquad \forall \quad r+s=\tilde{r}+\tilde{s},\label{multiplerealcoupling2}
 \end{align}
 Since MHS are irreps of the massive Little group $SO(2m-1)$, the antisymmetric indices $\tilde{r}$ have the following bound $m-1 \geq \tilde{r}\geq 0$. The Dirac bilinears are representations of the $Spin(2m-2)\otimes Spin(2m-2)$, which restricts the rank of the Dirac bilinear by $m-1 \geq r\geq 0$. The symmetric indices $s$ and $\tilde{s}$ have no constraints.

In odd $D=2m+1$ spacetimes, the independent bilinears are the even and the odd-ranked non-chiral bilinears. The total number of bilinears are $ 2\times \frac{N(N+1)}{2}$. The MHS transforms under the irreps of the Little group $SO(2m)$. The rank $r<m$ couplings of the bilinears with complex or real-valued MHS are given in \eqref{multiplecomplexcoupling1} and \eqref{multiplerealcoupling1}, respectively. The couplings for $\tilde{r}=m$ are of two types: the self-dual and the anti-self-dual couplings. Consider the rank $m$ tensors $\Phi^{(m,\tilde{s})}$ and $\widetilde{\Phi}^{(m,\tilde{s})}$ which follow the duality relation \eqref{dualitycovariance}. The possible couplings are of two types (with their hermitian conjugates)
 \begin{align}
 \frac{{\mathbf{d}_{ab}}}{2}~\mathbb{M}^{(m-1,\tilde{s}+1)}_{ab}{\Phi}^{(m,\tilde{s})}+\frac{\tilde{\mathbf{d}}_{ab}}{2}~\mathbb{M}^{(m-1,\tilde{s}+1)}_{ab}\widetilde{\Phi}^{(m,\tilde{s})}\label{multiplehodgedualcouplings}
\end{align}
The self dual couplings will have the couplings constant $\frac{1}{2}(\mathbf{d}_{ab}+\tilde{\mathbf{d}_{ab}})$. The anti self dual couplings will have the coupling constant $\frac{1}{2}(\mathbf{d}_{ab}-\tilde{\mathbf{d}_{ab}})$. Any MHS $\Phi^{(\tilde{r},\tilde{s})}$ with $1\leq \tilde{r}\leq m$ will have $2\times \frac{N(N+1)}{2}$ number of independent couplings. The number of independent couplings for $\tilde{r}=0$, i.e. scalars, is again one.

\subsection{Discussion on unconventional $CRT$}

As discussed, the $C,\, R_i, \, T$ transformations of these bilinears can also be defined in an unconventional manner. The $C,\, R_i,\, T$ transformations can come up with non-trivial matrices that intermixe the flavors but still keep any MHS couplings $CR_iT$ invariant (see chapter 2,  appendix C, \cite{Weinberg:1995mt} for discussion on unconventional CRT). The unconventional $C,\, R_i,\, T$ over the spinors can be defined as follows
\begin{align}
	C\Psi_{a}(x)C^{-1}&=(\zeta_{_C})_{ab}\, \mathcal{B}^{-1}\, \Psi_b(x)^*,\\
	R_i\Psi_{a}(x)R_i^{-1}&=(\zeta_{_R})_{ab}\, \Gamma^i\,\Psi_b(\mathcal{R}_ix)\\
	T\Psi_{a}(x)T^{-1}&=(\zeta_{_T})_{ab}\,\mathcal{C}\,\Psi_b(\mathcal{T}x)
\end{align}
Here \enquote{$a$}, \enquote{$b$} are the flavor indices and $(\zeta_{_C})_{ab},\, (\zeta_{_R})_{ab},\, (\zeta_{_T})_{ab}$ are the components of the matrix of intrinsic parities. The multiple flavor kinetic term of the Dirac spinors is diagonal\footnote{A non-diagonal kinetic term can always be made diagonal by field redefinitions preserving all its symmetries.}: $\delta_{ab}\bar{\Psi}_a\Gamma^\mu\partial_\mu\Psi_b$. For these transformations to keep the multiple Dirac spinor kinetic term invariant, the matrix of intrinsic parities must be unitary, i.e.
\begin{align}
	\zeta_{_C}^\dagger \zeta_{_C}=\mathbb{I}\qquad,\qquad \zeta_{_R}^\dagger \zeta_{_R}=\mathbb{I}\qquad,\qquad \zeta_{_T}^\dagger \zeta_{_T}=\mathbb{I},
\end{align}
where $\mathbb{I}$ is the $N\times N$ identity. In the case of conventional definitions of $C,\, R_i,\, T$, these matrices are all diagonal. We would need to do a $U(N)$ rotation in the field space to show that these matrices can be diagonalized simultaneously. Since the kinetic term has all the possible symmetries, any $U(N)$ transformation represented by the matrix $U$ over the field space will keep it invariant. This gives us the following transformations
  \begin{align}
 	U^{-1}\zeta_{_C}U^*=\zeta_{_C}'\qquad,\qquad U^{-1}\zeta_{_R}U=\zeta_{_R}'\qquad.\qquad U^{\mathbb{T}}\zeta_{_T}U=\zeta_{_T}'\label{unitaryintrinsictransformations}
 \end{align}
 where $'$ denotes the new transformed matrices and $U^\mathbb{T}$ denotes the transpose of $U$. Here, in the third identity, we have used the unitary property $U^* =(U^{-1})^\mathbb{T}$. The question is whether it is always possible to have such a unitary transformation that can diagonalize $\zeta_{_C},\, \zeta_{_R},\, \zeta_{_T}$ simultaneously. A non-diagonal matrix leaves many possibilities of the $C,\, R_i,\, T$ transformations. For this paper, we conclude by summarizing the unconventional $C,\, R_i,\, T$ transformations of the multiple flavor bilinears in table \ref{tab:CRTDiracmultiple}. 
 \renewcommand{\arraystretch}{2}
 \begin{table}[H]
 	\centering
 	\begin{tabular}{ ||c||c|c|| }
 		\hline
 		\hline
 		Bilinears            & 	$\mathbb{M}^{(r,s)}_{ab}$                   & 	$\mathbb{N}^{(r,s)}_{ab}$  \\
 		\hline
 $C$	& $(-1)^{s+1}t_1t_r(\zeta_{_C})^*_{ca}(\zeta_{_C})_{bd}\mathbb{M}^{(r,s)}_{cd}$     & $(-1)^{r+s+m+1}t_1t_r(\zeta_{_C})^*_{ca}(\zeta_{_C})_{bd}\mathbb{N}^{(r,s)}_{cd}$ \\
 		\hline
 	$R_i$ & $(-1)^{r+1}(\zeta_{_R})^*_{ca}(\zeta_{_R})_{bd}\mathbb{M}^{(r,s)}_{cd}$ & $(-1)^{r}(\zeta_{_R})^*_{ca}(\zeta_{_R})_{bd}\mathbb{N}^{(r,s)}_{cd}$  \\
 	\hline
 	$T$&$(-1)^{r+s}t_1t_r(\zeta_{_T})^*_{ca}(\zeta_{_T})_{bd}\mathbb{M}^{(r,s)}_{cd}$&$(-1)^{s+m+1}t_1t_r(\zeta_{_T})^*_{ca}(\zeta_{_T})_{bd}\mathbb{N}^{(r,s)}_{cd}$\\
 		\hline
 		\hline
 	\end{tabular}\caption{$C,R_i,T$ transformation properties of bilinears in any $D=2m/2m+1$ spacetime. See appendix \ref{App:CB} for the definition of $t_1,t_r$.}	\label{tab:CRTDiracmultiple}
 \end{table}
 The $C,\, R_i,\, T$ transformations of the three-point couplings can be figured out using table \ref{tab:CRTDiracmultiple}. In case of couplings with the real MHS, the transformations are same as in table \ref{tab:CRTDiracmultiple} including the intrinsic parities $\eta_{_C}^{(\tilde{r},\tilde{s})},\eta_{_R}^{(\tilde{r},\tilde{s})},\eta_{_T}^{(\tilde{r},\tilde{s})}$. Under $CR_iT$, the three-point couplings in \eqref{multiplerealcoupling1} and \eqref{multiplerealcoupling2} always transform upto the following factor 
\begin{align}
	CR_iT\qquad:\qquad  (\zeta_{_C})^*_{ca}\,(\zeta_{_R})^*_{\bar{c}c}\,(\zeta_{_T})^*_{\bar{\bar{c}}\bar{c}}\,(\zeta_{_C})_{bd}\,(\zeta_{_R})_{d\bar{d}}\,(\zeta_{_T})_{\bar{d}\bar{\bar{d}}}\,\eta_{_C}^{(\tilde{r},\tilde{s})}\, \eta_{_R}^{(\tilde{r},\tilde{s})}\,\eta_{_T}^{(\tilde{r},\tilde{s})}=+\delta_{a\bar{\bar{c}}} \,\delta_{b\bar{\bar{d}}}
\end{align}
From section \ref{sec:MultipleMHS}, we know that the intrinsic parity of an MHS always satisfies $\eta_{_C}^{(\tilde{r},\tilde{s})}\, \eta_{_R}^{(\tilde{r},\tilde{s})}\,\eta_{_T}^{(\tilde{r},\tilde{s})}=+1$. This implies that the intrinsic parities of a massless Dirac spinor are constrained through $(\zeta_{_C})^*_{ca}\,(\zeta_{_R})^*_{\bar{c}c}\,(\zeta_{_T})^*_{\bar{\bar{c}}\bar{c}}\,(\zeta_{_C})_{bd}\,(\zeta_{_R})_{d\bar{d}}\,(\zeta_{_T})_{\bar{d}\bar{\bar{d}}}=\delta_{a\bar{\bar{c}}} \,\delta_{b\bar{\bar{d}}}$. The only possible solutions to these constraints is $(\zeta_{_C})_{bd}\,(\zeta_{_R})_{d\bar{d}}\,(\zeta_{_T})_{\bar{d}\bar{\bar{d}}}=\pm e^{\iimg \theta}\, \delta_{b\bar{\bar{d}}}~ \forall ~b,\bar{\bar{d}}$.
With no loss of generality, let's consider $\theta=n\pi$ or $\theta=(2n+1)\frac{\pi}{2}$ which gives two solutions of the intrinsic parities
\begin{align}
	\zeta_{_C}\zeta_{_R}\zeta_{_T}=\pm \mathbb{I}\quad\text{or}\quad  \zeta_{_C}\zeta_{_R}\zeta_{_T}=\pm \iimg \,\mathbb{I}
\end{align}
Since $\zeta_{_C},\zeta_{_R},\zeta_{_T}$ are unitary, this shows that if one of the intrinsic parity matrices is diagonalized, the product of the other two also is diagonal by virtue of unitary transformations discussed in equation \eqref{unitaryintrinsictransformations}.

 Constraining the intrinsic parities for Dirac spinors and giving a classification of multiple flavor Dirac interactions with the MHS using $C,R_i,T$ is an interesting and open problem which is out of the scope of this paper. We leave it for the future work. Another interesting problem is to figure out square properties of $C,R_i,T$ and it's group structure in interacting theories of multiple flavors of Dirac spinors and MHS. See \cite{CPTwang, Wan:2023nqe} for the recent works in this direction.
\section{Conclusion and Future Directions}\label{sec:FutureDirections}
In this work, we classified the three-point couplings of the massive higher spin fields with the massless Dirac spinors. We also analyzed the action of $C$, $R_i,$ \& $T$ and the possible constraints of the space of such interaction with the help of these discrete transformations. 

The Dirac spinors are a complex representation of the real Clifford algebra; however, as a vector space, they are isomorphic to the irreducible representation of the complex Clifford algebras. The complex Clifford algebra exhibits a mod $2$ periodicity. One future goal is to classify the three-point interactions of massive higher spins interacting with two massless spinors, which transform under the irreducible representation of the real Clifford algebras. The real Clifford algebras (see appendix \ref{App:CliffordAlgebra}) have a mod $8$ isomorphism with the matrix algebras based on $\mathr,\,\mathc,\,\mathh$. In physics literature, these irreducible representations have different names (majorana, majorana-weyl, symplectic majorana) depending on the spacetime dimensions.

It would also be interesting to construct all massless three-point couplings of two spinors (Dirac or simplest spinors) with a massless boson. All massless three-point amplitudes are relevant for constructing on-shell recursion relations at the tree level, for example, the BCFW recursion relation in Yang-Mills \cite{Britto:2005fq, Elvang:2015rqa} and its generalization to gravity recursion relations \cite{Cachazo:2005ca} and effective field theories \cite{Cheung:2015ota}. The massless three-point couplings are kinematically restricted to vanish when all the particles are on-shell because the real-valued momenta of all the particles become proportional. Therefore, the couplings cannot be constructed using the on-shell approach used in this paper (see \cite{Buchbinder:2021qrg} for some works in this direction). In order to construct non-vanishing couplings, one must consider the momenta to be complex-valued, which would make them independent, keeping the momentum conservation independent. Another way to construct the on-shell massless couplings is to work with different signatures of the metric, for example, $(-,+,-,+)$, see page $17,27$ of \cite{Elvang:2015rqa}. Moreover, to complete the classification of the three-point interactions, it is also important to classify the space of all massive three-point couplings of fermions and bosons. Both of these works are under construction.

Another potential direction is to include the representations of supersymmetry algebra in various dimensions. In \cite{Balasubramanian2023}, the authors derived the supersymmetric constraints over the bosonic three-point couplings from various amounts of supersymmetry in $D=3+1$ spacetime. The form factors for four-point tree-level amplitudes for massless bosons mediated by massive higher spinning particles were shown to be the same as that of scalar particles. It would be interesting to impose supersymmetric constraints over the fermionic three-point couplings. It is expected that the form factors for the four-point massless bosonic amplitudes mediated by an MHS in any $D$ dimension should be given by Gegenbauer polynomials. It would also be interesting to explicitly show this for supersymmetric fermionic tree-level amplitudes mediated by MHS. In \cite{Chowdhury:2019kaq}, the authors classified the local four-point bosonic couplings. In order to constrain the tree level amplitudes with supersymmetries, it is also required to classify the four-point tree level for fermions.

This paper also classified the Dirac couplings based on their $C,\, R_i,\, T$ transformations. We have figured out the intrinsic parities of the MHS coupled with two massless Dirac spinors for the couplings invariant under $C, R_i, T$. Although this paper exhausts the $C, R_i, T$ transformations of the Dirac spinors and MHS, it would be interesting to figure out the space of $C, R_i, T$ transformations of the spinors transforming as irreducible representations of the real Clifford algebras. Although works have been done to understand $C, R_i, T$ transformations over the field space, not much is known about the Weinberg-like derivation of $C, R_i, T$ where its action is realized over the massive and massless physical Hilbert space in any $D$ dimension, see chapter $2$ of \cite{Weinberg:1995mt}. 

It is well known that the irreducible fields transform in the linear representation of the symmetry group $G$ and the projective representations of the symmetry group $Q$ such that there exists a short exact sequence 
\begin{align}
	1\rightarrow N\rightarrow G\rightarrow Q\rightarrow 1
\end{align}
where $N$ is the normal of $G$, see \cite{MooreQsymmetries}. There have been works in this direction recently, see \cite{Wan:2023nqe, CPTwang}, where the $CR_iT$ over Majorana fermions and its group extensions with the fermionic parity $(-1)^F$ has been studied for free theories in various spacetimes. It would also be interesting to figure out the $CR_iT$ group extension for interacting theories of simplest spinors and massive higher spins. Moreover, the group extensions were constructed for conventional $C, R_i, T$. The unconventional $C, R_i, T$, which, as we have studied in section \ref{sec:unconventionalCRT}, can lead to different possible group extensions which are absent in the case of conventional $C, R_i, T$. We hope to address some of these issues in the near future.\\
\newline
\textbf{Acknowledgment:} We thank Mahesh KNB, R. Loganayagam, and Arnab Priya Saha for their helpful discussions, without which this work would have been incomplete. We thank Mahesh KNB, Shaunak Patharkar, Raj Patil, Manav Shah, Arnab Priya Saha, Rahul Shaw, and Mrityunjay Verma for their valuable comments on the draft. KC acknowledges the support of the South African Research Chairs Initiative (SARChI) of the Department of Science and Innovation and the National Research Foundation. KC and AR would like to thank IMSC, Chennai, for their warm hospitality during the initial stage of this work. AK would like to thank ICTS for their hospitality during his visit, where discussions regarding the work initiated with R. Loganayagam. AK would also like to thank IISER Bhopal for supporting his research through the fellowship for PhD students. AY would like to thank IISERB for hospitality during his MS thesis, where some of the work had been done\footnote{Unpublished work \cite{AmeyMSthesis}. Expressions of bosonic tree amplitudes due to fermionic higher spin exchanges in $D=4$ spacetime were also included in the work.}. We thank the members of \href{https://sites.google.com/iiserb.ac.in/iiserbstrings/home}{strings@iiserb}, Department of Physics, IISERB, for providing a vibrant atmosphere. We thank the anonymous referees for suggesting various improvements.

Finally, we are grateful to the people of India for their generous funding for research in basic sciences.
\noindent
\newpage
\appendix
\section{Notations and Conventions}\label{sec:notations}
\vspace{-2em}
\begin{center}
	\begin{tabular}{l@{\hspace{2em}}l@{\vspace{-1em}}} \smallskip
		$\eta^{\mu\nu}=(-1,1,1,...,1)$ & Metric in flat spacetime\\
		$\iimg$&  $ \sqrt{-1}$ \\
		$\mu,\nu,\rho$&Spacetime indices\\
		$\alpha,\beta$&Spinor Indices\\
		$a,b$&Flavor indices\\
		$D=2m/2m+1$&Even or Odd spacetime dimension\\
			$\Lambda$ & Element of spin group \\
		$J^{\mu\nu}$ & Generator of spin group \\
		$\Gamma^\mu$& Gamma matrices in $D$ dimension\\ 
		$\Gamma_\star$& Chirality matrix in $D=2m$ dimensions \\ 
		$\Gamma^{(r)}$& Antisymmetric rank $r$ gamma matrices\\
			$C$ &  Charge Conjugation operator \\
		
		$P$ &  Parity operator \\
		
		$R_i$ &  Reflection operator along the $i^{th}$ spatial direction \\
		
		$T$ & Time reversal operator\\
				${(\mathcal{R}_i)}^{\mu}{}_\nu, $ 	$\mathcal{T}^{\mu}{}_\nu$ & Spacetime Representation of $R_i,T$\\
				$\mathcal{C},\mathcal{B},\beta$ &  Charge conjugation, Complex conjugation, and Dirac adjoint matrices\\
				$\zeta_{_C},\zeta_{_R},\zeta_{_T}$&Intrinsic parity of spinor under $C,R_i,T$\\
		$\Phi^{(\tilde{r},\tilde{s})}\equiv \Phi^{[\mu_1\cdots\mu_{\tilde{r}}]\{\nu_1\cdots\nu_{\tilde{s}}\}}$ & Massive boson with $\tilde{r}$ antisymmetric and $\tilde{s}$ symmetric indices\\ 
		$\Psi(x)$ & Dirac spinor\\ 
		$\mathbb{M}^{(r,s)}$, 	$\mathbb{N}^{(r,s)}$& Non-chiral, Chiral Dirac bilinear of rank $r$ with $s$ derivatives\\
		$\mathbb{M}^{(r,s)}_{ab}$, 	$\mathbb{N}^{(r,s)}_{ab}$& Non-chiral, Chiral Dirac bilinear for multiple flavors\\
		$\eta_{_C}^{(\tilde{r},\tilde{s})}$, $\eta_{_R}^{(\tilde{r},\tilde{s})}$, 	$\eta_{_T}^{(\tilde{r},\tilde{s})}$&Intrinsic parity of $\Phi^{(\tilde{r},\tilde{s})}$ under $C,R_i,T$\\
		$\mathbf{e},\, \mathbf{f},\, \mathbf{g}, \, \mathbf{h}$& Real Coupling constants\\
		$\mathcal{U}(\theta)=\exp(\iimg \theta)$&Vector transformation\\
		$\mathcal{U}_\star(\theta)=\exp(\iimg \theta\Gamma_\star)$&Axial transformation\\
		$\mathbf{m}$&Mass of the MHS\\
		$n=\lfloor {r/2}\rfloor,\tilde{n}=\lfloor {\tilde{r}/2}\rfloor$&Takes values in positive integers $\mathbb{Z}$\\
			MHS&Massive higher spin \\
		Irrep& Irreducible representations\\
	\end{tabular}
\end{center}

\section{Product of irreducible representations}\label{app:productofirreps}
In this section, we review the product of two spinor representations. Most of the content discussed in this section can be found in standard representation theory textbooks and articles \cite{georgi2018lie, fulton1991representation, goodman2009symmetry,das2014lie,Chakraborty:2020rxf}.

In $ D $-dimensional spacetime, the spinor irreps of the massless Little group $ SO(D-2) $ and the bosonic irreps of the massive Little group $ SO(D-1) $ are labeled by their highest weights $(w_1, w_2, \dots, w_l)$, where $ w_i $ are the eigenvalues of the rotation generators associated with the $ i^\text{th} $ plane, and $ l $ is the rank of the Little group. The weights $ w_i $ are integers for bosonic representations and half-integers for fermionic representations. To describe fermionic representations, we summarize the weight vectors for $ Spin(2l) $ and $ Spin(2l+1) $, the double covers of the orthogonal groups.
The weights of the bosonic (fermionic) irreducible representations of the Little group $ Spin(2l+1) $ are all positive integers (half-integers), subject to the condition:  
\begin{align}
	w_1 \geq w_2 \geq \cdots \geq w_l. \label{SO(2l+1)weights}
\end{align}  
There is a unique spinor representations of $Spin(2l+1)$ which we denote by $\rho$,
\begin{align}
\rho=(1/2,1/2,\cdots, 1/2).
\end{align}
A Dirac particle, which is a reducible representation of $Spin(2l+1)$, transforms under the direct sum $\rho\oplus \rho$. The irreducible state $\rho$ has $2^{l}$ degrees of freedom.

For $ Spin(2l) $, the weights $ w_1, \dots, w_{l-1} $ are all positive, while $ w_l $ can take both positive and negative values, subject to the condition:  
\begin{align}
	w_1 \geq w_2 \geq \cdots \geq |w_l|. \label{SO(2l)weights}
\end{align}  
The positive and negative values of $ w_l $ correspond to two inequivalent representations. For bosonic representations, these correspond to self-dual and anti-self-dual representations. For spinor representations, they correspond to two inequivalent fundamental spinor representations, associated with positive and negative handedness, respectively. As an example, consider $ Spin(4) \simeq SU(2) \otimes SU(2) $ whose representations are labeled by $\{j_1,j_2\}$, where $j_1,j_2$ are the spin labels of the two $SU(2)$'s. The relation between spin labels $j_1,j_2$ and the weights $w_1,w_2$ is as follows
\begin{align}
	w_1=j_1+j_2\qquad,\qquad w_2=j_1-j_2
\end{align}
 The left and the right handed spinor representations are : $\{1/2,0\}$ and $\{0,1/2\}$ respectively. The weight vectors for these representations are $(1/2,1/2)$ and $(1/2,-1/2)$. In general, we will denote the left handed representations by $\rho$ and the right handed representations by $\tilde{\rho}$
\begin{align}
	\rho=(1/2,1/2,\cdots, 1/2)\qquad,\qquad \tilde{\rho}=(1/2,1/2,\cdots, -1/2)\label{LeftRighthanded}
\end{align}
Both $\rho$ and $\tilde{\rho}$ have $2^{l-1}$ degrees of freedom. It should be noted that a Dirac particle, which is a reducible representation of $Spin(2l)$, transforms under the direct sum of $\rho\oplus \tilde{\rho}$.

For both $ Spin(2l+1) $ and $ Spin(2l) $, the product of two representations labeled by weights $(w_1, w_2, \dots, w_l)$ and $(q_1, q_2, \dots, q_l)$ decomposes into a direct sum of irreducible representations labeled by weights $ (v_1, v_2, \dots, v_m) $:  
\begin{align}
	(w_1, w_2, \dots, w_l) \otimes (q_1, q_2, \dots, q_l) = \bigoplus_{v_1, v_2, \dots, v_l} (v_1, v_2, \dots, v_l). \label{productofirreps}
\end{align}  
The irreps appearing in the direct sum on the RHS are determined by their characters. Each irrep is uniquely identified by a number (its character), which is defined as the trace of the representation matrix. The sum of the characters on the RHS of \eqref{productofirreps} must match the product of the characters on the LHS. In the following subsections we describe how to calculate the characters of the representations of $Spin(2l+1)$ and $Spin(2l)$.

\subsection{Representations of $Spin(2l+1)$}\label{app:productofoddirrpes}
For $Spin(2l+1)$ the character $\chi$ of a representation $(v_1,v_2,\cdots, v_l)$ can be calculated using the Weyl character formula \cite{goodman2009symmetry,fulton1991representation}
\begin{align}
	\chi(v_1,v_2,\cdots, v_l; \theta_1,\theta_2,\cdots,\theta_l)=\frac{Det(M)}{Det(N)}\label{odddimcharacter}
\end{align}
where $\theta_i$ are the rotation angles in the $i^{th}$ two plane, and $M$ and $N$ are $l\times l$ matrices  with the following entries
\begin{align}
	M_{ij}=(v_i+h_i) \sin\theta_j\qquad,\qquad N_{ij}=h_i \sin\theta_j, \qquad,\qquad h_i=l-i+\frac{1}{2}
\end{align}
Determining representations using the character formula is often a challenging task. However, it becomes more straightforward when dealing with spinor irreps by writing a simple mathematica code for \eqref{odddimcharacter}. For $ Spin(2l+1) $, the product of two spinor irreps is equal to the direct sum of all antisymmetric irreducible representations of rank $ 0 \leq r \leq l $. The weight vectors of the antisymmetric tensors have $v_i=1,0$ satisfying $v_i\geq v_{i+1}$. The odd rank tensors have odd number of $1$'s and remaining $0$'s. Whereas even rank tensors have even number of $1$'s and remaining $0$'s.
\begin{align}
	\rho\otimes \rho&=  (0,0,\dots,0)\oplus (1,0,\dots,0)\oplus \cdots \oplus(1,1,\dots,1)\\
	&=\bigoplus_{r=0}^{l}(v_1,\dots,v_{r}=1,~v_{r+1},\dots, v_{l}=0) \label{productirreps1}
\end{align}
 This can be verified using the character formula and by counting degrees of freedom. The antisymmetric rank-$ r $ tensor has $ \binom{2l+1}{r} $ degrees of freedom. The total degrees of freedom on the RHS of \eqref{productofirreps} is therefore  
\begin{align}
	\sum_{r=0}^{l} \binom{2l+1}{r} = 2^{2l},
\end{align}  
which matches the total degrees of freedom on the LHS, as the product of two spinor states, each with $ 2^l $ degrees of freedom, gives $ 2^l \times 2^l = 2^{2l} $.

As an example, consider the spinor representations of $ Spin(5) $, which are labeled by a two-dimensional weight vector $ (1/2, 1/2) $. The direct product $ (1/2, 1/2) \otimes (1/2, 1/2) $ decomposes into a scalar representation $ (0, 0) $, a vector representation $ (1, 0) $, and an antisymmetric rank-2 tensor representation $ (1, 1) $.
\begin{align}
	\big(1/2,1/2\big)\otimes \big(1/2,1/2\big)&=\big(1,1\big)\oplus\big(1,0\big)\oplus (0,0),\\
	4\times 4&= 10+5+1
\end{align}
\subsection{Representations of $Spin(2l)$}\label{app:productofevenirrpes}
For $Spin(2l)$ the character formula is slightly different
\begin{align}
	\chi(v_1,v_2,\cdots, v_l; \theta_1,\theta_2,\cdots,\theta_l)=\frac{Det(P)+\iimg^l Det(Q)}{Det(R)}
\end{align}
where $P,Q,R$ are $l\times l$ matrices  with the following entries
\begin{align}
	P_{ij}=(v_i+h_i) \cos\theta_j\quad,\quad Q_{ij}=(v_i+h_i) \sin\theta_j\quad,\quad R_{ij}=h_i \cos\theta_j, \quad,\quad h_i=l-i
\end{align}
Similar to $ Spin(2l+1) $, the product of two inequivalent spinor representations ($\rho$ and $\tilde{\rho}$, see eqn \eqref{LeftRighthanded}) of $ Spin(2l) $ can be determined using character decomposition. Interestingly, the result of this product depends on whether $ l $ is odd or even.

 For $ l = 2k $, the product of two spinor representations of $ Spin(2l) $ exhibits distinct behavior based on their handedness. The product of two representations with the same handedness results in the direct sum of even-rank tensors up to rank $ 2k $, while the product of representations with opposite handedness gives the direct sum of odd-rank tensors up to rank $ 2k-1 $. The weight vectors of even and odd ranked tensors of $Spin(2l)$ are same as that of $Spin(2l+1)$ except for the highest rank $l$ tensor. The highest rank $l$ tensor is either self-dual or anti-self-dual with weight vectors $(1,1,\dots,1)$ and $(1,1,\dots,-1)$ respectively. For $l=2k$, we have the following decomposition of $\rho$ and $\tilde{\rho}$,
 \begin{align}
 	{\rho}\otimes {\rho}&= (v_1,\dots,v_{l-1}=1, ~v_{2k}=1)~ \bigoplus_{r=0,rEven}^{2k-2} (v_1,\dots,v_r=1, ~v_{r+1},\dots, v_{2k}=0),	\label{productrepresentation2}\\
 	\tilde{\rho}\otimes \tilde{\rho}&= (v_1,\dots,v_{l-1}=1, ~v_{2k}=-1)~\bigoplus_{r=0,rEven}^{2k-2} (v_1,\dots,v_r=1, ~v_{r+1},\dots, v_{2k}=0),\label{productrepresentation3}	\\
 		\rho\otimes \tilde{\rho}&=\bigoplus_{r=1, r Odd}^{2k-1} (v_1,\dots,v_{r}=1, ~v_{r+1},\dots, v_{2k}=0)\label{productrepresentation4},\\
 			\tilde{\rho}\otimes {\rho}&=\bigoplus_{r=1, r Odd}^{2k-1} (v_1,\dots,v_{r}=1, ~v_{r+1},\dots, v_{2k}=0),\label{productrepresentation5}
 \end{align}
% \begin{align}
% 	(1/2, 1/2, \dots, \pm 1/2) \otimes (1/2, 1/2, \dots, \pm 1/2)  &=  (0,0,0,\dots,0)\oplus (1,1,0,\dots,0)\oplus \cdots \oplus(1,1,1,\dots,\pm 1),\label{productirreps2}\\
% 		(1/2, 1/2, \dots, 1/2) \otimes (1/2, 1/2, \dots, -1/2)  &=  (1,0,0,\dots,0)\oplus (1,1,1,\dots,0)\oplus \cdots \oplus(1,1,1,\dots,0),\label{productirreps3}
% \end{align}
	This decomposition can be verified by comparing the degrees of freedom on both sides. Each spinor representation has $ 2^{2k-1} $ degrees of freedom, so the product of two spinor representations yields $ 2^{4k-2} $ degrees of freedom. The total degrees of freedom for odd-rank tensors is given by  
	\begin{align}
		\sum_{r=1, \, r \, \text{odd}}^{2k-1} \binom{4k}{r} = 2^{4k-2}.
	\end{align}
 For even-rank tensors, the highest rank $ 2k $ tensor is either self-dual or anti-self-dual, each contributing $\frac{1}{2} \binom{4k}{2k}$ degrees of freedom. All lower even-rank tensors contribute $\binom{4k}{r}$ degrees of freedom. Thus, the total for even-rank tensors is  
	\begin{align}
		\frac{1}{2} \binom{4k}{2k} + \sum_{r=0, \, r \, \text{even}}^{2k-2} \binom{4k}{r} = 2^{4k-2}.
	\end{align}
	
	As an example, consider $ Spin(4) \simeq SU(2) \otimes SU(2) $ whose left and the right handed representations are : $\{1/2,0\}$ and $\{0,1/2\}$ respectively. Their products decompose as follows:  
	\begin{align}
		\begin{aligned}
			\{1/2,0\} \otimes \{1/2, 0\} &= \{1, 0\} \oplus \{0, 0\}, \\
			\{1/2,0\} \otimes \{0, 1/2\} &= \{1/2, 1/2\},\\
			\{0,1/2\} \otimes \{0,1/2\} &= \{0,1\} \oplus \{0, 0\},
		\end{aligned}
	\end{align}
where $\{1,0\}$ and $\{0,1\}$ are the rank $2$ self dual and anti-self dual tensors, respectively. The $\{0, 0\}$ represents the scalar representation, whereas $\{1/2, 1/2\}$ corresponds to the vector representation of $ SO(4) $.

 For $ l = 2k+1 $, the decomposition of spinor representation products contrasts with that in $ l = 2k $. The product of two spinor representations of the same chirality yields odd-rank tensors up to rank $ 2k+1 $, while the product of opposite chirality representations results in even-rank tensors up to rank $ 2k $. The highest odd-rank tensor, with rank $ 2k+1 $, is either self-dual or anti-self-dual.
 \begin{align}
	{\rho}\otimes {\rho}&= (v_1,\dots,v_{2k}=1, ~v_{2k+1}=1)~ \bigoplus_{r=1,rOdd}^{2k-1} (v_1,\dots,v_r=1, ~v_{r+1},\dots, v_{2k+1}=0),\label{productrepresentation6}	\\
	\tilde{\rho}\otimes \tilde{\rho}&= (v_1,\dots,v_{2k}=1, ~v_{2k+1}=-1)~
	\bigoplus_{r=1,rOdd}^{2k-1} (v_1,\dots,v_r=1, ~v_{r+1},\dots, v_{2k+1}=0),\label{productrepresentation7}	\\
	\rho\otimes \tilde{\rho}&=\bigoplus_{r=0, r Even}^{2k} (v_1,\dots,v_{r}=1, ~v_{r+1},\dots, v_{2k+1}=0),\label{productrepresentation8}\\
	\tilde{\rho}\otimes {\rho}&=\bigoplus_{r=0, r Even}^{2k} (v_1,\dots,v_{r}=1, ~v_{r+1},\dots, v_{2k+1}=0),\label{productrepresentation9}
\end{align}
	The degrees of freedom for these decompositions can be matched as follows: The odd-rank tensors, including the highest rank $ 2k+1 $ tensor, have  
	\begin{align}
		\frac{1}{2} \binom{4k+2}{2k+1} + \sum_{r=1, \, r \, \text{odd}}^{2k-1} \binom{4k+2}{r} = 2^{4k}.
	\end{align}
	The even-rank tensors contribute  
	\begin{align}
		\sum_{r=0, \, r \, \text{even}}^{2k} \binom{4k+2}{r} = 2^{4k},
	\end{align}  
	matching the total degrees of freedom, $ 2^{4k} $, for the product of spinor states.
	
	For example, consider the product of helicity $\pm 1/2$ states in $ SO(2) $. This gives rise to helicity $\pm 1$ and helicity $ 0 $ states:  
	\begin{align}
		\begin{aligned}
			\ket{\pm 1/2} \otimes \ket{\pm 1/2} &= \ket{\pm 1}, \\
			\ket{\pm 1/2} \otimes \ket{\mp 1/2} &= \ket{0}.
		\end{aligned}
	\end{align}

The discussion of both $Spin(2l+1)$ and $Spin(2l)$ concludes that the product of two irreducible spinor states gives rise to all possible antisymmetric irreducible tensors of rank $0\leq r\leq l $.\textit{ If the spinors are massless, the maximum rank of the antisymmetric tensor is equal to the rank of the massless Little group.} Therefore, in any $D=2m/2m+1$ dimensions, the maximum rank of an irreducible massless Dirac spinor bilinear is $m-1$ governed by the massless Little group $Spin(2m-2)$ or $Spin(2m-1)$. All the higher rank spinor bilinears are reducible to bilinears  of rank $0\leq r\leq m-1$.
\section{Branching rules}\label{app:Branchingrules}
In this section, we review the branching rules \cite{das2014lie, goodman2009symmetry,Chakraborty:2020rxf} for projecting the $Spin(2l)$ representations over $Spin(2l-1)$ representations, or projecting the $Spin(2l+1)$ representations over $Spin(2l)$ representations. 

A $Spin(2l)$ representation labeled by weight vector $(w_1,w_2,\dots, w_l)$ decomposes into a direct sum of $Spin(2l-1)$ representations labeled by $(v_1^a,v_2^a,\dots, v_{l-1}^a)$ such that the following conditions hold:
\begin{itemize}
	\item The weights $v_i^a$ and $w_i$ have the same integer property i.e. either integer or half integer.
	\item The weights $v_i^a$ and $w_i$ satisfy the following inequalities
	\begin{align}
		w_1\geq v_1^a\geq w_2\geq v_2^a \geq \cdots \geq v_{l-1}^a \geq |w_l|\label{branchingrule1}
	\end{align}
\end{itemize}
A $Spin(2l+1)$ representation labeled by weight vector $(w_1,w_2,\dots, w_l)$ decomposes into a direct sum of $Spin(2l)$ representations labeled by $(v_1^a,v_2^a,\dots, v_{l}^a)$ such that the following conditions hold:
\begin{itemize}
	\item The weights $v_i^a$ and $w_i$ have the same integer property i.e. either integer or half integer.
	\item The weights $v_i^a$ and $w_i$ satisfy the following inequalities
	\begin{align}
		w_1\geq v_1^a\geq w_2\geq v_2^a \geq \cdots \geq v_{l-1}^a \geq w_l\geq |v_{l}^a| \label{branchingrule2}
	\end{align}
\end{itemize}
\section{Projections of MHS representations}\label{app:Projectionofirreps}
In this section we describe how group theory enables us to compute all the independent couplings in table \ref{tab:oddcouplings} and \ref{tab:evencouplings}.
\subsection{Projections in \(D=2m+1\) Spacetime}  \label{app:Projectionofodddimirreps}

In \(D=2m+1\) spacetime dimensions, the massless Little group is \(Spin(2m-1)\), and the massive Little group is \(Spin(2m)\) (or \(SO(2m)\)). From the discussion in Appendix \ref{app:productofirreps}, the product $\rho\otimes \rho$ of two irreducible spinor states of \(Spin(2m-1)\) generates antisymmetric tensors up to rank \(m-1\). All possible kinematically allowed couplings of a bosonic irrep of \(Spin(2m)\) with these antisymmetric tensors are determined by the branching rule \eqref{branchingrule1}, where \(l = m\).  

The solutions to the branching rule are simplified by noting that the weights \(v_i^a\) of completely antisymmetric irreps are restricted to \(0\) or \(1\) (see \eqref{productirreps1}). This leads to three distinct boundary conditions for the weight vectors, along with their corresponding solutions for the higher spin weights:  
\begin{enumerate}
\item All zeros (\(v_1^a, \cdots, v_{m-1}^a = 0\)) :  
\begin{align}  
	w_1 \geq 0\quad, \quad w_2, \cdots, w_m = 0. \label{DOddprojections1}  
\end{align}  

\item \(r\) ones (\(v_1^a, \cdots, v_r^a = 1, \, v_{r+1}^a, \cdots, v_{m-1}^a = 0, \, r \leq m-2\)) :  
\begin{align}  
	w_1 \geq 1\quad, \quad w_2, \cdots, w_r = 1\quad, \quad w_{r+1} = 1 \text{ or } 0\quad, \quad w_{r+2}, \cdots, w_m = 0. \label{DOddprojections2}  
\end{align}  
\item All ones (\(v_1^a, \cdots, v_{m-1}^a = 1\)) :  
\begin{align}  
	w_1 \geq 1\quad, \quad w_2, \cdots, w_{m-1} = 1\quad, \quad w_m = \pm 1 \text{ or } 0. \label{DOddprojections3}  
\end{align}  
\end{enumerate}
The solutions \eqref{DOddprojections1}–\eqref{DOddprojections3} imply that states with \(w_2, \dots, w_m > 1\) cannot be projected over the product $\rho\otimes \rho$. Consequently, \(SO(2m)\) Young tableaux with more than one box in the $2^{nd}$ to \(m^\text{th}\) rows cannot couple to product of spinor states of $SO(2m-1)$. The same will be true for the projections over the product of Dirac particle states transforming under the direct sum $\rho\oplus \rho$. In this framework, the only MHS that can couple are represented by the Young tableaux \eqref{higherspinningboson}. The relation between the weights and the indices of MHS $\Phi^{(\tilde{r},\tilde{s})}$ is as follows
\begin{align}
	\tilde{r}=\text{Number of non-zero $w$'s}\qquad,\qquad \tilde{s}=\begin{cases}
		w_1-1, ~&w_1\geq 1\\
		0,~&w_1=0
	\end{cases}
\end{align}
  Using the solutions \eqref{DOddprojections1}–\eqref{DOddprojections3}, we categorize the couplings into three distinct types based on the properties of the MHS $\Phi^{(\tilde{r},\tilde{s})}$:
\begin{enumerate}
\item Scalar Couplings $\Phi^{(0,0)}$: The weight vector for a scalar MHS is  \((w_1, \cdots, w_m = 0)\). There is exactly one independent coupling governed by the branching rule \eqref{branchingrule1}. This corresponds to the projection of the scalar representation of \(Spin(2m)\) onto the scalar representation of \(Spin(2m-1)\). The corresponding coupling of a massive scalar \(\Phi^{(0,0)}\) with a rank-0 Dirac bilinear \(\bar{\Psi} \Psi \Phi\). This coupling is described in the first row of Table \ref{tab:oddcouplings} for \(\tilde{r} = \tilde{s} = 0\).  

\item MHS with rank $\tilde{r}< m,~ \tilde{s}\geq 0$ :
The weight vector for such a MHS is  $(w_1=\tilde{s}+1, w_2, \cdots, w_{\tilde{r}} = 1, w_{\tilde{r}+1}, \cdots, w_m = 0)$ which has two independent projections governed by the branching rule \eqref{branchingrule1}. 

The first projection is over the rank $r$ tensor $(v_1^a,\cdots, v_{\tilde{r}}^a=1, ~v_{\tilde{r}+1}^a,\cdots , v_{m-1}^a=0)$ which is achieved by contracting $\tilde{s}$ number of indices in the first row with derivatives acting over the spinors. The remaining $\tilde{r}$ antisymmetric indices are contracted with the gamma matrices. The corresponding coupling is listed in row $1$ of table \ref{tab:oddcouplings}. 

The second projection is over the rank $\tilde{r}-1$ tensor  $(v_1^a,\cdots, v_{\tilde{r}-1}^a=1, ~v_{\tilde{r}}^a,\cdots , v_{m-1}^a=0)$ which is achieved by contracting the entire top row with $\tilde{s}+1$ number of derivatives. The remaining $\tilde{r}-1$ antisymmetric indices are contracted with the gamma matrices. The corresponding coupling is listed in row $2$ of table \ref{tab:oddcouplings}.

\item MHS with rank $\tilde{r}= m,~ \tilde{s}\geq 0$ : Such a MHS transforms under the direct sum of self dual and anti-self dual tensor with weight vectors $(w_1=\tilde{s}+1, w_2, \cdots, w_m = 1)$ and $(w_1=\tilde{s}+1, w_2, \cdots, w_m = -1)$ respectively. The self dual and the anti-self dual MHS are given by equations \eqref{selfdual} and \eqref{antiselfdual}. For both of them, there exists a unique projection over the rank $m$ tensor $(v_1^a,\cdots, v_{m-1}^a=1)$. The projection is achieved by contracting the entire top row of the MHS with derivatives. The remaining indices $\tilde{r}-1$ antisymmetric indices are contracted with the gamma matrices. The details of these couplings are discussed in the paragraph below Table \ref{tab:oddcouplings}.  
\end{enumerate}
This exhausts all the kinematically allowed couplings of a MHS and two massless Dirac spinors in any odd spacetime. 
\subsection{Projections in $D=2m$ spacetime}\label{app:Projectionofevendimirreps}
In $D = 2m$ spacetime dimensions, the massless little group is $Spin(2m-2)$, while the massive little group is $Spin(2m-1)$ (or equivalently $SO(2m-1)$). As discussed in Appendix \ref{app:productofirreps}, the states $\rho$ and $\tilde{\rho}$ represent two independent spinor states of $Spin(2m-2)$, as defined in \eqref{LeftRighthanded}. The structures in the product of these representations depends on the rank $m-1$ of the massless little group:\\
- For $m-1 = 2k$ (even rank), the product of two states with the same (opposite) handedness decomposes into a direct sum of irreducible tensors of odd (even) rank.\\
- For $m-1 = 2k+1$ (odd rank), the product of two states with the same (opposite) handedness decomposes into a direct sum of irreducible tensors of even (odd) rank.

In both cases, the highest-rank tensor ($m-1$) is either self-dual or anti-self-dual. The coupling of these antisymmetric tensors of $Spin(2m-2)$ with a bosonic irrep of $Spin(2m-1)$ is determined by the branching rule \eqref{branchingrule2}. The weights in the product of the $Spin(2m-2)$ spinor states, as given in \eqref{productrepresentation2}-\eqref{productrepresentation5} and \eqref{productrepresentation6}-\eqref{productrepresentation9}, take the following values:
\begin{align}
	v_1^a, \dots, v_{m-2}^a \in \{1, 0\}, \quad v_{m-1}^a \in \{\pm 1, 0\}.
\end{align}
Here we write down the solutions to the branching rule \eqref{branchingrule2} for all possible weights of $Spin(2m-2)$:

1. All zeros $(v_1^a, \cdots, v_{m-1}^a = 0)$:  
\begin{align}
	w_1 \geq 0, \quad w_2 ,\cdots, w_{m-1} = 0. \label{DEvenprojections1}
\end{align}

2. $r$ ones $(v_1^a ,\cdots , v_r^a = 1, \, v_{r+1}^a , \cdots ,v_{m-1}^a = 0, \, r \leq m-2)$:  
\begin{align}
	w_1 \geq 1, \quad w_2,\cdots, w_r = 1, \quad w_{r+1} = 1 \text{ or } 0, \quad w_{r+2},\cdots,  w_{m-1} = 0. \label{DEvenprojections2}
\end{align}

3. All ones $(v_1^a ,\cdots , v_{m-2}^a = 1, \, v_{m-1} = \pm 1)$:  
\begin{align}
	w_1 \geq 1, \quad w_2 ,\cdots , w_{m-1} = 1. \label{DEvenprojections3}
\end{align}
Similar to $D=2m+1$ spacetime, the solutions \eqref{DEvenprojections1}-\eqref{DEvenprojections3} concludes that the only MHS in $D=2m$ spacetime which can couple have no more than one box in the $2^{nd}$ to $(m-1)^{th}$ row of the Young tableau i.e. $w_2,\cdots w_{m-1}\leq 1$. The same will be true for the projections over the product of Dirac states which transform under the direct sum $\rho\oplus \tilde{\rho}$. At the level of Lorentz group  all antisymmetric tensors are generated using the left Weyl spinor $\chi$ and the right Weyl spinor $\xi$,
\begin{align}
	\Gamma_\star\chi=+\chi\qquad,\qquad \Gamma_\star\xi=-\xi\label{chiralspinors}
\end{align}

All the even (odd) ranked tensors gives rise to bilinears of opposite (same) chirality Weyl spinors i.e. the only non-vanishing bilinears are
\begin{align}
	\bar{\chi}\Gamma^{(2n)}\xi\qquad,\qquad \bar{\xi}\Gamma^{(2n)}\chi,\qquad,\qquad
	\bar{\chi}\Gamma^{(2n+1)}\chi\qquad,\qquad \bar{\xi}\Gamma^{(2n+1)}\xi\label{chiralbilinears}
\end{align}
where $\bar{\chi}=\iimg \chi^\dagger \Gamma^{0}$ and  $\bar{\xi}=\iimg \xi^\dagger \Gamma^{0}$. All other bilinears vanish because of the chirality conditions \eqref{chiralspinors}. The even ranked bilinears are hermitian conjugates of each other upto a $\pm$ sign. Whereas the odd ranked bilinears are self conjugates again upto a $\pm$ sign. The Dirac spinor fields on the other hand can be written as a sum of left chiral $\chi$ and the right chiral $\xi$ Weyl spinor fields
\begin{align}
	\Psi=\chi+\xi
\end{align}
Following \eqref{chiralbilinears}, there are only two types of even and odd ranked Dirac spinor bilinears
\begin{align}
	\bar{\Psi}\Gamma^{(r)}\Psi\qquad,\qquad \bar{\Psi}\Gamma^{(r)}\Gamma_\star\Psi
\end{align}
The full hermitian form of the bilinears is given in \eqref{DiracBilinears1} and \eqref{DiracBilinears2}. Following the branching rule \eqref{branchingrule2}, we again make three possible classes of the couplings depending on the properties of the MHS $\Phi^{(\tilde{r},\tilde{s})}$:
\begin{enumerate}
	\item Scalar Couplings $(w_1,\cdots, w_m=0)$: The scalar representations of $Spin(2m-1)$ can only be projected over the scalar representations of $Spin(2m-2)$. Depending on $m$, there are two scalar representations of $Spin(2m-2)$ generated out of either the product of two same handed states (${\rho}\otimes 
	{\rho}$ and $\tilde{\rho}\otimes 
	\tilde{\rho}$) or two oppositely handed states  (${\rho}\otimes 
	\tilde{\rho}$ and $\tilde{\rho}\otimes 
	{\rho}$). Note that  This gives rise to two independent couplings given known as the Yukawa coupling of a scalar and Pseudoscalar : $\bar{\Psi}\Psi\Phi$ and $\iimg \bar{\Psi}\Gamma_\star\Psi\Phi$ which are different linear combinations of the Weyl spinor couplings: $(\bar{\chi}\xi+\bar{\xi}\chi)\Phi$ and $\iimg (\bar{\chi}\xi-\bar{\xi}\chi)\Phi$. These couplings are enlisted in the row $1$ and $3$ of table \ref{tab:evencouplings} for $\tilde{r}=\tilde{s}=0$. The couplings in rows $2$ and $4$ do not exist for massive scalars. 
	
	\item  MHS with rank $\tilde{r}< m-1,~ \tilde{s}\geq 0$ : The weight vector for such a MHS is $(w_1=\tilde{s}+1, w_2,\cdots,w_{r}=1, w_{r+1},\cdots, w_{m-1}=0)$ which has two possible projections in this case governed by the branching rule \eqref{branchingrule2}. Similar to the previous case, each projection gives rise to two possible couplings : with and without a $\Gamma_\star$.
	
	 The first projection is over the rank $\tilde{r}$ tensor $(v_1^a,\cdots,v_{\tilde{r}}^a=1,~ v_{\tilde{r}+1},\cdots, v_{m-1}^a=0)$ obtained by contracting $\tilde{s}$ indices in the first row of MHS with derivatives. The remaining indices contract with the gamma matrices in the bilinear. The corresponding couplings are listed in row $1$ and $3$ of table \ref{tab:evencouplings} for $\tilde{r}< m-1, ~\tilde{s}\geq 0$.
	 
	  The second projection is over the rank $(\tilde{r}-1)$ tensor $(v_1^a,\cdots,v_{\tilde{r}-1}^a=1,~ v_{\tilde{r}},\cdots, v_{m-1}^a=0)$ obtained by contracting the entire first row of MHS with derivatives. The corresponding couplings are listed in row $2$ and $4$ of table \ref{tab:evencouplings} for $\tilde{r}< m-1, ~\tilde{s}\geq 0$.

	\item MHS with rank $\tilde{r}= m-1,~ \tilde{s}\geq 0$ : The weight vector for such a MHS is $(w_1=\tilde{s}+1, w_2,\cdots, w_{m-1}= 1)$. In this case there are two possible projections corresponding each giving rise to a chiral and a non-chiral coupling. 
	
	The first projection is over the rank $m-2$ tensor $(v_1^a,v_2^a,\cdots, v_{m-2}^a=1, ~v_{m-1}^a=0)$. The corresponding couplings are listed in row $2$ and $4$ of tale \ref{tab:evencouplings} which are obtained by contracting the entire first row of MHS with derivatives.
	
	The other projection is over the rank $m-1$ tensor which transforms as a direct sum of self dual tensor $(v_1^a,\cdots, v_{m-2}^a=1, ~v_{m-1}^a=1)$ and anti-self dual tensor $(v_1^a,\cdots, v_{m-2}^a=1, ~v_{m-1}^a=-1)$ of $SO(2m-2)$. One of these couplings is given by contracting the first row of MHS with $\tilde{s}$ number of derivatives. The remaining antisymmetric indices are contracted with the Dirac bilinear
	\begin{align}
		\bar{\Psi}\Gamma^{\mu_1\cdots \mu_{m-1}}\Psi\label{selfdualbilinear}
	\end{align}
	 This gives rise to the coupling listed in row $1$ of table \ref{tab:evencouplings}. The other coupling is obtained by contracting the antisymmetric indices with the rank $m-1$ bilinear involving the $SO(2m-2)$ epsilon tensor. In the momentum space, the covariant version of the $SO(2m-2)$ epsilon tensor is \cite{Chakraborty:2020rxf}
	\begin{align}
		\epsilon_{\mu\nu\mu_1\cdots \mu_{2m-2}}k_1^\mu k_2^\nu
	\end{align}
	The corresponding coupling is given by contracting the first row of the MHS with $\tilde{s}$ derivatives and the remaining antisymmetric indices with rank $m-1$ bilinear,
	\begin{align}
			\epsilon_{\mu\nu\mu_1\cdots \mu_{m-1}\nu_1\cdots\nu_{m-1}}	(\partial^\mu	\bar{\Psi})\Gamma^{\nu_1\cdots \nu_{m-1}}(\partial^\nu\Psi)\Phi^{[\mu_1\cdots \mu_{\tilde{m-1}}]\{\nu_1\cdots \nu_{\tilde{s}}\}}\label{epsilonbilinear}
	\end{align}
	However, it can be shown using the gamma identity \eqref{evengammarelation} that the rank $m-1$ bilinear with the epsilon tensor is same as the rank $m-1$ bilinear \eqref{selfdualbilinear} with a $\Gamma_\star$. For example, in $D=4$ spacetime, the coupling with the bilinear \eqref{epsilonbilinear} is $	\epsilon_{\mu\nu\rho\sigma}(\partial^{\mu}\bar{\Psi})\Gamma^{\rho}(\partial^{\nu}\Psi)\Phi^\sigma$ which can be reduced as follows
	\begin{align}
		\epsilon_{\mu\nu\rho\sigma}(\partial^{\mu}\bar{\Psi})\Gamma^{\rho}(\partial^{\nu}\Psi)\Phi^\sigma	&= 3! \iimg (\partial^{\mu}\bar{\Psi})\Gamma_{\mu\nu\rho}\Gamma_\star(\partial^{\nu}\Psi)\Phi^\rho	,\\
		&= \iimg (\partial^{\mu}\bar{\Psi})(
		 \Gamma_\mu \Gamma_\nu\Gamma_\rho
		-\Gamma_\mu \Gamma_\rho\Gamma_\nu
		-\Gamma_\nu \Gamma_\mu\Gamma_\rho \nonumber\\
		&\qquad\qquad +\Gamma_\nu \Gamma_\rho\Gamma_\mu
		-\Gamma_\rho \Gamma_\nu\Gamma_\mu
		+\Gamma_\rho \Gamma_\mu\Gamma_\nu)\Gamma_\star(\partial^{\nu}\Psi)\Phi^\rho,\\
	&=\iimg (\partial^{\mu}\bar{\Psi})(
-\Gamma_\nu \Gamma_\mu\Gamma_\rho +\Gamma_\nu \Gamma_\rho\Gamma_\mu
-\Gamma_\rho \Gamma_\nu\Gamma_\mu)\Gamma_\star(\partial^{\nu}\Psi)\Phi^\rho	,\\
&= 6 \iimg (\partial^\mu\bar{\Psi})\Gamma^\nu\Gamma_\star(\partial_\mu \Psi)\Phi_\nu \label{reduciblecoupling}
	\end{align}
	where in the third equality we have used the equation of motion $ (\partial^\mu\bar{\Psi})\Gamma_\mu=0=\Gamma_\nu(\partial^\nu\Psi) $, and in the fourth equality we have used the identity 
	\begin{align}
		\Gamma_{\mu_1}\Gamma_{\mu_2}\Gamma_{\mu_3}=2\eta_{\mu_1\mu_2}\Gamma_{\mu_3}-2\eta_{\mu_1\mu_3}\Gamma_{\mu_2}+2\eta_{\mu_2\mu_3}\Gamma_{\mu_1}-\Gamma_{\mu_3}\Gamma_{\mu_2}\Gamma_{\mu_1}
	\end{align}
	Moreover, note that the coupling in \eqref{reduciblecoupling} has two derivatives contracted with each other. In the momentum space, these are proportional (upto the Mandelstam factor $k_1\cdot k_2$) to the coupling which has no derivatives contracted with each other i.e.
	\begin{align}
	\bar{\Psi}\Gamma^\nu\Gamma_\star\Psi\Phi_\nu
	\end{align} 
	The same procedure can be applied to any epsilon coupling in $D=2m$ spacetime. The corresponding coupling with a $\Gamma_\star$ is listed in row $3$ of table \ref{tab:evencouplings}.
\end{enumerate} 
This exhausts all the kinematically allowed couplings of a MHS and two massless Dirac spinors in any $D=2m$ spacetime.
\section{Clifford Algebra}\label{App:CliffordAlgebra}
A Clifford algebra $Cl(p,q)$ over a $D=p+q$ dimensional vector space $(V,\mathbb{F})$, equipped with a quadratic form $Q(\vec{v})\, \forall\, \vec{v}\in V$, is generated by the elements $\Gamma^\mu$ which satisfy the following algebra
\begin{align}
	\{	\Gamma^\mu,\Gamma^\nu\}=2\eta^{\mu\nu}
\end{align}
where the metric $\eta^{\mu\nu}$ is the metric of $Q(\vec{v})$ with $p$ number of $+$ and $q$ number of $-$. It should be made clear that $\Gamma^\mu$ $\forall ~1\leq \mu\leq p+q$ are the orthogonal elements which spans the $p+q$ dimensional vector space. Any vector $\vec{v}\in V$ is given  by 
\begin{align}
	\vec{v}=v_1\Gamma^1+v_2\Gamma^2+\cdots +v_{p+q}\Gamma^{p+q}
\end{align}
where $v_i\in \mathbb{F}$ $\forall ~ 1\leq i\leq p+q$. The quadratic form $Q(\vec{v})$ is given by $Q(\vec{v})=\eta^{\mu\nu}v_{\mu}v_{\nu}$. On the other hand, the Clifford algebra basis is spanned by the tensors of $\Gamma^{\mu}$'s. Any rank $r$ tensor of Clifford basis is given by $\Gamma^{(r)}\equiv\Gamma^{\mu_1}\cdots\Gamma^{\mu_r}$. Any general element $x\in  Cl(p,q)$ is a linear combination of it's tensor elements $\Gamma^{(r)}$
\begin{align}
x
%	\sum_{r=0}^{D}v_{\mu_1}v_{\mu_2}\cdots v_{\mu_r}\Gamma^{\mu_1\mu_2\cdots \mu_r}
=x_0\mathbb{I}+x_\mu\Gamma^\mu+x_{\mu\nu}\Gamma^{\mu}\Gamma^\nu+\cdots +x_{\mu_1\cdots\mu_D}\Gamma^{\mu_1}\cdots \Gamma^{\mu_D}
\end{align}
where the coefficients $x_{\mu_1\cdots\mu_r}\in \mathbb{F}$ $\forall ~0\leq r\leq D$. There exists an automorphism called the grade involution $(\, \Hat{x}\, )$ defined as
\begin{align}
	\hat{\Gamma}^{(r)}= (-1)^{r} \Gamma^{(r)}
\end{align}
Under the grade involution, any Clifford algebra has a $\mathbb{Z}_2$ grading
\begin{align}
	Cl(p,q)\simeq Cl^{(0)}(p,q)\oplus Cl^{(1)}(p,q)
\end{align}
where $Cl^{(0)}(p,q)$ forms a sub-Clifford-algebra which consists of all the even tensors. The $Cl^{(1)}(p,q)$ contains all the odd tensors of $Cl(p,q)$ but doesn't form a subalgebra.

For vector spaces based on the real fields i.e. $\mathbb{F}=\mathbb{R}$, the Clifford algebras are called real Clifford algebras. There's a mod $8$ isomorphism between the real Clifford algebras and the matrix algebras over $\mathr,\mathc$, and $\mathh$. The real quadratic spaces $V=\mathr^{p,q}$ have the quadratic form given by
\begin{align}
	Q(\vec{v})=v_1^2+\cdots +v_p^2-v_{p+1}^2-\cdots-v_{p+q}^2,\,\forall \, \vec{v}\in \mathr^{p,q}
\end{align}
For arbitrary $p$ and $q$, these isomorphism are summarized in table \eqref{tab:RealCliffordalgebra}, where $\mathr[N],\mathc[N],$ and $\mathh[N]$ denotes that the matrix algebras of $N\times N$ matrices with real, complex, and quaternionic entries.

\renewcommand{\arraystretch}{2}
\begin{table}[H]
	\centering
	\begin{tabular}{ ||c||c|c|c|c|c|c|c|c|| }
		\hline
		\hline
		$(p$ - $q)$ mod $8$ & $0$         & $1$                        & $2$         & $3$         & $4$         & $5$                         & $6$         & $7$         \\ \hline
		$Cl(p,q)$     & $\mathr[N]$ & $\mathr[N]\oplus\mathr[N]$ & $\mathr[N]$ & $\mathc[N]$ & $\mathh[N]$ & $\mathh[N]\oplus \mathh[N]$ & $\mathh[N]$ & $\mathc[N]$ \\
		\hline
		\hline
	\end{tabular}\caption{Real Clifford algebras in different conventions in $D$-spacetime.}	\label{tab:RealCliffordalgebra}
\end{table}
 For $\mathr[N]$ and $\mathc[N]$, $N=2^{\lfloor{\frac{D}{2}}\rfloor}$ whereas for $\mathh[N]$, $N=2^{\lfloor{\frac{D}{2}}\rfloor-1}$. Note that the real Clifford algebras in different signatures need not be isomorphic to each other. For example, $Cl(3,1)\simeq \mathr[4]$ whereas $Cl(1,3)\simeq \mathh[2]$. However, the even subalgebras of \enquote{opposite} signatures are isomorphic to the same matrix subalgebras. for example,  $Cl^{(0)}(3,1)\simeq \mathc[4]$ whereas $Cl^{(0)}(1,3)\simeq \mathc[4]$. \footnote{It should be emphasized that this doesn't imply an isomorphism between real Clifford algebras of opposite signature. Traditionally, Clifford algebras of opposite signatures have been related through Wick rotation of Clifford generators which is not an isomorphism. In the literature, there also exists various other techniques. For example: tilt of the opposite metric.}

The isomorphism between even real Clifford algebras and the matrix algebras are summarized in table \eqref{tab:EvenRealCliffordalgebra}.
\renewcommand{\arraystretch}{1.5}
\begin{table}[H]
	\centering
	\begin{tabular}{ ||c||c|c|c|c|c|c|c|c|| }
		\hline
		\hline
		$p-q$ mod $8$   & $0$                        & $1$         & $2$         & $3$         & $4$                         & $5$         & $6$         & $7$         \\ \hline
		$Cl^{(0)}(p,q)$ & $\mathr[N]\oplus\mathr[N]$ & $\mathr[N]$ & $\mathc[N]$ & $\mathh[N]$ & $\mathh[N]\oplus \mathh[N]$ & $\mathh[N]$ & $\mathc[N]$ & $\mathr[N]$ \\
		\hline
		\hline
	\end{tabular}\caption{Real Clifford algebras in different conventions in $D$-spacetime.}	\label{tab:EvenRealCliffordalgebra}
\end{table}
For $\mathr[N]$ and $\mathc[N]$, $N=2^{\lfloor{\frac{D}{2}}\rfloor}$ whereas for $\mathh[N]$, $N=2^{\lfloor{\frac{D}{2}}\rfloor-1}$. The complex Clifford algebras on the other hand are based on the complex fields $\mathbb{F}=\mathbb{C}$. The complex quadratic spaces $V=\mathc^{p+q}$ have the quadratic form
\begin{align}
	Q(z)=z_1^2+z_{2}^2+\cdots+z_{p+q}^2
\end{align}
Unlike real Clifford algebras, the complex Clifford algebras are independent of the metric signature and are therefore denoted as $\mathc l(D)$. The complex Clifford algebras have mod $2$ isomorphism with the complex matrix algebras which are summarized in table \eqref{tab:ComplexCliffordalgebra}.
\renewcommand{\arraystretch}{1.5}
\begin{table}[H]
	\centering
	\begin{tabular}{ ||c||c|c|| }
		\hline
		\hline
		$D$ mod $2$   & $0$         & $1$                        \\ \hline
		$\mathc l(D)$ & $\mathc[N]$ & $\mathc[N]\oplus\mathc[N]$ \\
		\hline
		\hline
	\end{tabular}\caption{Complex Clifford algebras in different conventions in $D$-spacetime. Here $N=2^{\lfloor{\frac{D}{2}}\rfloor}$.}	\label{tab:ComplexCliffordalgebra}
\end{table}
One easier way to generate complex Clifford algebras is by the complexification of real Clifford algebras
\begin{align}
	\mathc l(p+q)\simeq Cl(p,q)\otimes_\mathr \mathc
\end{align}
 This can be easily checked using the following matrix algebra isomorphism $	\mathr[N] \otimes_\mathr \mathc \simeq\mathc[N],\, 	\mathc[N] \otimes_\mathr \mathc \simeq\mathc[N]\oplus \mathc[N],\,	\mathh[N] \otimes_\mathr \mathc \simeq\mathc[2N]$.
The even subalgebra $\mathc l^0(D)$ are isomorphic to $ \mathc l(D-1)$.
\subsection{Pinors and Spinors}
All the invertible elements of the Clifford algebra forms a subgroup known as the Lipschitz group. (see \cite{lounesto2001clifford})
\begin{align}
	Lip(p,q)\equiv \{x\in Cl(p,q)| \, \forall \, \vec{v}
	\in \mathbb{R}^{p,q}, \, x\, \vec{v} \,\hat{x}^{-1}\in \mathr^{p,q}\}
\end{align}
where $\hat{x}$ is the grade involution of the element $x\in Cl(p,q)$.
 There are two important subgroups of the real Lipschitz group known as the real $Pin$ and the real $Spin$ groups.
In general for any metric $\eta^{\mu\nu}$ over a field $\mathbb{F}$, the pinors are the representations of the $Pin$ group, whereas spinors are the representations of the $Spin$ group. 

 The real Pin group is the subgroup of the Lipschitz group such that the invertible element has norm $\pm 1$. The real spinor group consist of all the even rank elements of the Lipschitz group with norm $\pm 1 $.
\begin{align}
	Pin(p,q)  & =\{x\in Lip(p,q)|\, x\tilde{x}=\pm1 \}               \\
	Spin(p,q) & =\{x_{(2r)}\in Lip(p,q)|\, x\tilde{x}=\pm1 \}
\end{align}
where $\tilde{x}$ is defined as the reversion of the element $x$ under which any rank $r$ tensor of the Clifford algebra picks up a factor $(-1)^{\frac{r(r-1)}{2}}$.

However, the Dirac pinors and spinors are based on complex Clifford algebras whose Lipschitz group is defined as
\begin{align}
	\mathbb{L}ip(p+q)\equiv \{x\in \mathc l(p,q)| \, \forall \, z\in \mathbb{C}^{p+q}, \, xz \hat{x}^{-1}\in \mathc^{p+q}\}
\end{align}
The complex pinor group is defined as the subgroup of $\mathbb{L}ip(p+q)$ such that the invertible elements have norm $+1$. Whereas, the complex spin group consist of all the even rank elements of $\mathbb{L}ip(p+q)$ with norm $+1$.
\begin{align}
	\mathbb{P}in(p+q)  & =\{x\in \mathbb{L}ip(p+q)|\,x\tilde{x}=+1 \}               \\
	\mathbb{S}pin(p+q) & =\{x_{(2r)}\in \mathbb{L}ip(p+q)|\,x\tilde{x}=+1 \}
\end{align}
The Dirac pinors are the representations of the complex Pin group $\mathbb{P}in(p+q)$, whereas the Dirac spinors are the representations of the complex Spin group $\mathbb{S}pin(p+q)$.
\vspace{-1em}
\section{Charge Conjugation and Complex conjugation matrices}\label{App:CB}
In this appendix, we write down the properties of the charge conjugation matrix $\mathcal{C}$ and the complex conjugation matrix $\mathcal{B}$ which are defined as follows
\begin{align}
	\mathcal{C}d(\Lambda)\mathcal{C}^{-1}\equiv d(\Lambda^{-1})^T\qquad :\qquad \mathcal{B}d(\Lambda )\mathcal{B}^{-1}\equiv d(\Lambda)^*
\end{align}
i.e. the matrix $\mathcal{C}$ maps the spinor representation to it's inverse-transpose whereas the matrix $\mathcal{B}$ maps the representation to it's complex conjugate. 
In Euclidean signature, the finite dimensional spinor representations $d(\Lambda)$ are unitary: $d(\Lambda^{-1})^T=d(\Lambda)^*$. Therefore, the complex conjugation and the charge conjugation matrices are same. For non unitary representations i.e. in Minkowski spacetime, $\mathcal{C}$ and $\mathcal{B}$ are related by the parity matrix $\beta=\iimg \Gamma^0$ which maps the representation to it's inverse-hermitian adjoint
\begin{align}
	\beta d(\Lambda)\beta^{-1}=d(\Lambda^{-1})^\dagger\qquad \text{such that} \qquad \mathcal{B}=t_0\mathcal{C}\beta
\end{align}
Since we are working in the mostly $+ve$ signature, $\beta$ is hermitian. In this paper, we work with the conventions followed in \cite{Freedman:2012zz}. The symmetric properties of unitary matrices $\mathcal{C}$ and $\mathcal{B}$ are controlled by the parameters $t_0$ and $t_1$ such that
\begin{align}
	\mathcal{C}^T=-t_0\mathcal{C}\qquad,\qquad 	\mathcal{B}^T=-t_1\mathcal{B}
\end{align}
The charge conjugation matrix maps the $\Gamma^{(r)}$ to it's transpose such that
\begin{align}
	(\mathcal{C}\Gamma^{(r)})^T=-t_r \mathcal{C}\Gamma^{(r)}\label{Cproperty}
\end{align}
which implies that
\begin{align}
	t_{2n}=(-1)^{n}\, t_0\qquad,\qquad t_{2n+1}=(-1)^{n}\, t_1 \quad ,\qquad n\in \mathbb{Z}\label{t_0t_1}
\end{align}
 Using this, we can write down the symmetric and conjugate properties of $\Gamma^\mu$ as
\begin{align}
	\mathcal{C}\Gamma^\mu\mathcal{C}^{-1}=t_0t_1(\Gamma^\mu)^T\qquad ,\qquad 	\mathcal{B}\Gamma^\mu\mathcal{B}^{-1}=-t_0t_1(\Gamma^\mu)^*\label{CBproperty}
\end{align}
In Minkowski spacetime with mostly $+ve$ signature the $t_0,t_1$ values have a mod $8$ classification which can be arranged over the Clifford clock\cite{BaezClock}.

In even dimensions there can be other possibilities for $t_0$ or $t_1$, see \cite{Freedman:2012zz}. For the purpose of this paper we choose to work with these set of $t_0,\, t_1$ values. Under the reflection along the $t_0$ axis (horizontal axis), the same clock gives the $t_0,t_1$ in the mostly $-ve$ signature.

 %\providecommand{\href}[2]{#2}
%\addcontentsline{toc}{section}{References}
%\bibliographystyle{JHEP}
\bibliographystyle{utphys}
\bibliography{diracbiblio} 

\end{document}